\documentclass[aps,apl,twocolumn,showpacs]{revtex4}
\usepackage{graphicx}
\usepackage{hyperref}

\begin{document}

\title{Spin-polarized quasiparticle  control in a double spin-filter tunnel junction}

\author{P. K. Muduli}\email{muduli.ps@gmail.com}
\affiliation{Department of Materials Science and
Metallurgy,University of Cambridge, 27 Charles Babbage Road,
Cambridge CB3 0FS,United Kingdom}

\date{\today}
\begin{abstract}
Spin-polarized quasiparticles can be easily created during
spin-filtering through a ferromagnetic insulator (FI) in contact
with a  superconductor due to pair breaking effects at the
interface. A combination FI-N-FI sandwiched between two
superconductors can be used to create and analyze such
spin-polarized quasiparticles through their nonequilibrium
accumulation in the middle metallic (N) layer. We report
spin-polarized quasiparticle regulation in a double spin-filter
tunnel junction in the configuration NbN-GdN1-Ti-GdN2-NbN. The
middle Ti layer provides magnetic decoupling between two
ferromagnetic GdN and a place for nonequilibrium quasiparticle
accumulation. The two GdN(1,2) layers were deposited under
different conditions to introduce coercive contrast. The
quasiparticle tunneling spectra has been measured at different
temperatures to understand the tunneling mechanism in these double
spin-filter junctions. The conductance spectra were found to be
comparable to an asymmetric SINI'S-type tunnel junction. A
hysteretic R-H loop with higher resistance for the antiparallel
configuration compared to parallel state was observed asserting
the spin-polarized nature of quasiparticles. The hysteresis in the
R-H loop was found to disappear for sub-gap bias current. This
difference can be understood by considering suppression of the
interlayer coupling due to nonequilibrium spin-polarized
quasiparticle accumulation in the Ti layer.
\end{abstract}
\maketitle
\clearpage
\section{Introduction}

In superconductors, below the critical temperature $T_C$ the
electrons with opposite momentum and spin are bound in (singlet)
Cooper pairs, therefore, they can transport only charge but not
spin. At finite temperature a fraction of Cooper pairs is broken
into excited states called (Bogoliubov) quasiparticles which is
capable of transporting both charge and spin. Quasiparticles can
be created inside a superconductor while injecting current through
a tunnel barrier or by irradiating electromagnetic radiation with
energy, $h\nu
>>\Delta$, where $\nu$ is the frequency of radiation and $\Delta$ is
the  superconducting energy gap (binding energy of Cooper
pairs)\cite{wolf-book}. Eventually with time the quasiparticles
recombine to from Cooper pairs after emitting a phonon maintaining
equilibrium. In the presence of extra disturbances quasiparticle
concentration can be increased and driven out of equilibrium which
follow non-Fermi Dirac distribution function. The number and
dynamics of these nonequilibrium quasiparticles has been the
subject of intense research lately as they are primary source of
decoherence in almost all superconducting
electronics\cite{Martinis}. However, these nonequilibrium
quasiparticles can be very advantageous for spintronics purposes
as they have very large mean free path ($\lambda_Q$) compared to
ordinary electrons\cite{vasenko}.

\emph{Quasiparticle spintronics} is not new and has been there
since 1970s. Meservay and Tedrow have shown that spin polarization
of various ferromagnets can be determined by injecting
spin-polarized quasiparticles from a ferromagnet into a
superconductor\cite{tedrew}. Recently, quasiparticle spintronics
have got renewed interest and most of the study has been focused
on spin transport inside superconductors through quasiparticle
excitations\cite{Yang,hubler,Quay,Chevallier,poli,Wakamura-prl}.
It is now believed that spin and charge are transported by
separate quasiparticle excitations in a superconductor\cite{Quay,
Kivelson}. Signatures of spin transport over distances up to
several $\mu$m has been observed in Zeeman split superconductors
in proximity to a ferromgnetic
insulator\cite{Wolf,mason,bobkova,Krishtop,virtanen}. Many
spintronics phenomena like quasiparticle mediated spin Hall effect
(SHE)\cite{wakamura,takahasi}, Seebeck effect induced by
spin-polarized quasiparticles\cite{kolenda}, quasiparticle spin
resonance\cite{Quay,Quay-prb}, etc., has been experimentally
observed. However, many fundamental aspects of the
\emph{Quasiparticle spintronics} remains poorly understood. Most
interesting prospect of quasiparticle spintronics would be to
explore possibility to take quasiparticles out of superconductor
into a normal metal and introduce spintronics functionality. One
obvious system for this type of study is double spin-filter device
of the type S-FI-N-FI-S (Here FI is ferromagnetic insulator, N is
normal metal and S is the superconductor) which is analogous to
conventional SINIS-type devices\cite{miao-natcom,Yang}. Double
barrier superconducting tunnel junctions of the
S-I-N(\emph{s})-I-S structure have been extensively studied to
cool down the electron in the normal metal (N) from 300 to 100 mK
or to enhance superconductivity in the middle \emph{s}
layer\cite{leivo,Chaudhuri,pekola}. Operation of these devices are
based on the modification of the quasiparticle distribution
function in the N region of the junction which can have non-Fermi
Dirac form leading to measurable out come. Blamire \emph{et. al.}
have observed enhancement in the superconductivity of Al  up to 4
K in a symmetric Nb-AlO$_x$-Al-AlO$_x$-Nb double barrier
junction\cite{blamire-prl}. Enhancing superconductivity by means
of nonequilibrium effects got substantial theoretical interest but
still remains controversial
experimentally\cite{heslinga,Nevirkovets,Klapwijk}.

Spin-filter tunnel junction comprising superconductors produces a
great amount of spin-polarized quasiparticles by enforcing Cooper
pairs to split while tunneling\cite{Tokuyasu}. Therefore double
spin-filter devices of the type S-FI-N-FI-S provide unique
opportunity to explore quasiparticle spintronics through
nonequilibrium quasiparticle accumulation in the middle N
layer\cite{gxmiao,kawabata}. In this kind of devices when the two
spin-filter layers are parallel to each other no spin accumulation
happens as the number of injected spin-up electrons in the N layer
is same as the number of spin-up electrons leaving it. Whereas in
the antiparallel case finite nonequilibrium  spin accumulation in
the middle layer is expected which relaxes through spin-flip
processes. Double spin-filter device with
superconductivity-induced nonequilibrium has been predicted to
show huge TMR $\sim$10$^2$-10$^6$$\%$ which can be tuned with
biasing voltage and temperature\cite{giazotto, Worledge}.

In this paper, we report fabrication of double spin-filter devices
in which a metallic Ti layer is symmetrically connected to two
identical superconductors through ferromagnetic (GdN) tunnel
barriers. We present  quasiparticle tunneling spectra measurements
on the NbN-GdN-NbN, NbN-Ti-GdN-NbN and NbN-GdN1-Ti-GdN2-NbN tunnel
junctions measured at different temperatures. We explore
possibility of creating nonequilibrium quasiparticle accumulation
in the Ti layer and its effect on the magnetic coupling between
the two GdN layers. The R-H loops of the double spin-filter tunnel
junctions were measured at different bias currents and temperature
to explore these effects.

\section{Experimental}
Multilayer structures NbN-GdN1-Ti-GdN2-NbN were grown by DC
sputtering in an ultrahigh vacuum (UHV) chamber at room
temperature. The NbN and GdN layers were deposited under similar
conditions as described in the
references\cite{kartikprb,senapati,muduli,muduli-arxive,muduli-Co}.
It have been observed that the magnetic and electrical property of
GdN is sensitive to deposition condition and can be tuned by
changing different Ar and N$_2$ gas mixture and deposition
power\cite{kartikprb}. The two GdN(1,2) layers were grown with
different gas mixture in order to introduce coercive contrast. The
GdN1 and GdN2 layers were deposited with 8 $\%$ and 4$\%$ Ar -
N$_2$ gas mixture, respectively. The Ti layer was grown in a pure
Ar gas environment with a pressure 1.5 Pa and sputtering power of
40 W. The thickness of top and bottom NbN layers were kept fixed
at 50 nm while thickness of GdN and Ti layers were varied in
different depositions. Eight multilayer stack with different
thickness of Ti were grown in the same deposition in the sequence
NbN-GdN1-Ti-GdN2-NbN from left to right.

The double junctions were fabricated in a mesa structure in which
junction area (7 $\mu$m $\times$ 7 $\mu$m ) was defined by CF$_4$
plasma etching and Ar-ion milling. The fabrication process is
similar to described in the references\cite{muduli} except these
devices were Ar-ion milled for 14 min instead of 4 min to ensure
complete milling of Ti till bottom NbN layer. Fig. 2(d) shows
schematic of the double tunnel junction in the mesa structure with
measurement scheme. The electrical characterization of the devices
up to 4.2 K were done in a custom made dip-stick. For 300 mK
measurements a He-3 sorption insert form \emph{Cryogenics Lmt.}
was used. The differential conductance $dI/dV$ of the junctions at
300 mK were obtained by numerically differentiating measured I-V
curves. Conductance spectra at 4.2 K were obtained with standard
lock-in technique. The $R-H$-loops were measured with a DC current
source and nanovoltmeter. In this report we show the results of
one representative double junction. Measurements done on other
junctions on the same chip and devices with different thickness of
Ti are shown in the supplementary material. All the data reported
in the manuscript were found to be extremely reproducible as shown
in supplementary figures.

\section{Results and discussion}
\begin{figure}[!h]
\begin{center}
\abovecaptionskip -10cm
\includegraphics [width=8 cm]{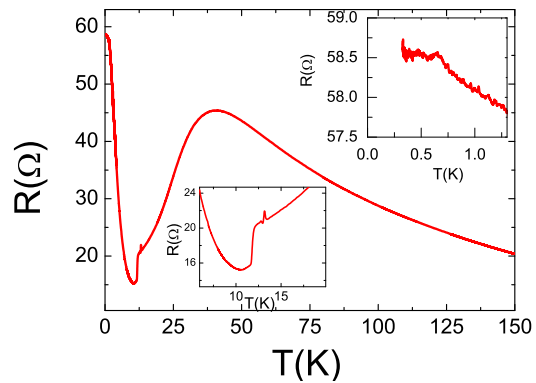}
\end{center}
\caption{\label{fig1} (Color online) Temperature dependence of
resistance of the NbN(50nm) - GdN(2nm-8$\%$) - Ti(8nm) -
GdN(2nm-4$\%$) - NbN(50 nm) double spin-filter tunnel junction.
The measurement was done using a current $I$ = 10 $\mu$A. The
upper inset shows $R(T)$ in the range 0.3 to 1.3 K. Lower inset
shows $R(T)$  close to $T_C$.}
\end{figure}

Figure 1 shows temperature dependence of resistance of a double
spin-filter tunnel junction with 8 nm thick Ti spacer. A
semiconducting behavior can be seen till 35 K and metallic-like
behavior below it due to onset of spin filtering at the Curie
temperature, $T_{Curie}$ $\approx$35 K of GdN layers. The $R(T)$
is similar to a single NbN-GdN-NbN spin-filter tunnel
junction\cite{senapati,pal}. The superconducting transition of NbN
can be seen to start at $T_C$ $\sim$ 13 K. The transition was
found to be broad with a width of $\sim$1.7 K as shown in the
lower inset of Fig. 1. This is due to the difference in the $T_C$
of top and bottom NbN in the double tunnel junction. The $R(T)$ of
some other double tunnel junctions are shown in the supplementary
material (SFig. 3). For measurements done with a bias voltage
smaller than gap voltage, i.e., $eV < 2\Delta$, the resistance was
found increase rapidly below $T_C$ of NbN. The electrical
transport below $T_C$ is determined by quasiparticles. For bias
voltage in the sub-gap region the tunneling current is weakly
dependent on bias voltage and scales with temperature dependent
quasiparticle density $ n(T) \propto \sqrt T e^{ - \frac{\Delta
}{{k_B T}}} $\cite{Muhonen}. Therefore, temperature dependence of
sub-gap resistance follows an exponential dependence, $R(T)
\propto e^{ - \Delta /k_B T}$, with a constant parallel leakage
resistance\cite{blamire}. The upper inset in Fig. 1 shows $R(T)$
in the range 1.3 to 0.3 K. Bulk Ti is known to be a superconductor
with $T_C$ $\sim$0.49 K. However, we could not observe any
superconducting transition of Ti in our devices till 0.3 K. This
might be due to large suppression of $T_C$ of the thin Ti layer
sandwiched between two magnetic GdN layers.

\subsection{Tunneling behavior}

A double tunnel junction is essentially made of two tunnel
junction in series. In our NbN-GdN1-Ti-GdN2-NbN double tunnel
junction devices there are  two tunnel junctions NbN-GdN1-Ti (Jn1)
and Ti-GdN2-NbN (Jn2) in series. As the two tunnel junctions are
deposited with opposite sequence they most likely have different
resistances; $R_{Jn1}$ and  $R_{Jn2}$. Besides  NbN-GdN interface
is expected to be more resistive than Ti-GdN interface due to
different Schottky barrier height; $ \Phi _{Sh} = W - E_g^{GdN}$.
Where $E_g^{GdN}$ is the band gap of GdN and $W$ is the work
function of the metal. As work function of NbN $\sim$4.7 eV
\cite{Gotoh}is larger than that of  Ti $\sim$4.3 eV\cite{crc},
NbN-GdN interface have lower transparency than Ti-GdN interface.
Therefore, the double tunnel junction NbN-GdN1-Ti-GdN2-NbN is most
likely to develop asymmetry even with ideal interface without
considering fabrication issues. Traditionally double tunnel
junction with superconductors has been studied with structure
Nb-Al$_2$O$_3$-Al-Al$_2$O$_3$-Nb or  Nb-NbO$_x$-Al-AlO$_x$-Nb
\cite{cassel,blamire-sust}. The Al spacer is most popular due to
its tendency to form high quality pin-hole free native oxide. In
this kind of tunnel junctions AlO$_x$ provides a large barrier
height $\sim$1.7 to 2.5 eV\cite{Cimpoiasu} which makes it possible
to create potential well and observe fascinating effects like
resonant tunneling in double barrier tunnel junctions. But in the
case of GdN the barrier height is usually small $\sim$10-100 meV
\cite{pal}, therefore, more transparent tunnel barrier is
expected.

\begin{figure}[!h]
\begin{center}
\abovecaptionskip -10cm
\includegraphics [width=8 cm]{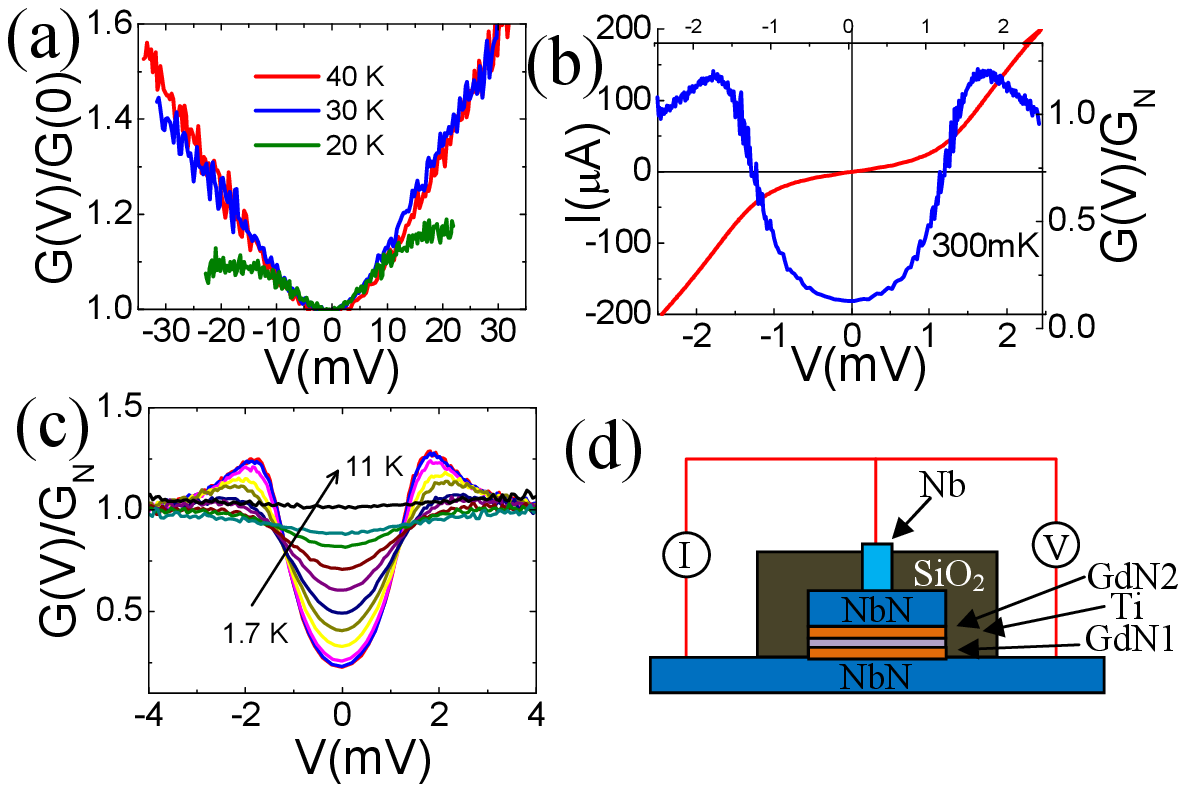}
\end{center}
\caption{\label{fig2} (Color online) (a) Normalized conductance
spectra $G(V)/G(0)$ of the double spin-filter tunnel junction
measured at 40, 30 and 20 K. (b) The I-V and normalized
conductance spectra of the same junction measured at 300 mK. (c)
Normalized conductance spectra of the junction measured in the
temperature range 1.7-11 K. (d) Schematic of the double tunnel
junction in the mesa structure with measurement scheme.}
\end{figure}

The I-V and $dI/dV-V$ measurements were done at different
temperatures to understand tunneling nature of the double
junctions. Fig. 2 shows conductance spectra $G(V)$ (=$dI/dV$)
normalized to its value at $V = 0$ measured at different
temperatures above the $T_C$ of NbN. Parabolic conductance spectra
suggest tunneling type transport in these devices. For the $dI/dV$
measurements at 20 K deviation from parabolic behavior above
$\pm$10 mV is probably due to exchange splitting of the GdN tunnel
barrier below the $T_{Curie}$. A small asymmetry can also be seen
in the conductance spectra which suggest  the two tunnel junctions
involved in the double tunnel junction have different resistances;
$R_{Jn1}$ and $R_{Jn2}$. The conductance spectra of the same
junctions were also measured below $T_C$ of NbN. The normalized
$dI/dV$ spectra measured in the temperature range 1.7 to 11 K are
shown in the Fig. 2(c). Clear appearance of superconducting gap
validate a tunneling type transport in these double tunnel
junctions. Fig. 2(b) shows IV and $dI/dV$ measurement done on the
same junction at 300 mK. Two conductance peaks separated by
4$\Delta$ $\sim$3.3 meV can be observed. The superconducting gap
$\Delta$ of NbN is suppressed along with smearing of gap edges
probably due to magnetic GdN \cite{Giaever}. In SINIS tunnel
junctions nonequilibrium effects usually leads to sub-gap step
structures whose position and amplitude strongly depends on the
temperature\cite{Klapwijk,gazi}. In some cases much sharper
gap-edge structure is considered as an evidence of the
nonequilibrium effects\cite{blamire-sust}. However, none of these
features can be seen in the conductance spectra as shown in Fig.
2(b,c). The reason for this is discussed below.

\begin{widetext}

\begin{center}
\begin{figure}[h*]
\begin{center}
\abovecaptionskip -10cm
\includegraphics [width=12 cm]{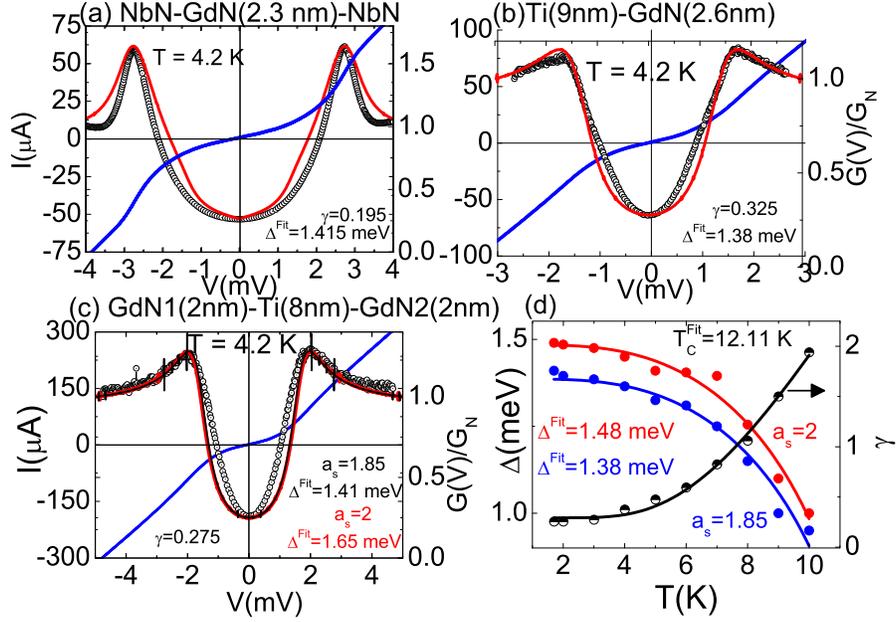}
\end{center}
\caption{\label{fig1} (Color online)(a) I-V and normalized
conductance spectra $G(V)/G_N$ of NbN-GdN(2.3 nm)-NbN tunnel
junction. The red solid line represents fitting to the S-I-S
tunneling model with fitting parameter $\Delta$ = 1.415 meV and
$\gamma$ = 0.195. (b) I-V and normalized conductance spectra
$G(V)/G_N$ of NbN-Ti(9 nm)-GdN(2.6 nm)-NbN tunnel junction. The
red solid line represents fitting to the N-I-S tunneling model
with fitting parameter $\Delta$ = 1.38 meV and $\gamma$ = 0.325.
(c) I-V and normalized conductance spectra $G(V)/G_N$ of
NbN-GdN1(2 nm)-Ti(8 nm)-GdN(2 nm)-NbN tunnel junction. The red
solid line represents fitting to the S-I-N-I-S tunneling model
with asymmetry parameter $a_s$ = 2 (red) and 1.85 (black).(d)
Temperature evolution of the fitting parameters  $\Delta$  and
$\gamma$ found from fitting Eq. (1) to the conductance spectra
shown in Fig. 2(c). The red ($a_s$ = 2) and blue ($a_s$ = 1.85)
solid lines are the fitting to the BCS type temperature
dependence; $ \Delta (T) = \Delta (0)\tanh (1.74\sqrt {(T_C  -
T)/T} ) $ with $T_C$ = 12.11 K. Black solid line is the fitting to
an exponential of the from; $ \gamma \propto e^{ - \zeta/
T}$\cite{note}. }
\end{figure}
\end{center}

Below the $T_C$ of NbN the conductance spectra of our devices can
be understood in terms of an asymmetric SINIS tunnel model with an
asymmetry parameter $a_s = \frac{{2x}}{{1 + x}}$ ($ 1 \le a_s  \le
2$) with $ x = \frac{{R_{Jn1} }}{{R_{Jn2} }}$. Normalized
conductance of an asymmetric SINIS tunnel junction can be written
as\cite{Courtois};

\begin{equation}
\frac{{G(V)}}{{G_N }} = \frac{1}{{a_s
}}\frac{d}{{d(eV)}}\int\limits_{ - \infty }^{ + \infty } {N_S
(E)\left[ {f_N (E - a_s \frac{{eV}}{2}) - f_N (E + a_s
\frac{{eV}}{2})} \right]} dE,
\end{equation}

where $f_N(E)$ is the non-equilibrium distribution function inside
the Ti layer and can be expressed as;

\begin{equation}
f_N (E) = \frac{{N_s (E - a_s \frac{{eV}}{2})f_0 (E - a_s
\frac{{eV}}{2}) + N_s (E - a_s \frac{{eV}}{2})f_0 (E - a_s
\frac{{eV}}{2}) + \frac{{f_0 (E)}}{{\tau _E \Gamma }}}}{{N_s (E -
a_s \frac{{eV}}{2}) + N_s (E - a_s \frac{{eV}}{2}) +
\frac{1}{{\tau _E \Gamma }}}}.
\end{equation}
\end{widetext}

Here $f_0 (E,T) = \frac{1}{{1 + \exp (E/k_B T)}}$ is the
Fermi-Dirac function at temperature $T$.  Superconducting
quasiparticle density of state with Dynes parameter $\gamma$ is
given by $N_S(E) = N(0)\left| {{\mathop{\rm Re}\nolimits} \left(
{\frac{{E/\Delta  - i\gamma }}{{\sqrt {(E/\Delta  - i\gamma )^2  -
1} }}} \right)} \right|$. Here $\gamma$ incorporates finite
life-time of quasiparticles in the superconductor. In Eq. (2),
$\tau _E $ is the relaxation time representing time scale for
interchange of energy between the quasiparticle and the rest of
the system. This energy relaxation rate is determined by
electron-electron interactions, electron-phonon interactions and
magnetic impurities which can induce decoherence. Here $ \Gamma ^{
- 1}$ refers to the mean residency time of quasiparticles inside
Ti layer and is given by; $ \Gamma = \frac{2}{{N_N (E_F )R_N ALe^2
}}$. Here $N_N (E_F )$ is the normalized density of states of Ti,
$R_N$ is the normal state resistance of Ti, $A$ and $L$ are the
cross-section area and length of the normal metal (Ti),
respectively. Clearly, the residency time and thereby influence of
non-equilibrium processes grows with decreasing tunnel junction
volume and tunneling resistance.

Now we discuss conditions for nonequilibrium. The distribution
function of electron inside the normal metal is mainly determined
by the ratio of the relaxation rate and rate of injection the
electron into it.  Usually for $ \tau _E \Gamma  >
> 1$ (injection rate exceed relaxation rate) the distribution
function in the normal metal deviates from the thermal equilibrium
Fermi distribution function $f_0 (E,T)$ and if $ \tau _E \Gamma <<
1$ (equilibrium), the normal metal follows a Fermi distribution
function. The conditions $\tau _E \Gamma \to 0$ and $\tau _E
\Gamma  \to \infty$ correspond to complete equilibrium and
nonequilibrium, respectively.

Although intuitively it seems in low resistance GdN tunnel barrier
the nonequilibrium effects will be enhanced. But when the barrier
transparency is decreased or effective life time of the
quasiparticle in the interlayer is increased, the amount of
quasiparticles that is scattered inelastically also increases due
to magnetic nature of the tunnel
barriers\cite{brinkman,Lemberger,Anthore}. Therefore, driving the
middle Ti layer far from equilibrium in double spin-filter tunnel
junction is not trivial like a SINIS-tunnel junctions with
nonmagnetic elements. Our double spin-filter tunnel junctions can
be reasonably  assumed as a series connection of SIN and NIS
junctions where the energy distribution function in the interlayer
is the equilibrium Fermi distribution function, i.e., $ f_N (E,T)
\approx f_0 (E,T)$.

Figure 3(a,b,c) shows $IV$ and conductance spectra of different
type of tunnel junction measured at 4.2 K in the same experimental
set-up (measured with lock-in technique). In a typical NbN-GdN-NbN
tunnel junction conductance spectra shows a superconducting gap
$\Delta$(4.2 K) $\sim$1.4-1.5 meV depending on the tunnel barrier
thickness and transparency. See supplementary material (SFig. 13)
for conductance spectra of NbN-GdN-NbN tunnel junctions with
different thickness of GdN. Fig. 3(a) shows conductance spectra of
a NbN-GdN (2.3 nm)-NbN tunnel junction measured at 4.2 K. The red
solid line is the fit to the typical SIS tunneling model with
fitting parameter $\Delta = 1.415$ meV and $\gamma$ = 0.195 (See
supplementary material for SIS tunnel model used for fitting).
Fig. 3(b) shows conductance spectra of a NbN-Ti(9 nm)-GdN(2.6
nm)-NbN tunnel junction. As the thickness of the Ti ($\sim$9 nm)
in this type of device is larger than both the superconducting
coherence length $\xi _{NbN}$ $\sim$4.1 nm \cite{muduli-Co} and
$\xi^{N} _{Ti}$ $\sim$3.6 nm\cite{Fritzsch}($\xi^{N} _{Ti}$;
Normal state coherence length of Ti), this type of tunnel junction
can be considered as NIS-type tunnel junction. The red solid line
shows fitting of the NIS tunnel model to the conductance spectra
with fitting parameter $\Delta  = 1.38$ meV and $\gamma$ = 0.325
(see supplementary material for more detailed study of NIS-type
tunnel junctions)\cite{Giaever}. Fig. 3(c) shows conductance
spectra of the  double tunnel junction NbN-GdN1(2 nm)-Ti(8
nm)-GdN2(2 nm)-NbN. The conductance spectra looks more like a
NIS-type tunnel junction. The red and black solid lines show
fitting to Eq. (1) with asymmetry parameter $a_s$ =2 and $a_s$
=1.85, respectively. Note that the asymmetry parameter is limited
to have values in the range $ 1 \le a_s  \le 2$. For $a_s = 1$,
Eq. (1) correspond to a symmetric SINIS tunnel junction while for
$a_s  = 2$ it reduces to a single SIN tunnel junction. Fitting
Eq.(1) with asymmetry parameter $a_s$ =2 to the conductance
spectra shown in Fig. 3(c), gives $\Delta = 1.65$ meV. This  is
much larger than the value of $\Delta$ found from SIS and NIS-type
tunnel junction as shown in Fig.3 (a,b). Therefore our double
tunnel junction is not a single NIS-type and most likely a SINIS
type double tunnel junction with a large asymmetry. The black
solid line shows fitting to Eq. (1) with $a_s$ = 1.85. This gives
$\Delta = 1.41$ meV which is more reasonable. Note that $a_s$ =
1.85 means the resistance ratio between the two tunnel junctions;
$ R_{Jn1} /R_{Jn2} \sim12.3 $. This can easily happen considering
different deposition condition. Fig. 3(d) shows temperature
dependence of $\Delta$ and $\gamma$ obtained from fitting
conductance spectra measured at different temperature shown in
Fig. 2(c). The conductance spectra was fitted for two asymmetry
parameter $a_s$ = 1.85 (blue) and $a_s$ = 2 (red) with same
smearing parameter $\gamma$. The red and blue solid lines show
fitting to BCS type temperature dependence; $ \Delta (T) = \Delta
(0)\tanh (1.74\sqrt {(T_C  - T)/T} ) $. For both the asymmetry
parameter $T_C$ = 12.11 K was found. The smearing parameter
$\gamma$ was found to decrease rapidly with temperature. Black
solid line shows fitting to a exponential decay; $ \gamma \propto
e^{ - \zeta/ T}$, where $\zeta$ is the decay constant.

\subsection{Spin-valve behavior}

In a SINIS-type tunnel junction when the bias voltage $eV$ exceeds
2$\Delta$, quasiparticle current is produced from the energy
gained primarily from applied bias voltage. Besides even for
voltages less than 2$\Delta$ at a finite temperature thermally
excited quasiparticles above the gap are present whose number
exponentially reduces as temperature is lowered below $T<<T_C$.
However, in a double spin-filter tunnel junction additional
spin-polarized quasiparticles are present due to pair breaking
processes which equally populate electron and hole-like excitation
spectrum\cite{Tokuyasu}. In double spin-filter tunnel junction the
spin-polarized quasiparticle current can be turned ON and OFF by
reorienting magnetization of the two spin-filter barrier parallel
and antiparallel with resect to each other, respectively.

\begin{widetext}

\begin{center}

\begin{figure}[h*]
\begin{center}
\abovecaptionskip -10cm
\includegraphics [width=11 cm]{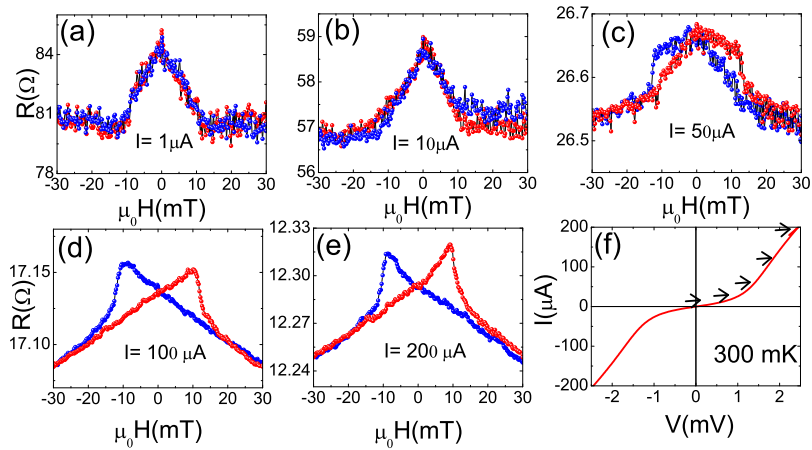}
\end{center}
\caption{\label{fig1} (Color online)(c) R-H loops measured with
different bias current (a) 1 $\mu$A (b) 10 $\mu$A(c) 50
$\mu$A(d)100 $\mu$A, and (e) 200$\mu$A. (f) The I-V curve of the
junction at the same temperature (T = 300 mK). The arrows indicate
different points in the I-V curve where R-H loop was measured.
Gradual disappearance of the hysteresis in the R-H loop can be
seen as bias current is reduced to below sub-gap value.  }
\end{figure}
\end{center}

\end{widetext}

The presence of the superconducting gap in the conductance spectra
induces an energy selectivity of quasiparticle tunneling.
Therefore, in double spin-filter devices the number and the energy
of quasiparticles can be drastically altered if a bias voltage
above and below the gap is applied. The R-H loops measured above
and below the gap voltage can provide valuable information about
non-equilibrium quasiparticle accumulation in the middle metallic
(Ti) layer\cite{Luo}. Fig. 4(a-e) shows R-H loops of the double
spin-filter tunnel junction measured at 300 mK with different bias
current. The I-V curve measured at the same temperature is shown
in the Fig. 4(f). The conductance spectra gap edges can be seen at
$\sim$1.68 meV which corresponds to a bias current $I = 99 \mu$A.
At bias current $I = 200 \mu$A a clear hysteretic R-H loop with
resistance peaks near $\pm$10 mT can be observed. As coercive
field of a single GdN layer is typically $\sim$5
mT\cite{senapati,zhu}, the hysteretic R-H loop observed in these
double tunnel junctions is due to relative magnetization
orientation of the two GdN(1,2) layers. A broad switching is
observed in this case due to multi-domain nature of GdN layers.
One striking thing to note is that the hysteresis in the R-H loops
was found to disappear as current is decreased from 200 to 1
$\mu$A. However, an overall high resistance state can be seen in
the magnetic field range $\pm$15 mT when the two magnetic GdN(1,2)
layers are not parallel to each other. The number of charge
carriers (quasiparticles) that can transport charge through the
S-FI-N-FI-S structure is reduced  drastically as the bias voltage
is reduced below gap-voltage. This can be seen as increase in the
resistance of the double spin-filter tunnel junction from 12 to 81
$\Omega$ as bias current is reduced from 200 to 1 $\mu$A. The R-H
loops were also measured at different temperatures and similar
behavior was found at all temperatures below $T_C$ of NbN. The
hysteresis in the R-H loop was found to disappear above 15 K (See
supplementary figure SFig. 4). Although, $T_{Curie}$ of GdN
$\sim$35 K absence of hysteretic R-H loop above 15 K suggest
absence of well established parallel and antiparallel state.  A
linear decrease in resistance with magnetic field can still be
observed above 15 K confirming magnetic nature of individual GdN
layers above 15 K.

The switching behavior can be understood by considering
spin-polarized quasiparticle accumulation and relaxation inside
the Ti layer. A finite spin-polarized quasiparticle accumulation
is expected inside Ti layer when the two GdN(1,2) layers are
antiparallel  to each other. Therefore conductance is reduced and
the resistance for the antiparallel state is expected to be higher
than that for the parallel configuration. Also spin-polarized
quasiparticle accumulation can modify interlayer exchange
coupling. The absence of hysteretic R-H loop at sub-gap bias
current is most likely due to the suppression of interlayer
exchange coupling between two GdN(1,2) layers. This is expected as
magnetic coupling is usually suppressed in F-S-F trilayer system
below the critical temperature $T_C$ of the
superconductor\cite{sipr,melo}. Suppressed magnetic coupling has
been observed in Fe$_4$N-NbN-Fe$_4$N\cite{matson}, (100)-oriented
GdN/W/NbN/W multilayers\cite{osgood} and GdN-NbN-GdN
trilayers\cite{kartikapl}. Recently, a different kind of
interlayer exchange coupling mechanism in GdN-Nb-GdN has been
proposed\cite{zhu}. Interlayer exchange coupling between
ferromagnetic metallic layers separated by superconducting spacer
has been investigated extensively in many systems and a detailed
discussion is beyond the scope of this
paper\cite{melo,rcb,sipr,halterman,Moraru}. A more detailed
experimental study with different thickness of the normal-metal
spacer is needed to understand the interlayer exchange mechanism
in presence of nonequilibrium quasiparticles in these double
spin-filter tunnel junctions.

\section{Conclusions}

In conclusion, we have fabricated  double spin-filter tunnel
junction in the configuration NbN-GdN1-Ti-GdN2-NbN. The
conductance spectra in these double spin-filter tunnel junctions
were found to be analogous to a highly asymmetric SINIS-type
tunnel junction. We have demonstrated spin-polarized quasiparticle
control in these double spin-filter tunnel junction with R-H
measurements done at different bias voltage above and below gap
voltage $eV = 2\Delta$. Hysteresis in the R-H loop was found to be
absent for sub-gap bias current.  Absence of hysteresis in R-H
loop may be considered as an experimental signature of
non-equilibrium spin-polarized quasiparticle accumulation.
Although nonequilibrium effects cannot be inferred conclusively
from these experiments, these preliminary experimental results are
of fundamental importance and calls for further experimental and
theoretical investigation. Magnetic manipulation of quasiparticles
is pivotal for the advancement of \emph{quasiparticle
spintronics}\cite{linder}.

\noindent  \textbf{Acknowledgments}\\
The idea of double spin-filter tunnel junction based on GdN is a
part of the proposal: ERC Advanced Investigator Grant SUPERSPIN.
Experimental data presented in this manuscript was collected by
PKM during June 2012 to April 2015.  PKM acknowledge Dr David
Gustafsson for assistance during 300 mK measurement.

\newpage
\setcounter{figure}{0}
\renewcommand{\figurename}{SFig.}
\begin{widetext}
\begin{center}
\textbf{\huge\underline{Supplementary Information}}
\end{center}
\begin{center}
{\large\emph{Spin-polarized quasiparticle  control in a double
spin-filter tunnel junction}

P. K. Muduli}

\emph{Department of Materials Science and Metallurgy,University of
Cambridge, 27 Charles Babbage Road, Cambridge CB3 0FS,United
Kingdom}
\end{center}

\begin{widetext}
\begin{center}
\textbf{CONTENTS}
\end{center}

Series of tunnel junctions with different configuration were
fabricated from multilayer stacks NbN-GdN-NbN, NbN-Ti-GdN-NbN and
NbN-GdN1-Ti-GdN2-NbN. The thickness variation in different stacks
was achieved by controlling rotation speed of the sample stage
during deposition. Below a detailed summary of all measurements
done with these tunnel junctions are shown.

\begin{tabular}{ l l r }
\hline
  No & Contents & Page\\
\hline
  1 & Tunneling spectra of NbN-GdN1-Ti(t)-GdN2-NbN tunnel junctions & 9\\
  2 & The R(T) of NbN-GdN1-Ti(t)-GdN2-NbN tunnel junctions & 11\\
  3 & The R-H loops of NbN-GdN1-Ti(8 nm)-GdN2-NbN tunnel junction at 2,5,10 and 15 K & 12\\
  4 & Reproducibility of spin-valve behavior in NbN-GdN1-Ti(t)-GdN2-NbN tunnel junctions & 13 \\
  5 & Tunneling spectra of  NbN-Ti-GdN(t)-NbN tunnel junctions & 14\\
  6 & Tunneling spectra of NbN-GdN(t)-NbN tunnel junctions & 16\\
\hline

\end{tabular}

\newpage
\begin{center}
\textbf{NbN-GdN1-Ti(t)-GdN2-NbN tunnel junction}
\end{center}
\begin{figure}[!h]
\begin{tabular}{ll}
  \centering
  \includegraphics[width= 6 cm]{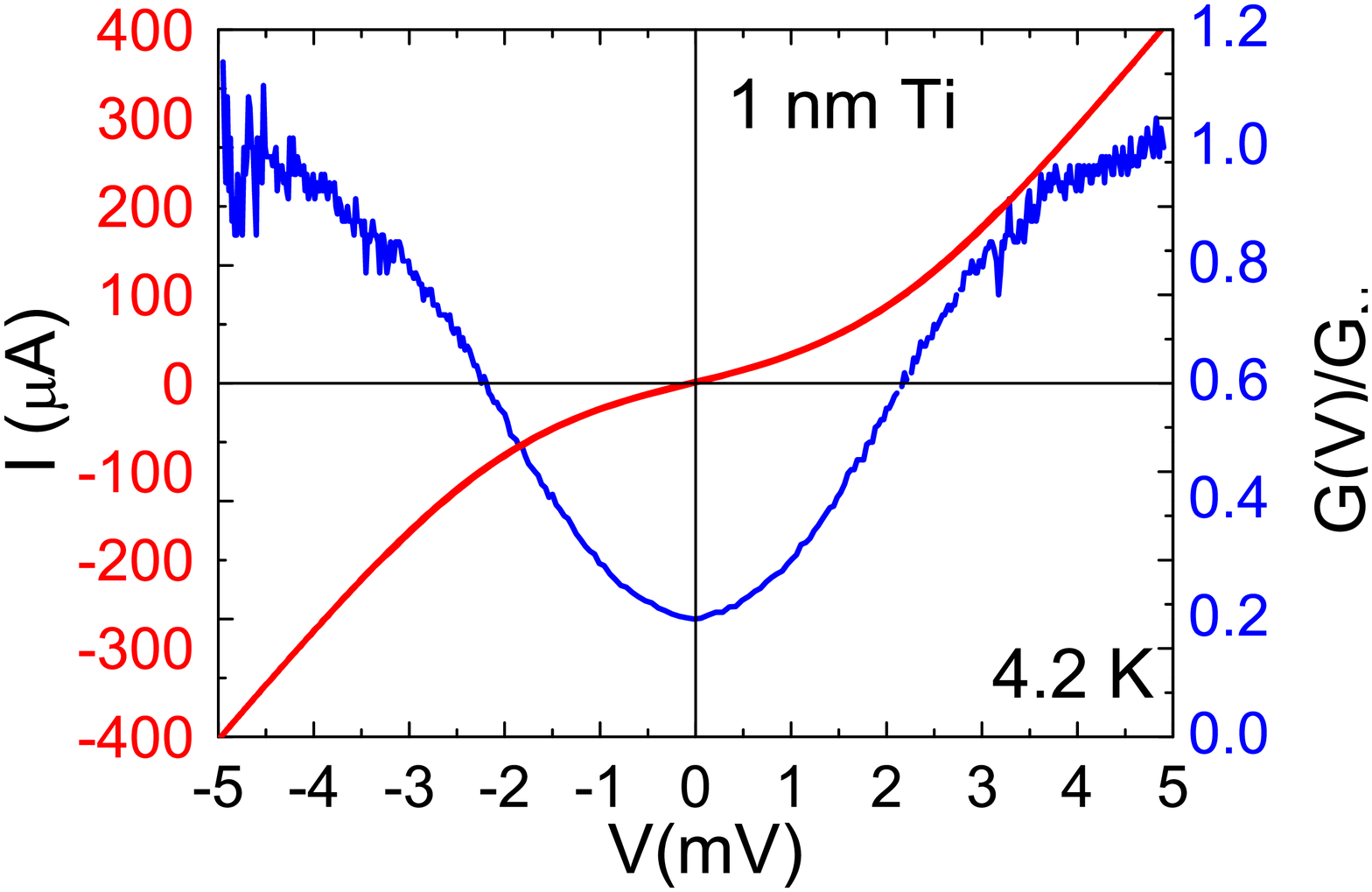}
&

  \includegraphics[width= 6 cm]{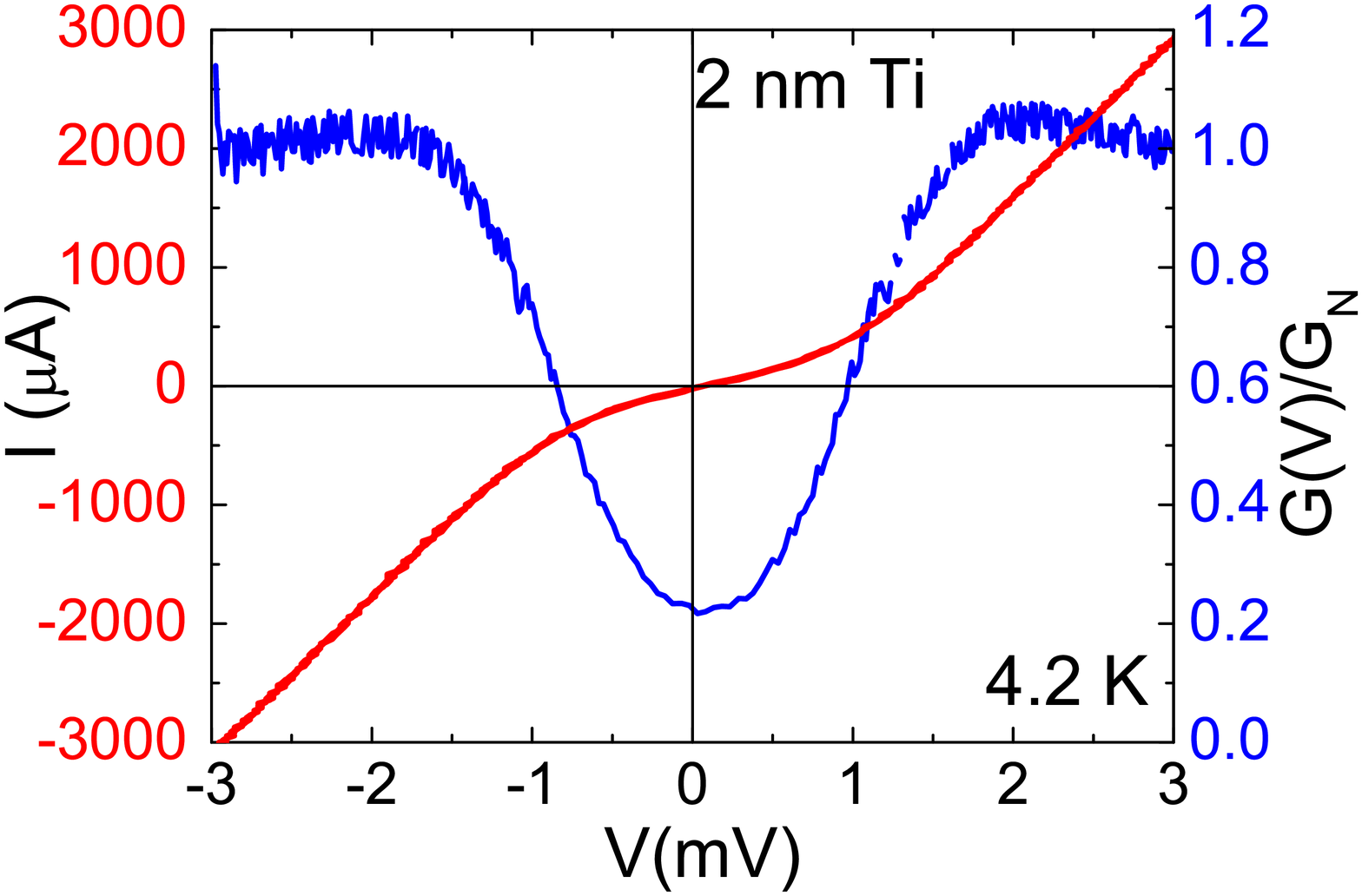}\\
 \centering
 \includegraphics[width= 6 cm]{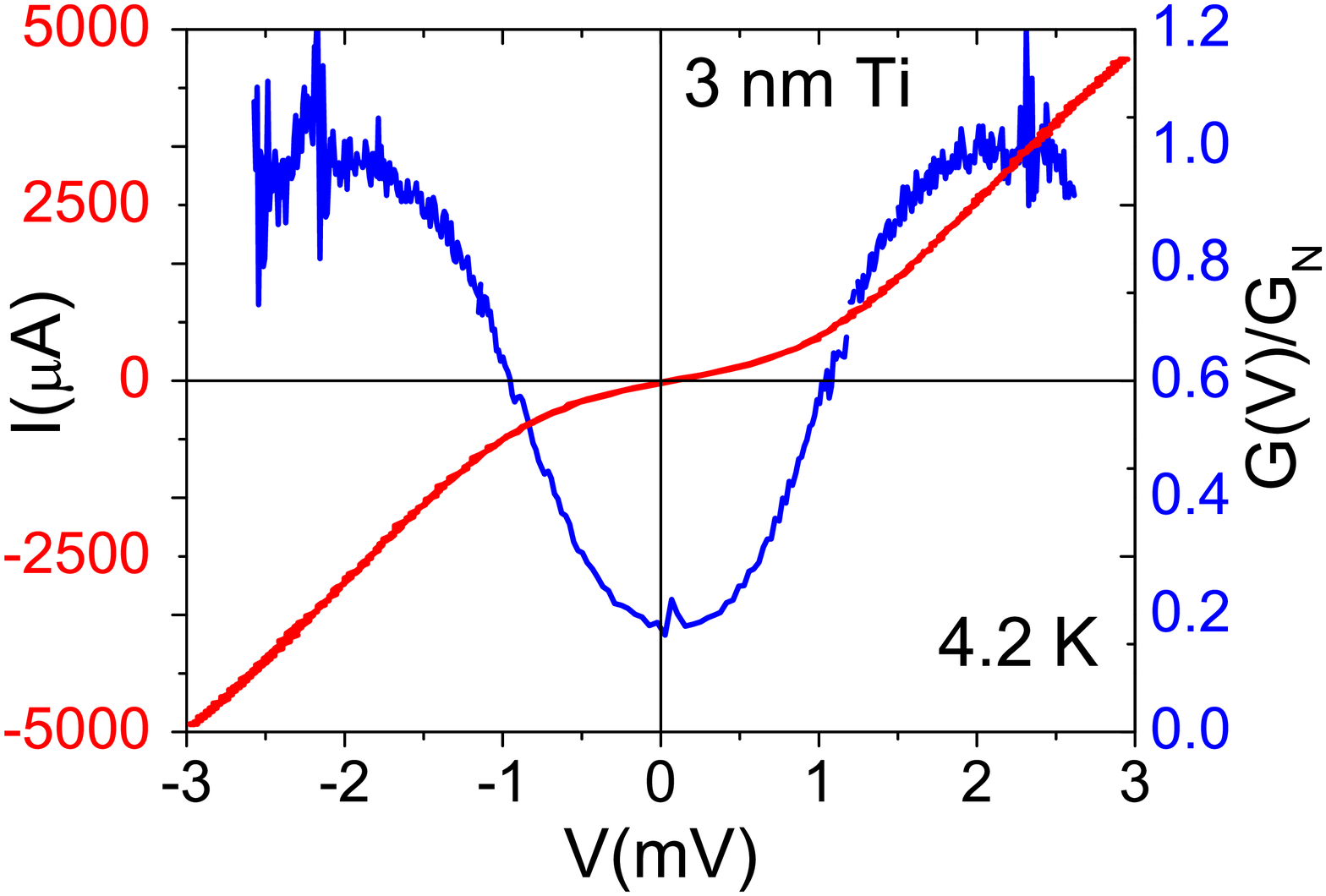}
&

 \includegraphics[width= 6 cm]{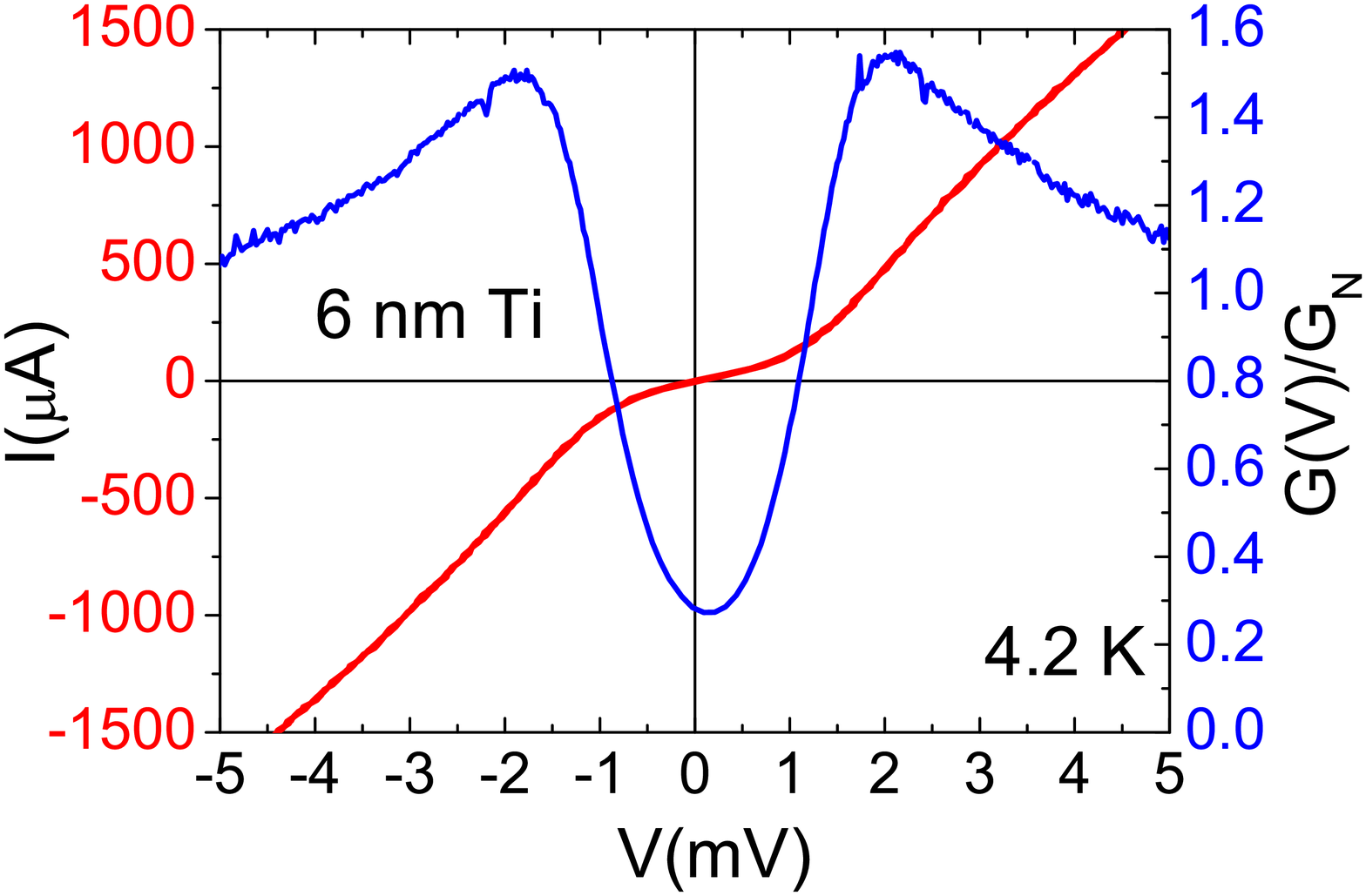}\\

 \centering
 \includegraphics[width= 6 cm]{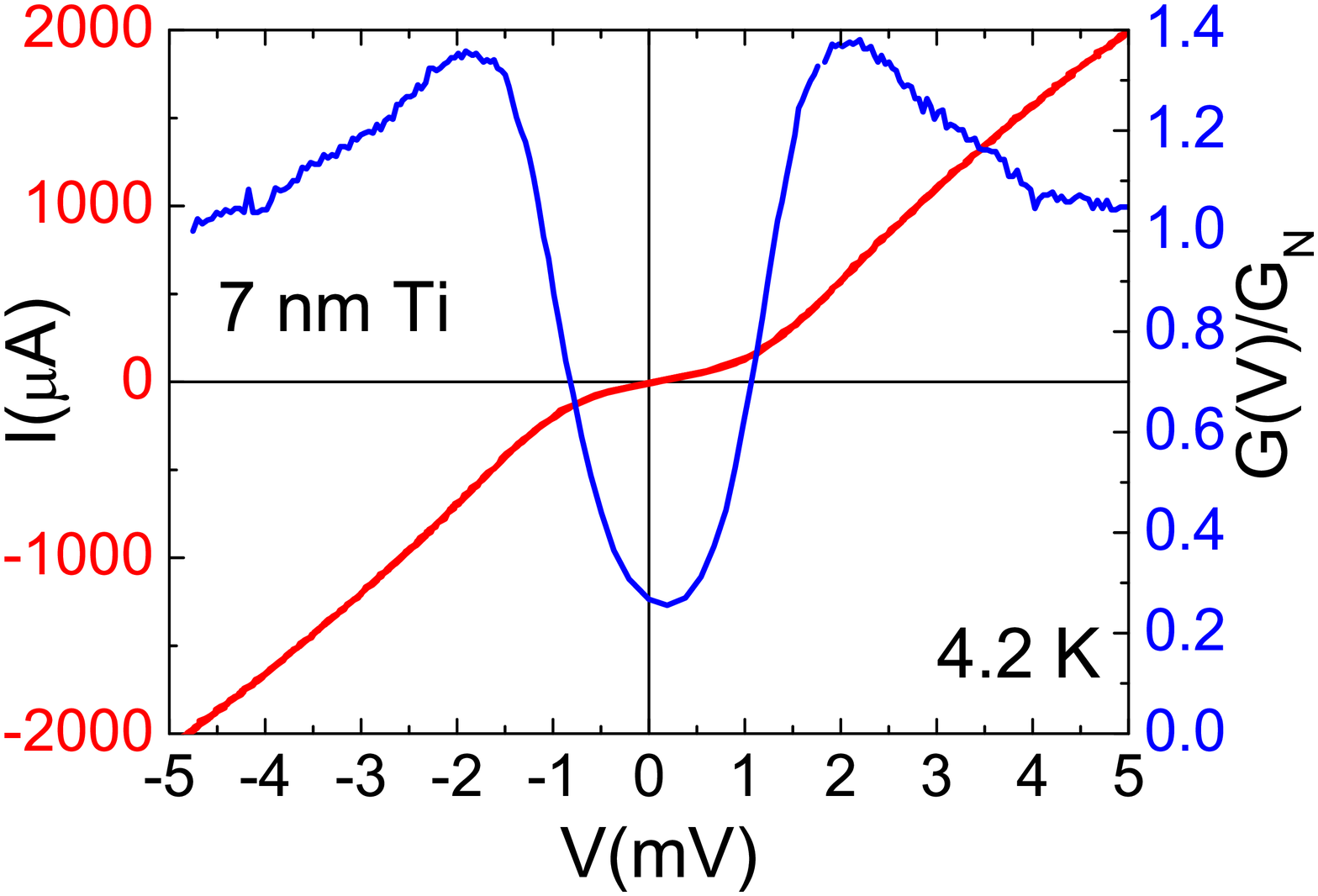}
&

 \includegraphics[width= 6 cm]{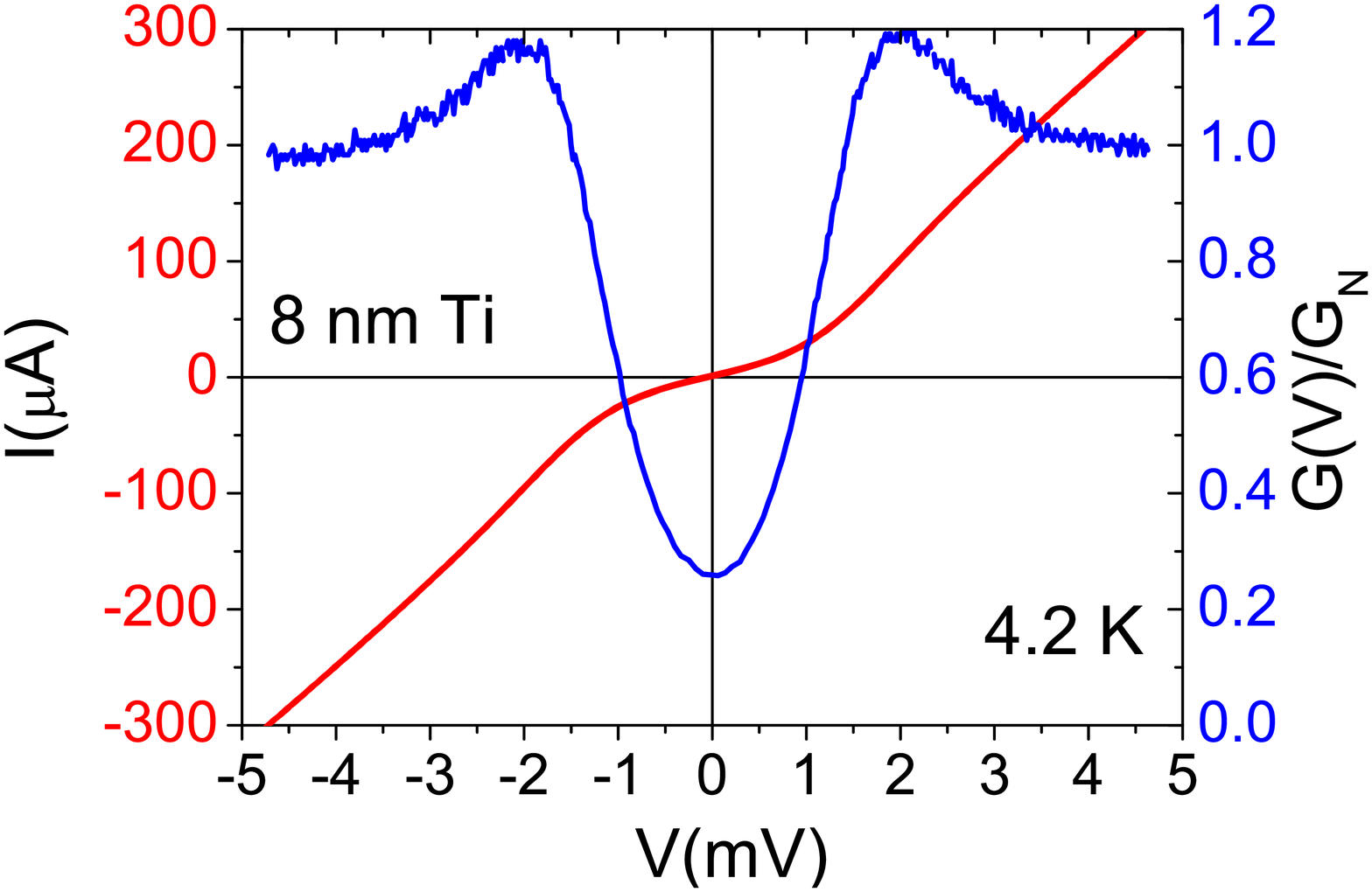}
\end{tabular}
\caption{SERIES-1 :I-V and normalized conductance spectra
$G(V)/G_N$ of the NbN(50 nm)-GdN1(2 nm)-Ti($t$)-GdN2 (2 nm)-NbN(50
nm) double spin-filter device with Ti thickness in the range 1-8
nm. The GdN1 and GdN2 layers were deposited with 8 $\%$ and 4 $\%$
N$_2$ and Ar gas mixture.The conductance spectra at 4.2 K were
measured with a standard lock-in technique.}
\end{figure}

\newpage
\begin{figure}[!h]
\begin{tabular}{ll}
  \centering
  \includegraphics[width= 6 cm]{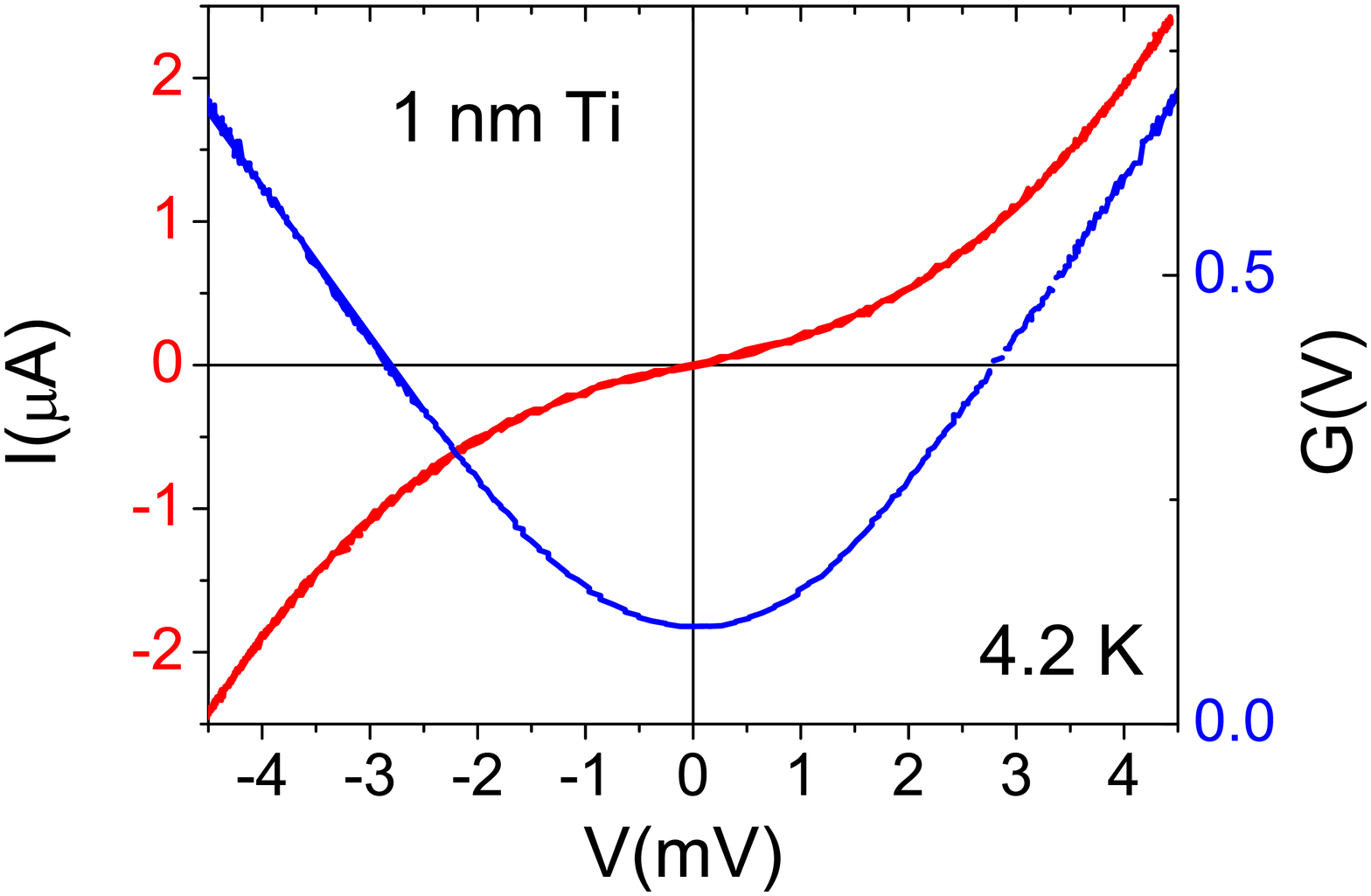}
&

  \includegraphics[width= 6 cm]{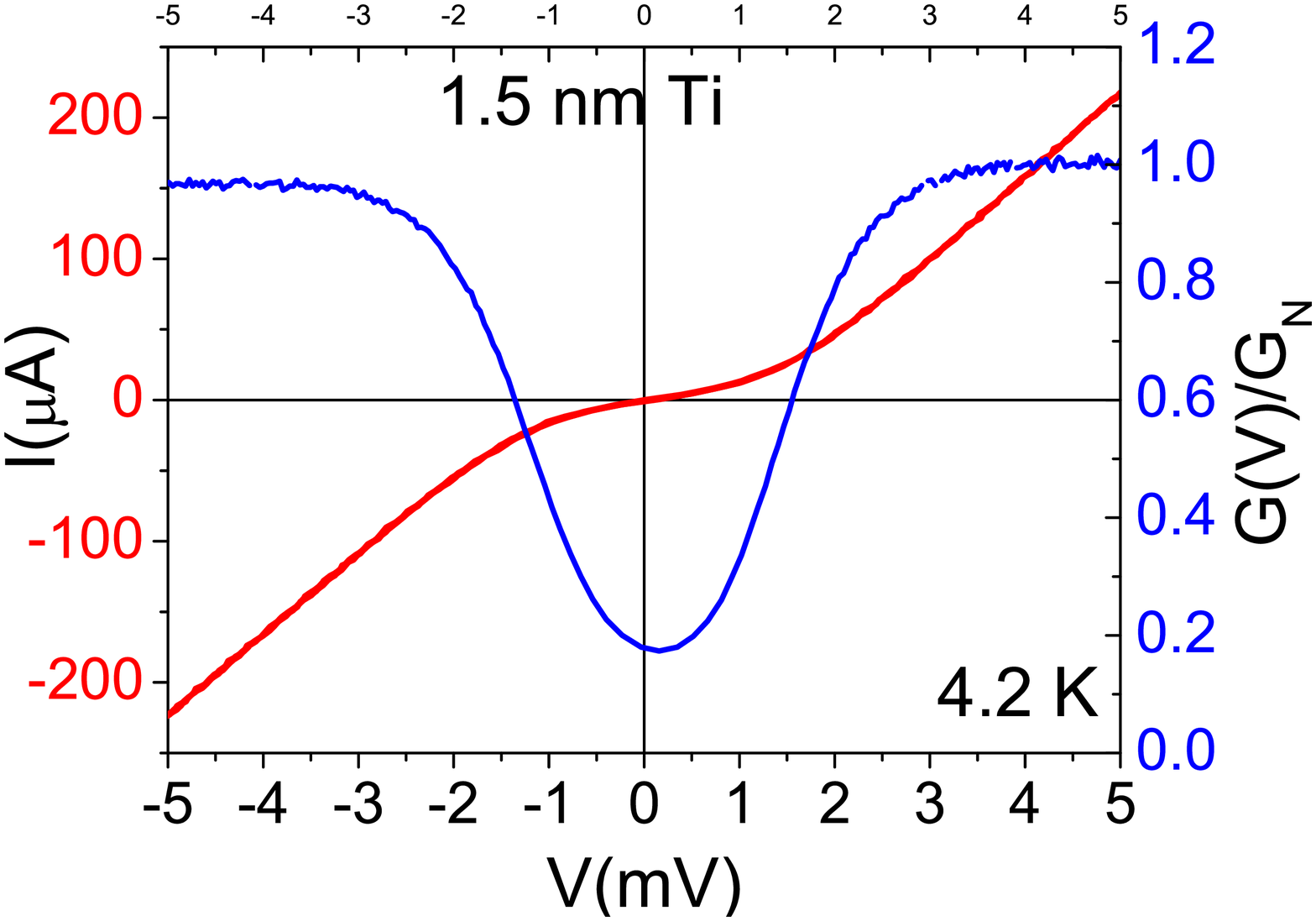}\\

 \centering
 \includegraphics[width= 6 cm]{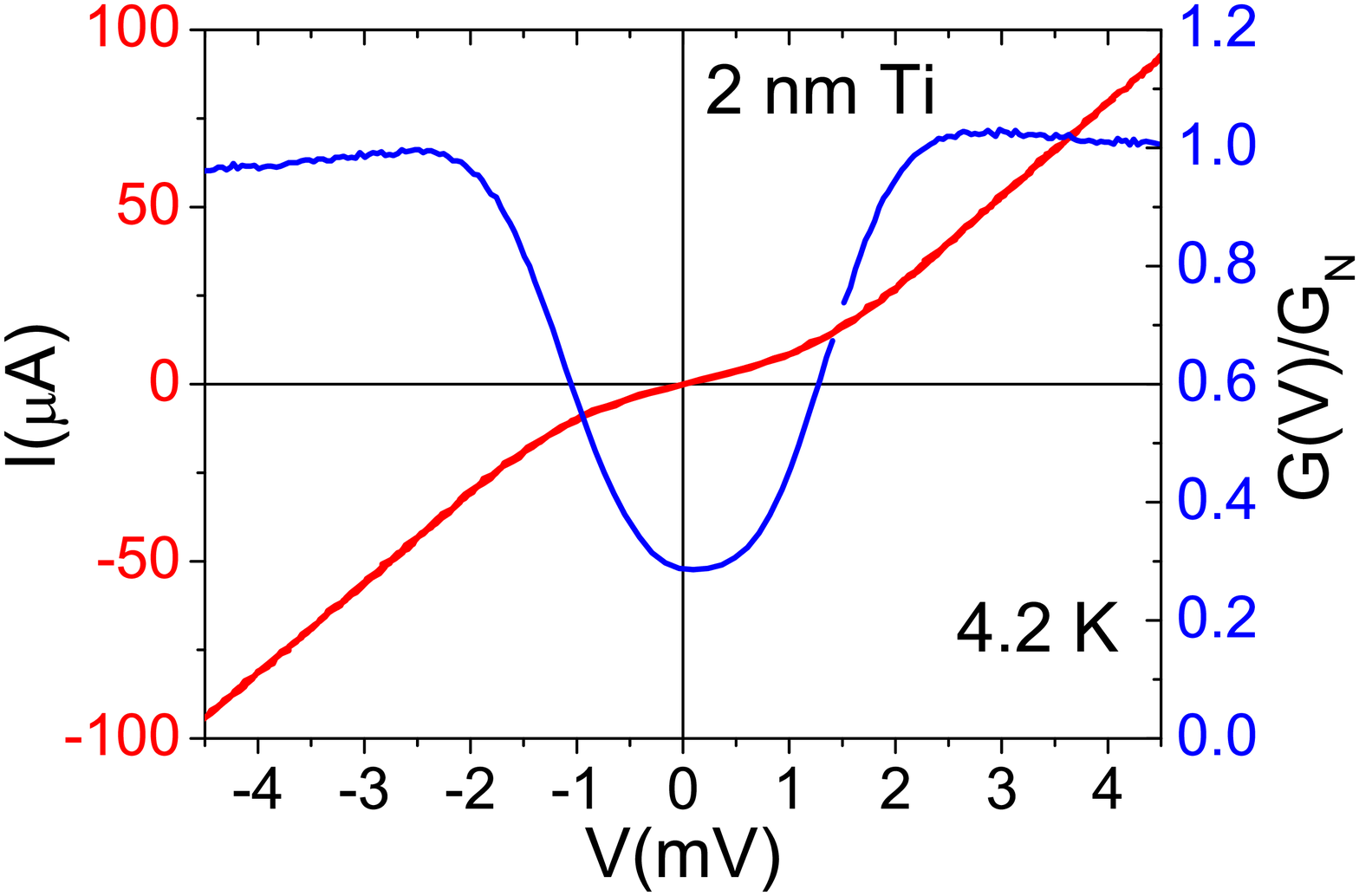}
&

 \includegraphics[width= 6 cm]{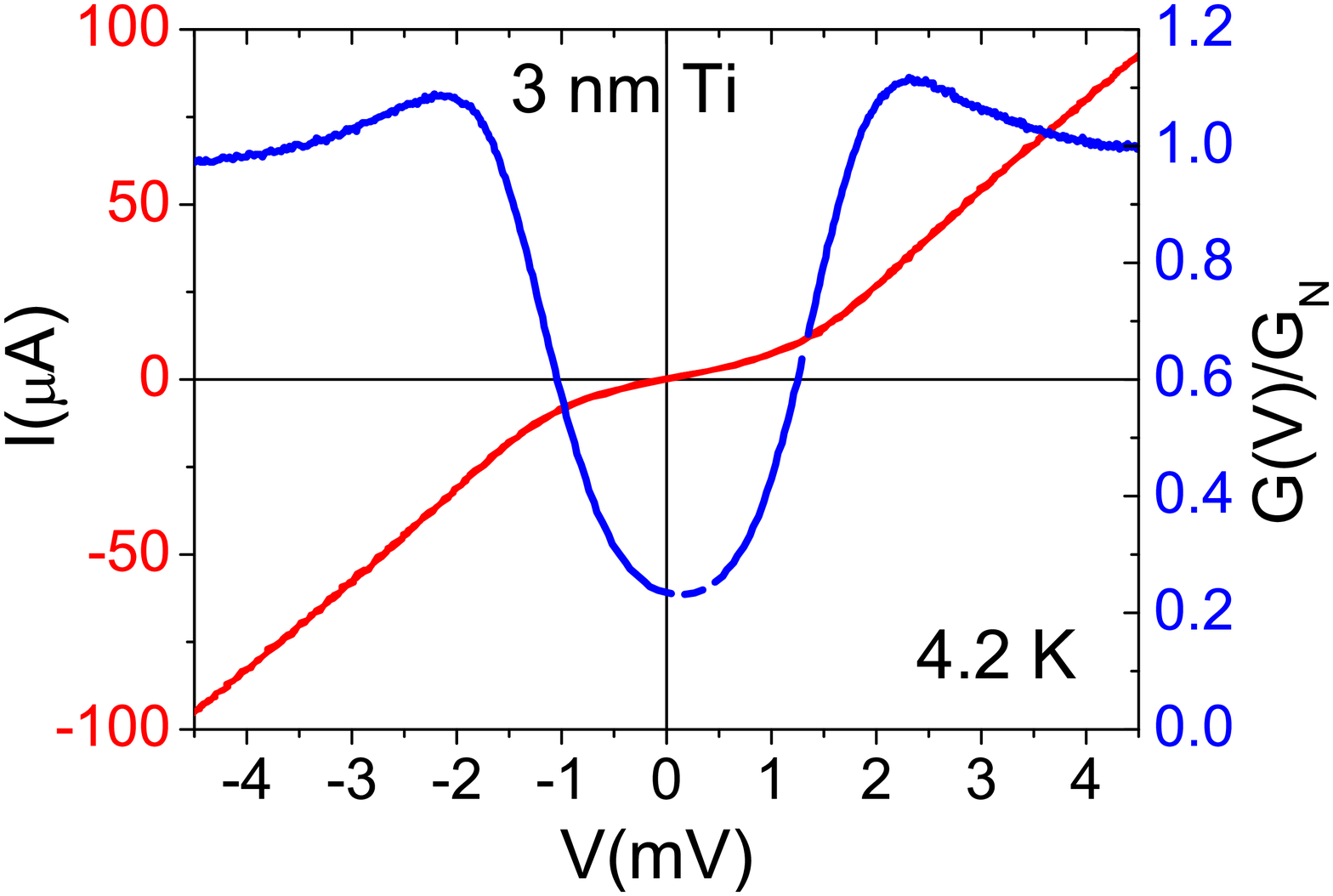}\\

 \centering
 \includegraphics[width= 6 cm]{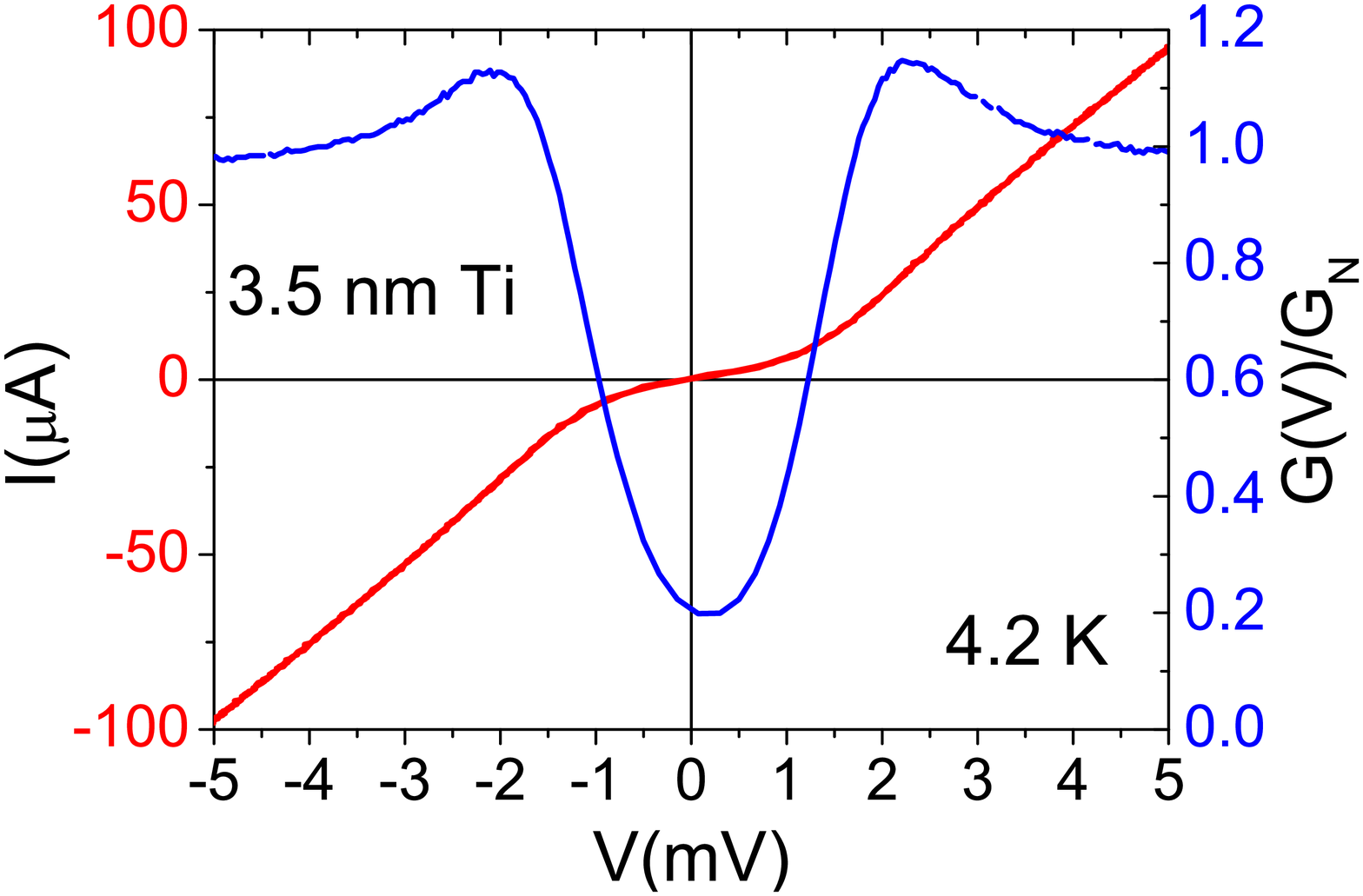}
&

 \includegraphics[width= 6 cm]{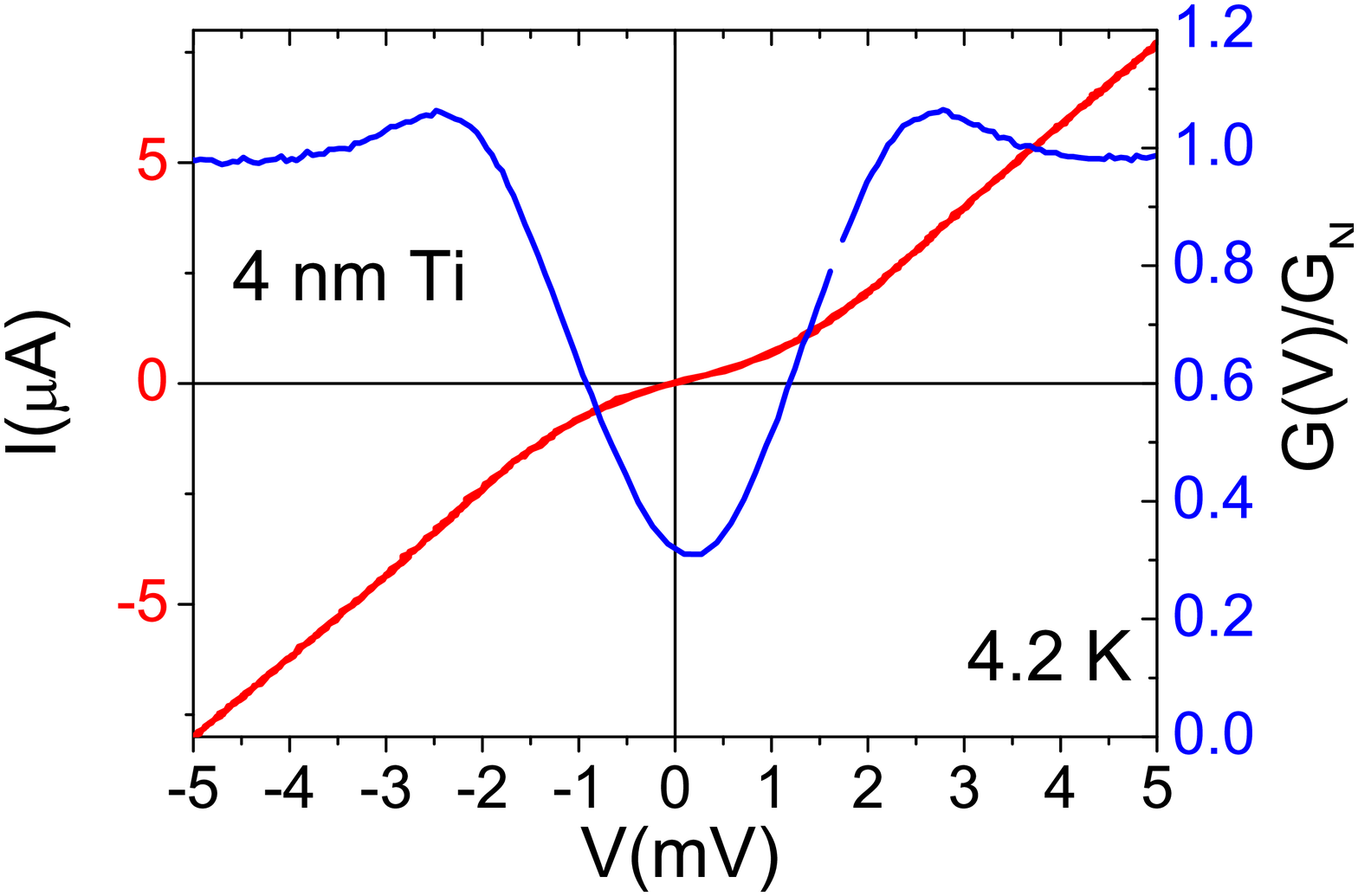}
\end{tabular}
\caption{SERIES-2: I-V and normalized conductance spectra
$G(V)/G_N$ of the NbN(50 nm)-GdN1(2 nm)-Ti-GdN2 (2 nm)-NbN(50 nm)
double spin-filter device with Ti thickness in the range 1-4 nm.
Both the GdN1 and GdN2 layers were deposited with 8 $\%$ N$_2$ and
Ar gas mixture. The conductance spectra at 4.2 K were measured
with a standard lock-in technique. The junction with $>$1 nm Ti is
equivalent to a NbN(50 nm)-GdN(4 nm)-NbN(50 nm) tunnel junction
which is at the thickness limit of  tunneling-type transport. }
\end{figure}

\newpage
\begin{figure}[!h]
\begin{tabular}{ll}
  \centering
  \includegraphics[width= 6 cm]{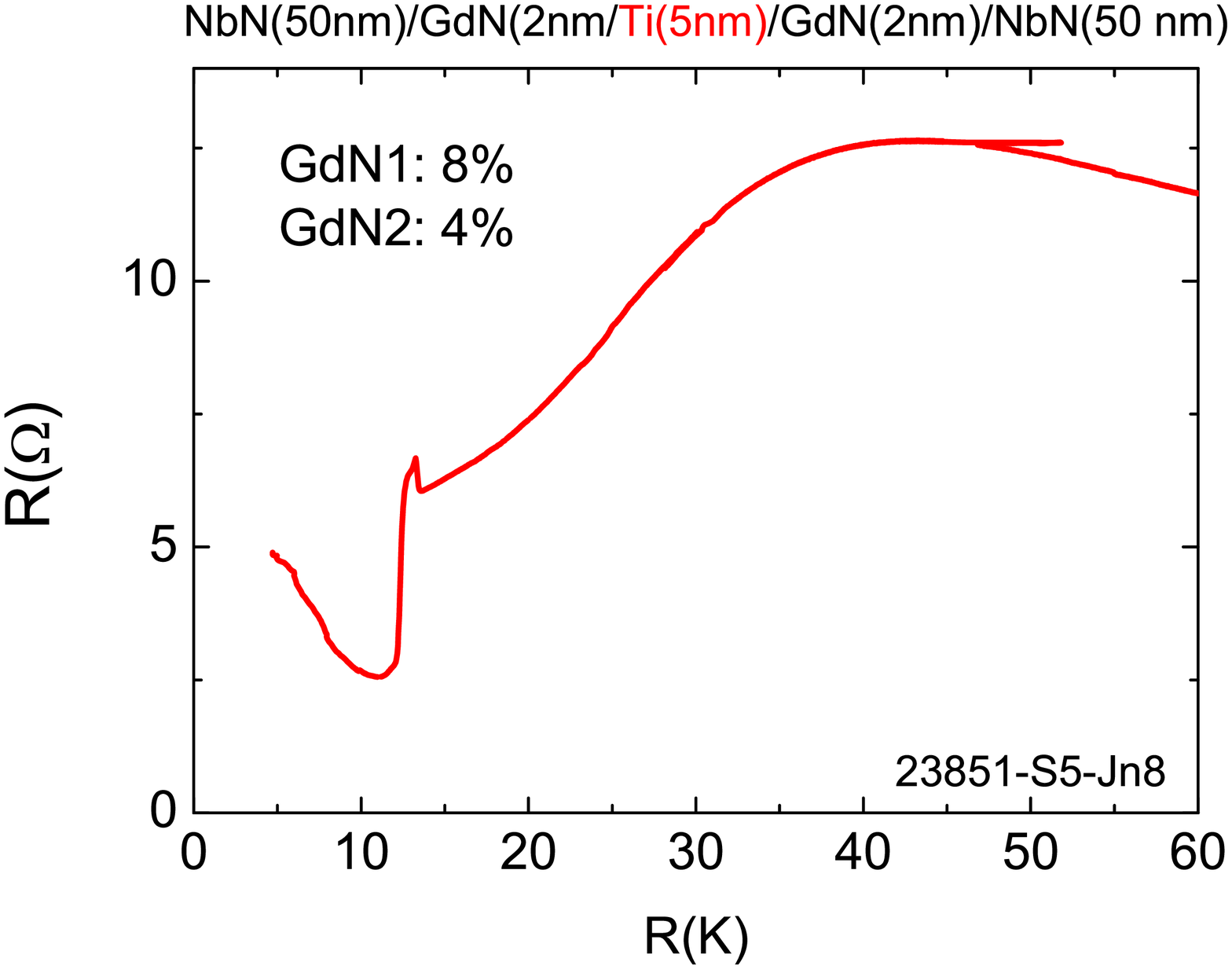}
&

  \includegraphics[width= 6 cm]{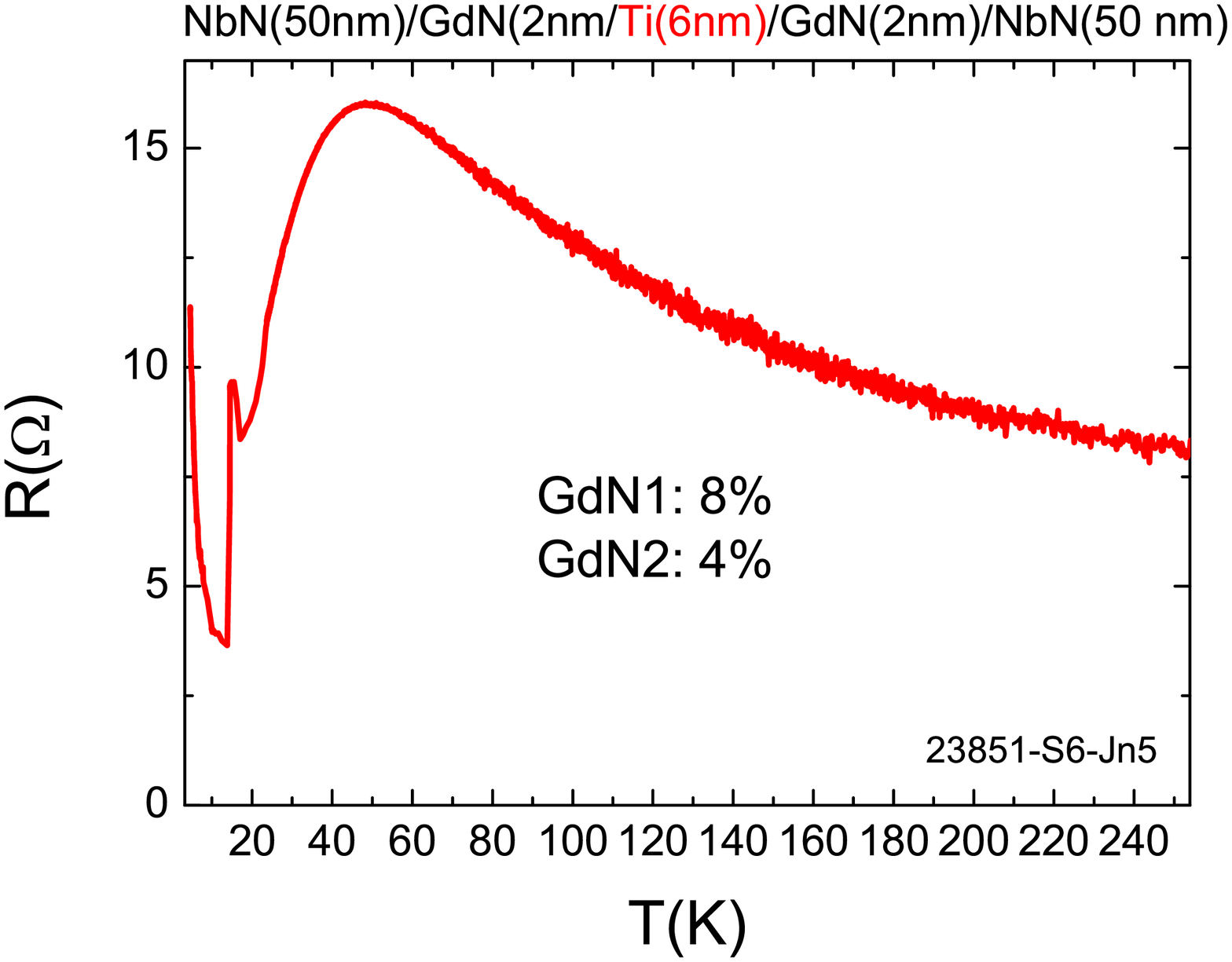}\\

 \centering
 \includegraphics[width= 6 cm]{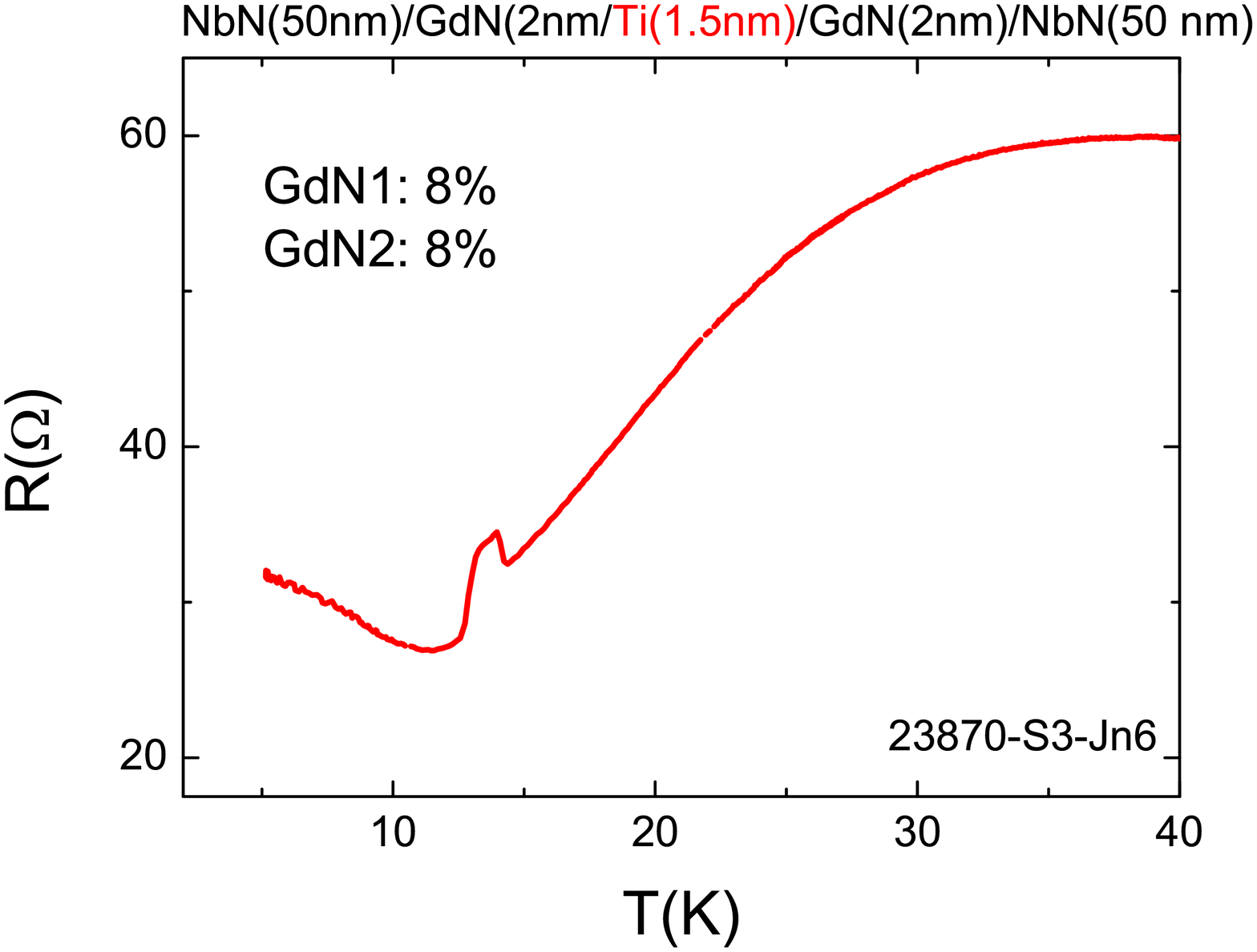}
&

 \includegraphics[width= 6 cm]{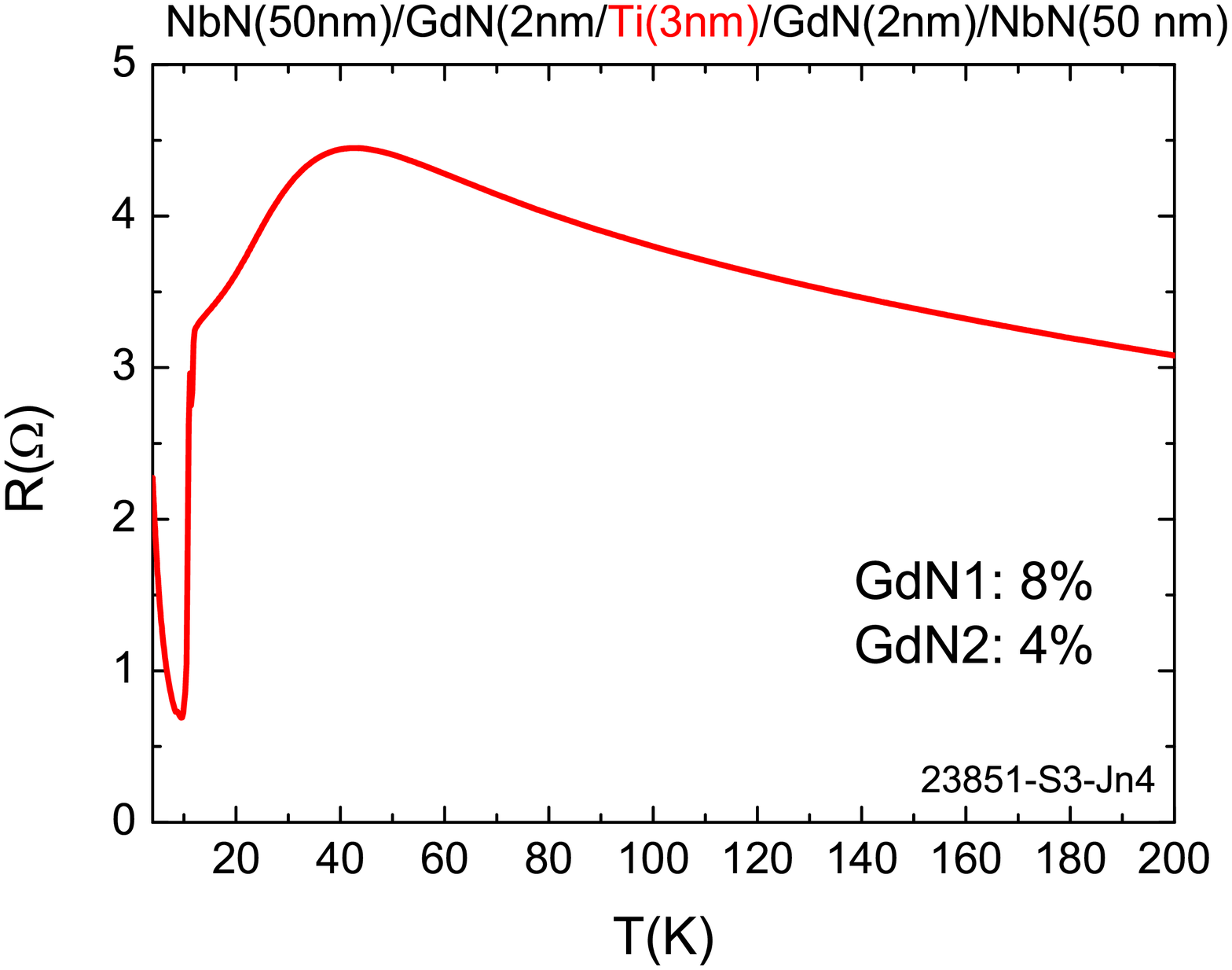}\\

 \centering
 \includegraphics[width= 6 cm]{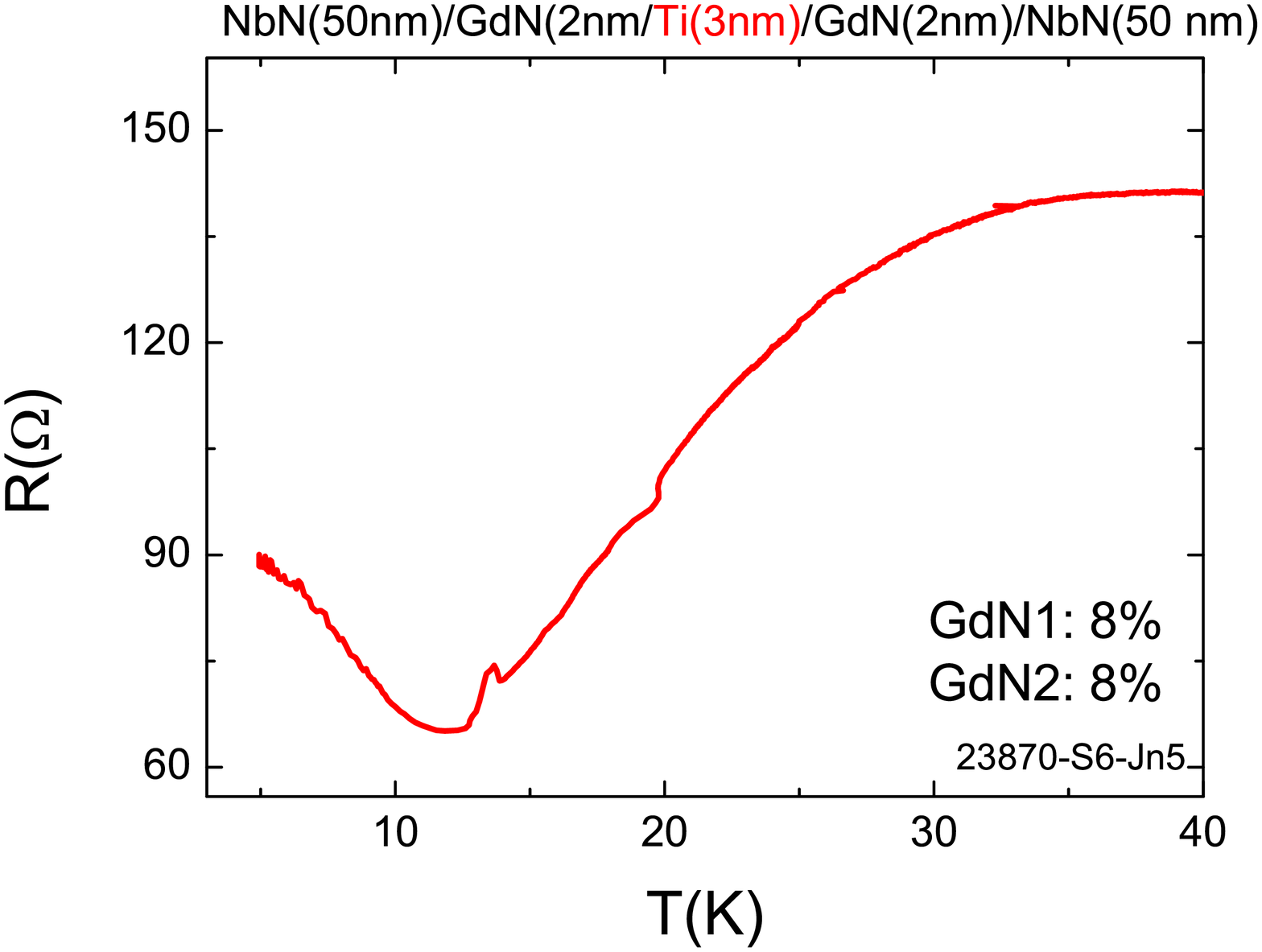}
&

 \includegraphics[width= 6 cm]{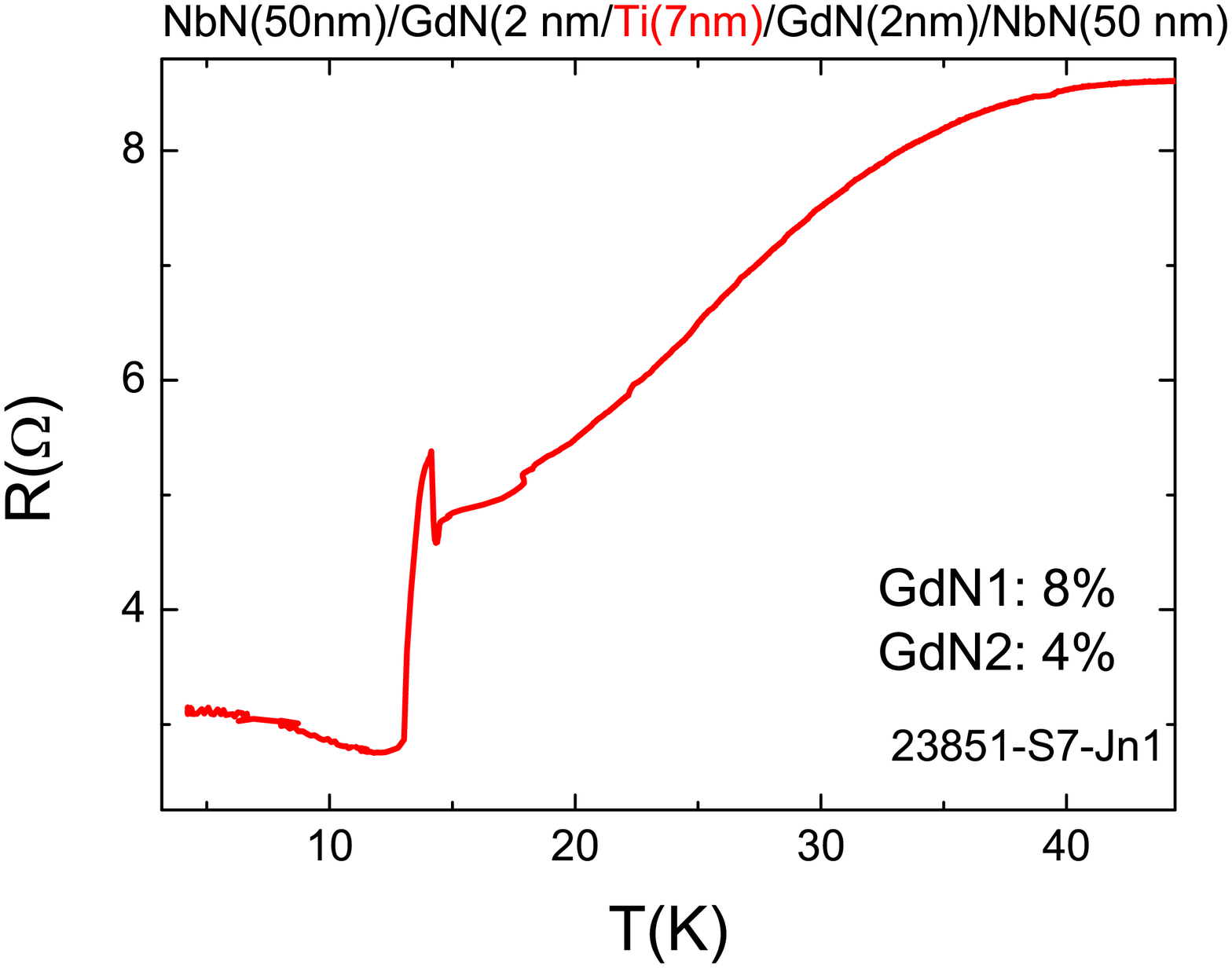}
\end{tabular}
\caption{The $R(T)$ of few representative double spin-filter
tunnel junction with different thickness of Ti spacer and
deposition gas mixture for GdN(1,2). The magnitude of resistance
may not correspond to the thickness of GdN in these double
junctions as all the $R(T)$ measurements were done with different
bias current. Note that these tunnel junctions show nonlinear IV.
 }
\end{figure}

\newpage

\begin{figure}[!h]
\begin{center}
\abovecaptionskip -10cm
\includegraphics [width=8 cm]{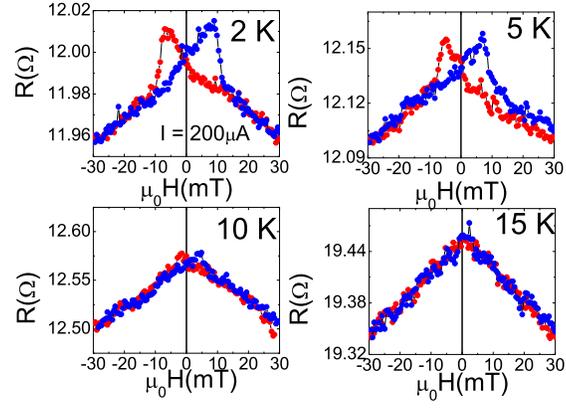}
\end{center}
\caption{\label{fig1} (Color online) R-H loops of Jn1 (the same Jn
as reported in the manuscript) measured at different temperatures
with a bias current of $I= 200 \mu$A.}
\end{figure}

\newpage
\begin{center}
\textbf{Reproducibility of spin-valve behavior in
NbN-GdN1-Ti(t)-GdN2-NbN tunnel junction}
\end{center}

\begin{figure}[!h]
\begin{center}
\abovecaptionskip -10cm
\includegraphics [width=8 cm]{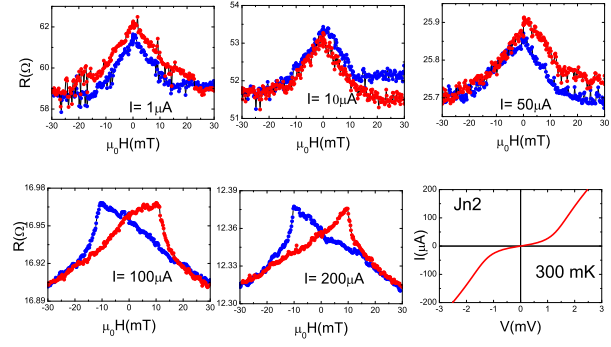}
\end{center}
\caption{\label{fig1} (Color online) R-H loops and I-V curve of
\textbf{Jn-2 }(Note that each chip contain 8 identical junctions
and measurements done on Jn-1 on the same chip is reported in the
manuscript) measured at 300 mK with different bias current.}
\end{figure}

\begin{figure}[!h]
\begin{center}
\abovecaptionskip -10cm
\includegraphics [width=8 cm]{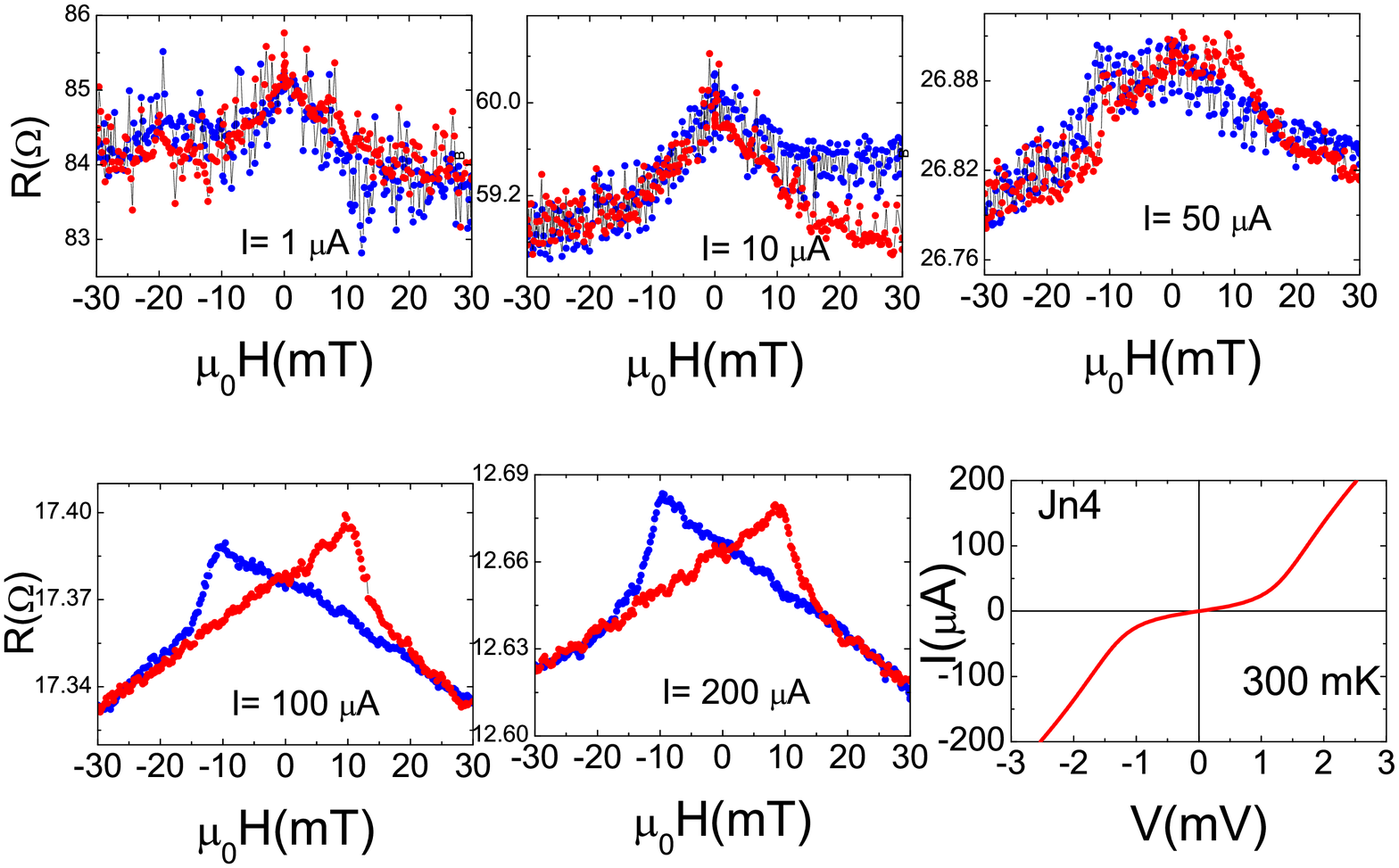}
\end{center}
\caption{\label{fig1} (Color online) R-H loops and I-V curve of
\textbf{Jn-4} on the same chip measured at 300 mK with different
bias current.}
\end{figure}

\begin{figure}[!h]
\begin{center}
\abovecaptionskip -10cm
\includegraphics [width=8 cm]{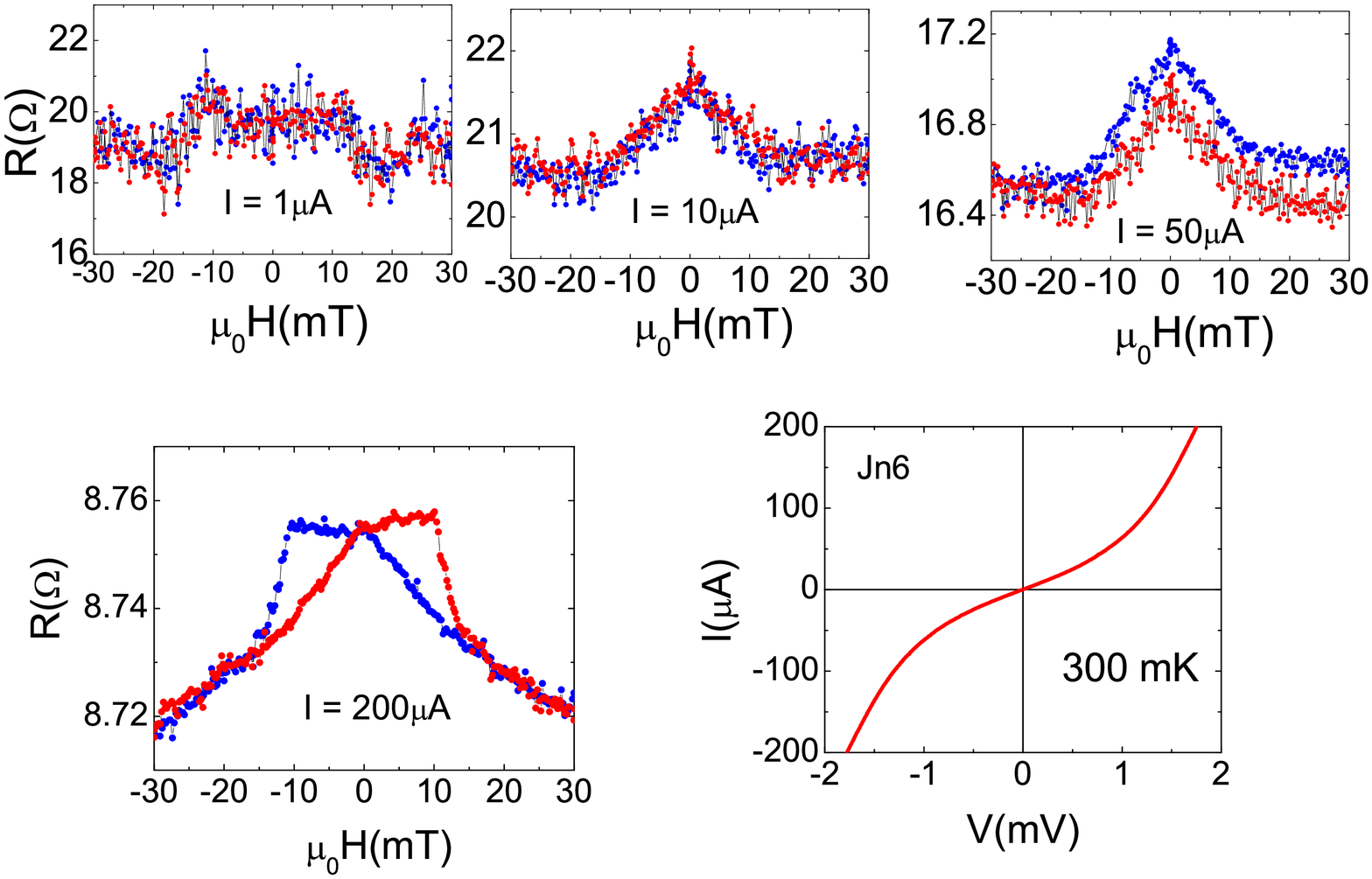}
\end{center}
\caption{\label{fig1} (Color online) R-H loops and I-V curve of
\textbf{Jn-6} on the same chip measured at 300 mK with different
bias current.}
\end{figure}

\begin{figure}[!h]
\begin{center}
\abovecaptionskip -10cm
\includegraphics [width=8 cm]{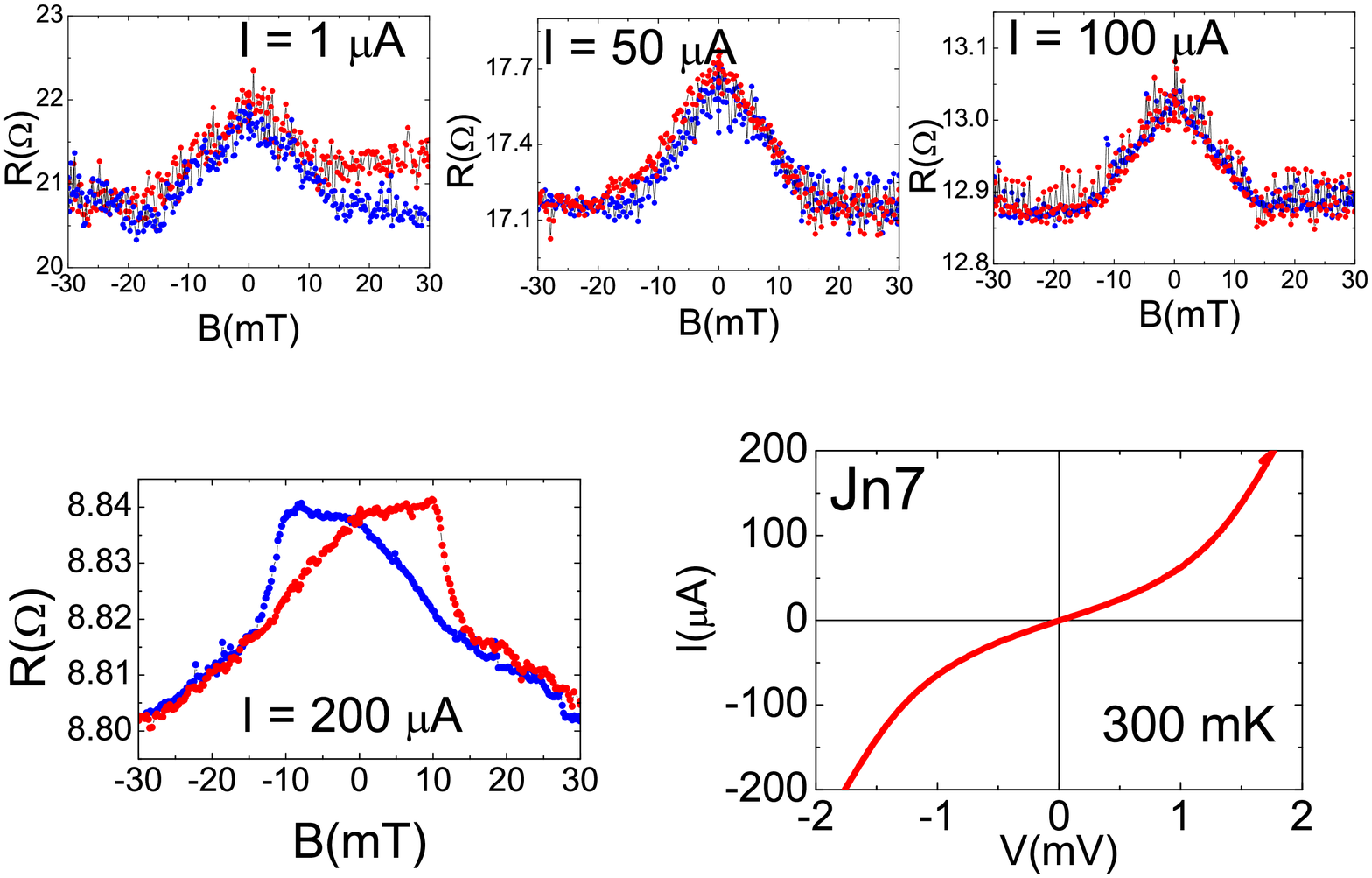}
\end{center}
\caption{\label{fig1} (Color online) R-H loops and I-V curve of
\textbf{Jn-7} on the same chip measured at 300 mK with different
bias current.}
\end{figure}

\begin{figure}[!h]
\begin{center}
\abovecaptionskip -10cm
\includegraphics [width=8 cm]{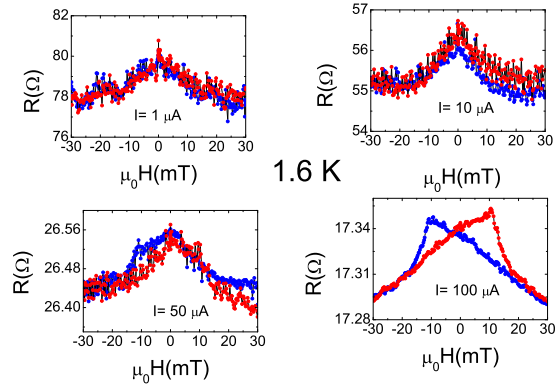}
\end{center}
\caption{\label{fig1} (Color online) The R-H loop of the Jn-1
measured at 1.6 K with different bias current. The R-H loops
measured below sub-gap current were found to be extremely
sensitive to temperature stability during the measurement.}
\end{figure}


\newpage

\begin{center}
\emph{\textbf{Tunneling spectra of  NbN-Ti-GdN-NbN tunnel
junction:}}
\end{center}

The NbN-Ti-GdN-NbN tunnel junction with thick $\sim$9 nm Ti can be
regarded as a N-I-S-type tunnel junction. Normalized tunneling
conductance of a NIS junction at a bias voltage $V$ can be written
as:
\begin{equation}
\frac{{G_s (V)}}{{G_N (V)}} = \frac{d}{{d(eV)}}\int\limits_{ -
\infty }^\infty  {N(E)} [f(E) - f(E + eV)]dE,
\end{equation}
where $f(E)$ is Fermi-Dirac distribution function and $N(E)$ is
the normalized BCS density of state of the superconductor. Here
$G_N(V)$ is the normal state conductance of the junction.
Following Dynes approach\cite{dynes} the quasiparticle density of
states can be written as, $ N(E) = N(0)\left| {{\mathop{\rm
Re}\nolimits} \left( {\frac{{E/\Delta - i\gamma }}{{\sqrt
{(E/\Delta  - i\gamma )^2  - 1} }}} \right)} \right|$. Here the
smearing parameter $\gamma$ is included to consider finite
lifetime of quasiparticles. SFig. 9(b) shows fitting of Eq. (3)to
conductance spectra of the NbN-Ti(9 nm)-GdN(2.6 nm)-NbN tunnel
junction.

\begin{figure}[!h]
\begin{tabular}{ll}

  \centering
  \includegraphics[width= 6 cm]{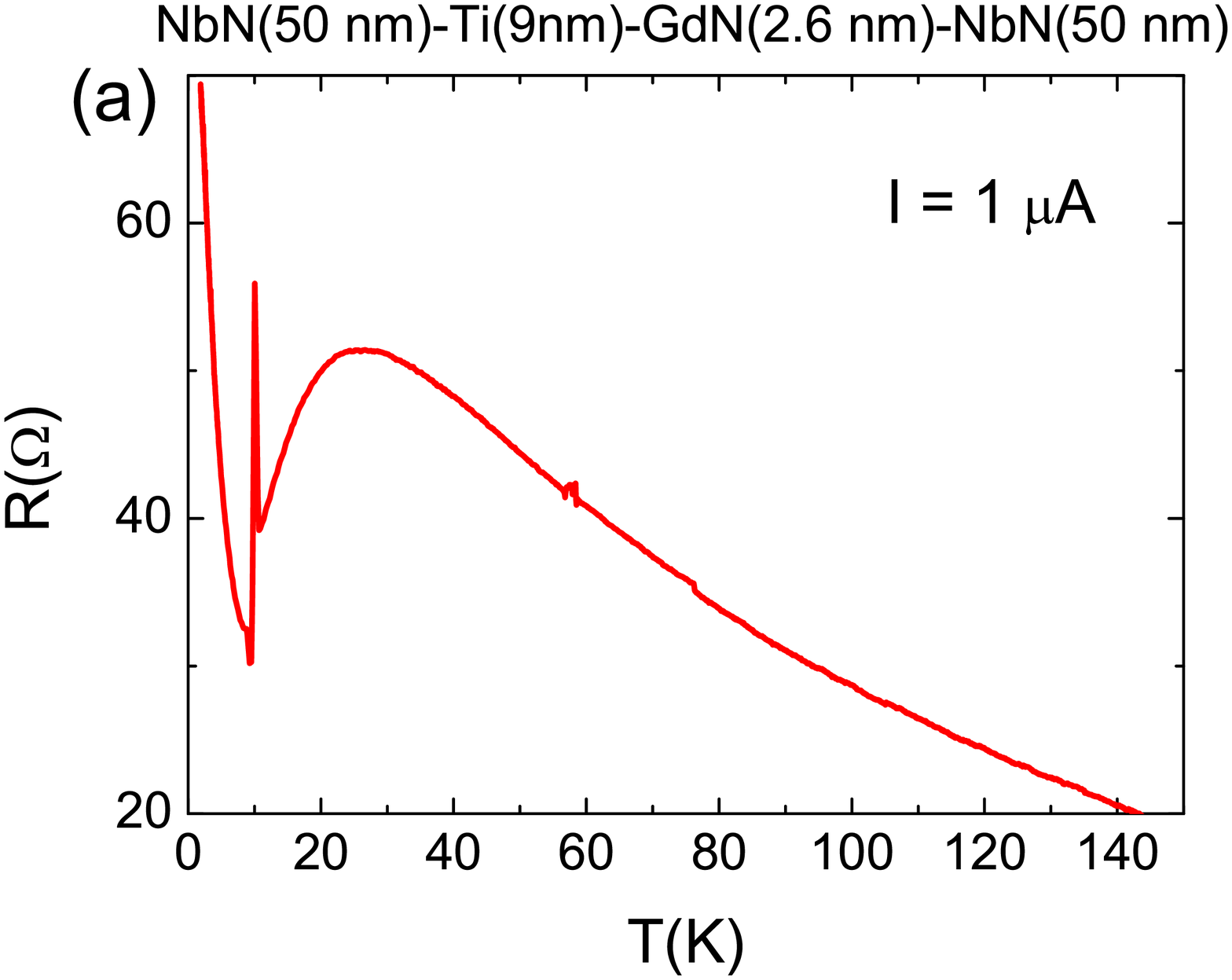}
&

  \includegraphics[width= 6 cm]{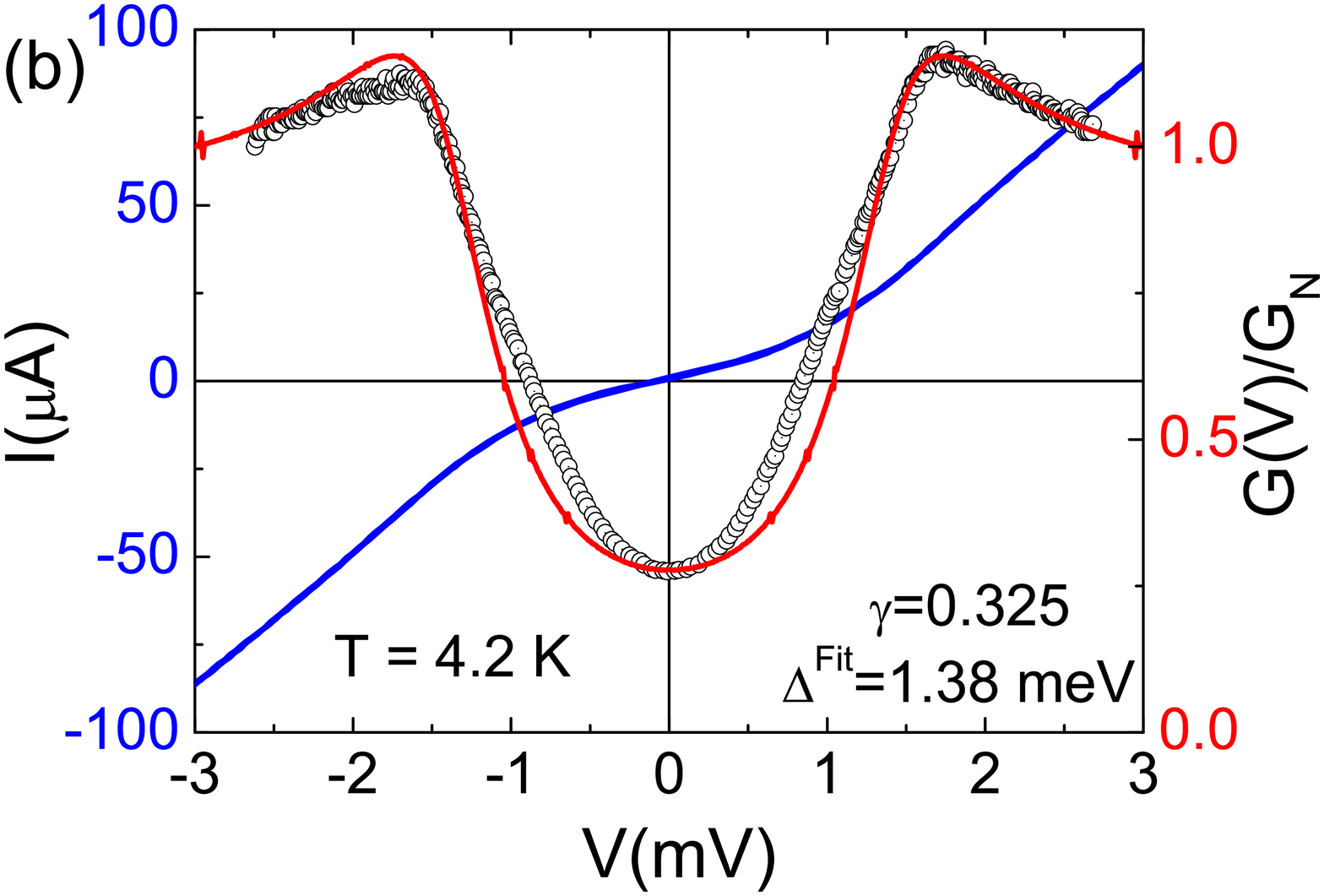}\\

 \centering
 \includegraphics[width= 6 cm]{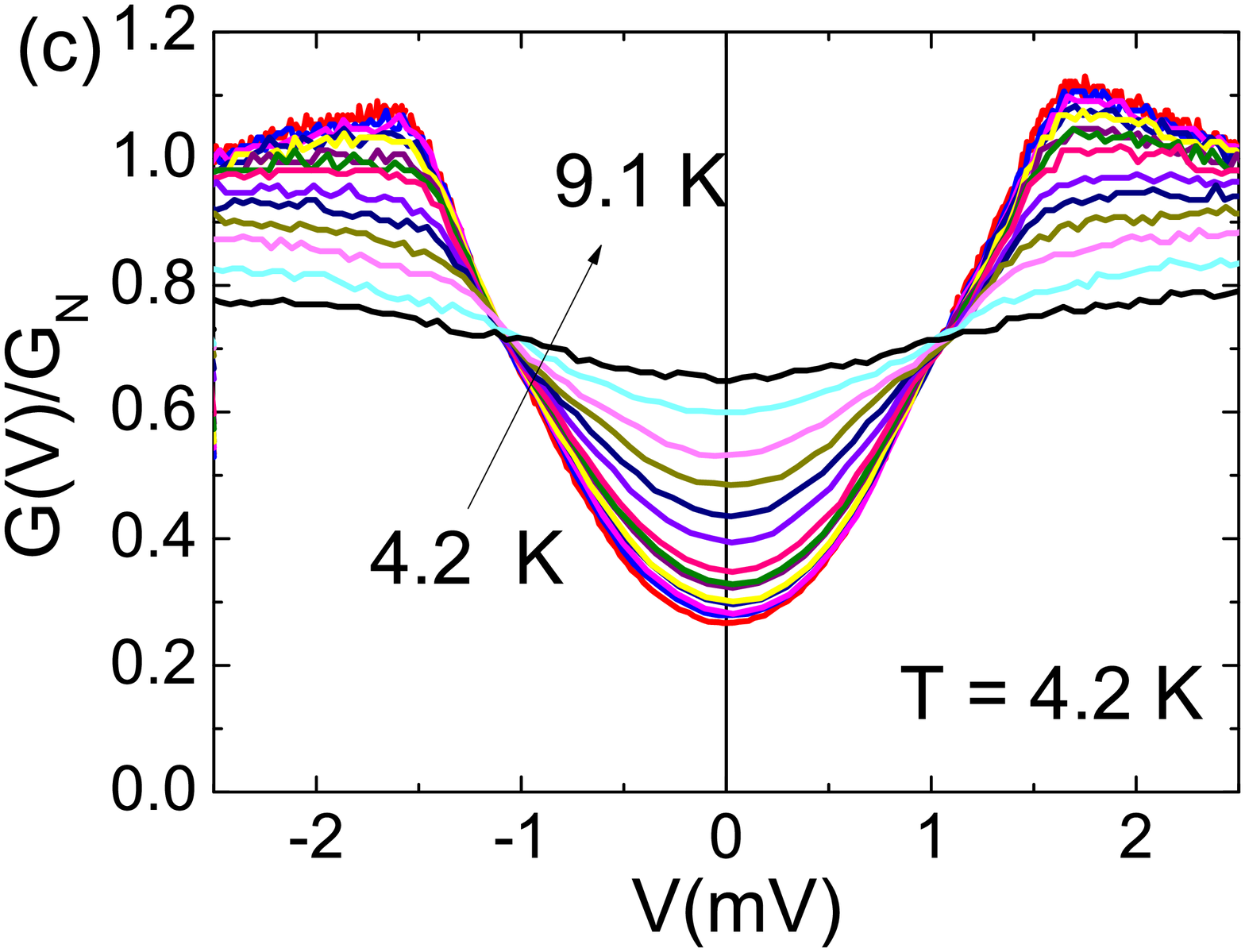}
 &

 \includegraphics[width= 6 cm]{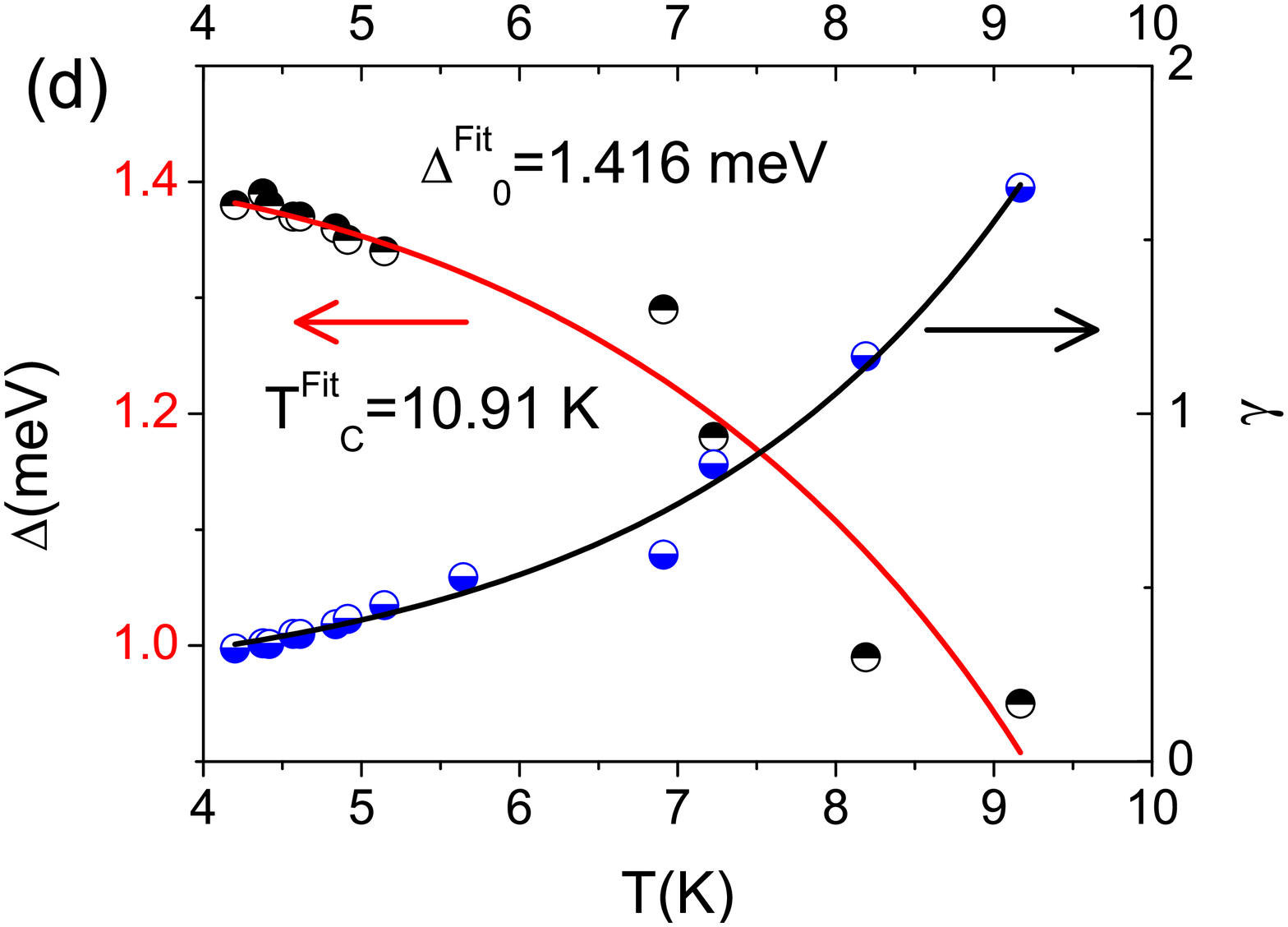}\\

\end{tabular}
\caption{(a) Temperature dependence of resistance of the NbN-Ti(9
nm)-GdN(2.6 nm)-NbN tunnel junction. measured with I = 1 $\mu$A.
(b) I-V and normalized conductance spectra $G(V)/G_N$ of the
tunnel junction measured at 4.2 K. The red solid line shows
fitting to Eq. (3) with fitting parameter $\Delta^{Fit}$ = 1.38
meV and $\gamma$ = 0.325. (c) Temperature evolution of conductance
spectra from 4.2 to 9.1 K. (d) Temperature dependence of
$\Delta^{Fit}$  and $\gamma$. A BCS model fit (red solid line);$
\Delta (T) = \Delta (0)\tanh (1.74\sqrt {(T_C  - T)/T} ) $ gives
$\Delta (0)$ = 1.416 meV and $T_C$ = 10.91 K. The smearing
parameter $\gamma$ can be seen to increase exponentially (black
solid line) with temperature.}
\end{figure}


\newpage

\begin{center}
\emph{\textbf{Tunneling spectra of  NbN-GdN ($t$)-NbN tunnel
junction:}}
\end{center}
The quasiparticle tunneling conductivity $G(V)= dI/dV$ in a S-I-S
junction can be written as;
\begin{equation}
\begin{array}{l}
 \frac{{G_s (V)}}{{G_N (V)}} = \frac{d}{{d(eV)}}\int\limits_{ - \infty }^\infty  {N(E + eV)N(E)} [f(E) - f(E + eV)]dE \\
  + \frac{V}{{R_{sh} }} \\
 \end{array}
\end{equation}

Where ${N(E)}$ is density of states of the supercondutor and
$f(E)$ is the Fermi distribution function. Here  ${R_{sh} }$ is a
shunt resistance in series with the S-I-S junction. The modified
density of states with Dynes parameter can be expressed as, $ N(E)
= N(0)\left| {{\mathop{\rm Re}\nolimits} \left( {\frac{{E/\Delta -
i\gamma }}{{\sqrt {(E/\Delta  - i\gamma )^2  - 1} }}} \right)}
\right|$. Here $\Delta$  is the superconducting gap and $\gamma$
is the smearing parameter. SFig. 11 and SFig. 12 shows typical
$dI/dV-V$ spectra of NbN-GdN-NbN tunnel junction with two
different barrier transparency.  The temperature dependence of
$\Delta$ was fitted to the BCS-type temperature dependence, $
\Delta (T) = \Delta (0)\tanh (1.74\sqrt {(T_C  - T)/T} ) $.

\newpage

\newpage
\begin{figure}[!h]
\begin{tabular}{ll}

  \centering
  \includegraphics[width= 6 cm]{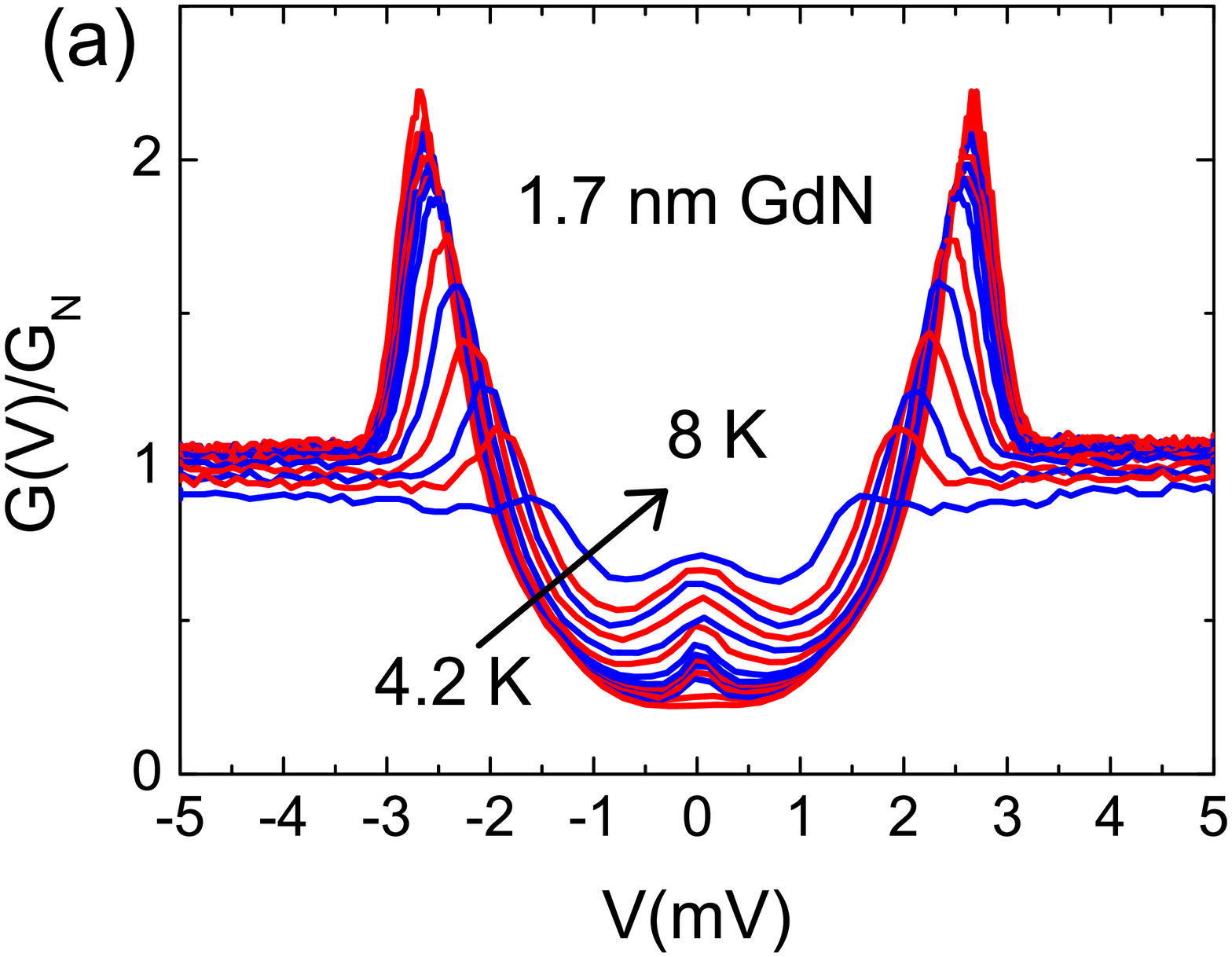}
&

  \includegraphics[width= 6 cm]{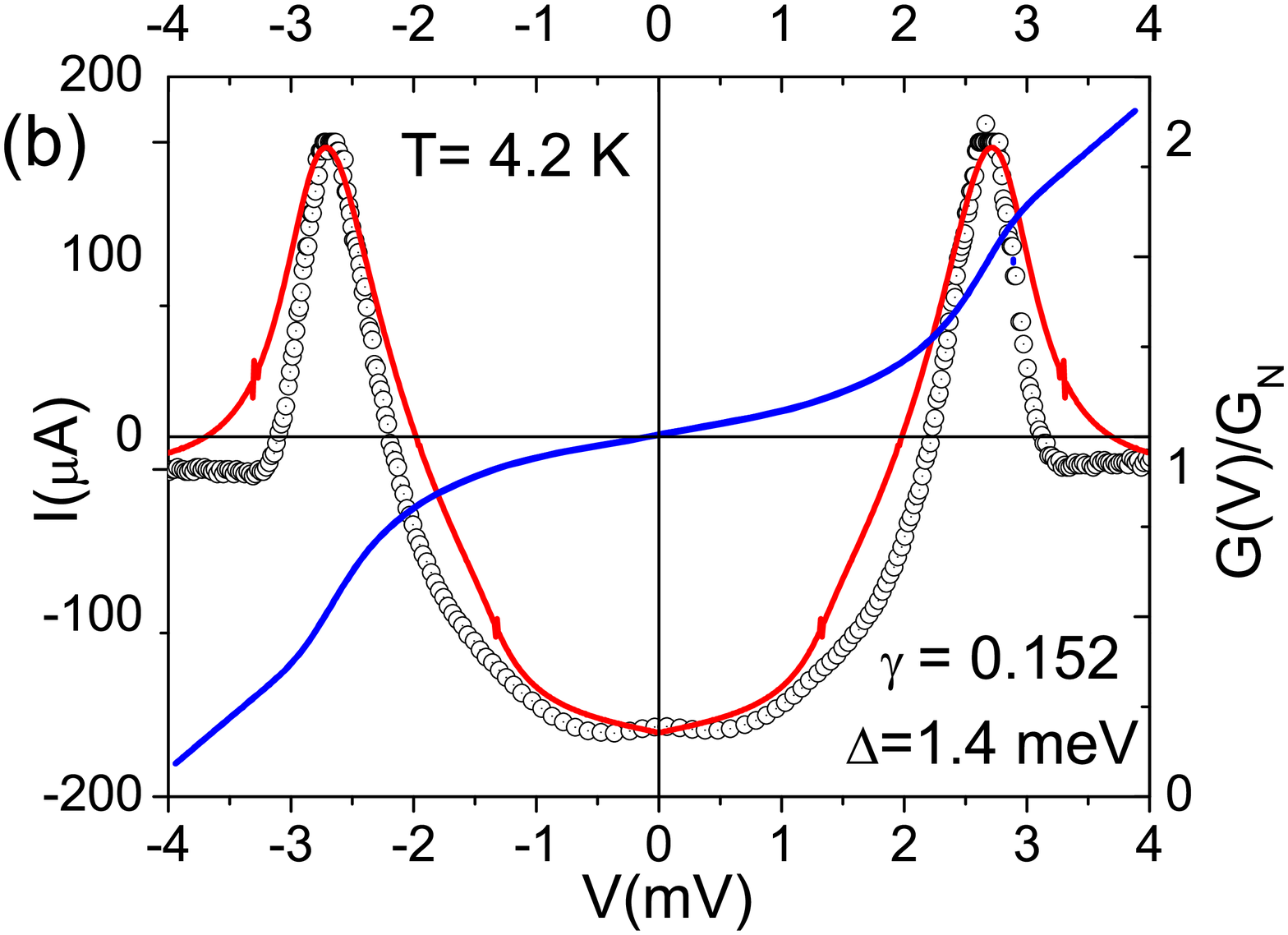}\\

 \centering
 \includegraphics[width= 6 cm]{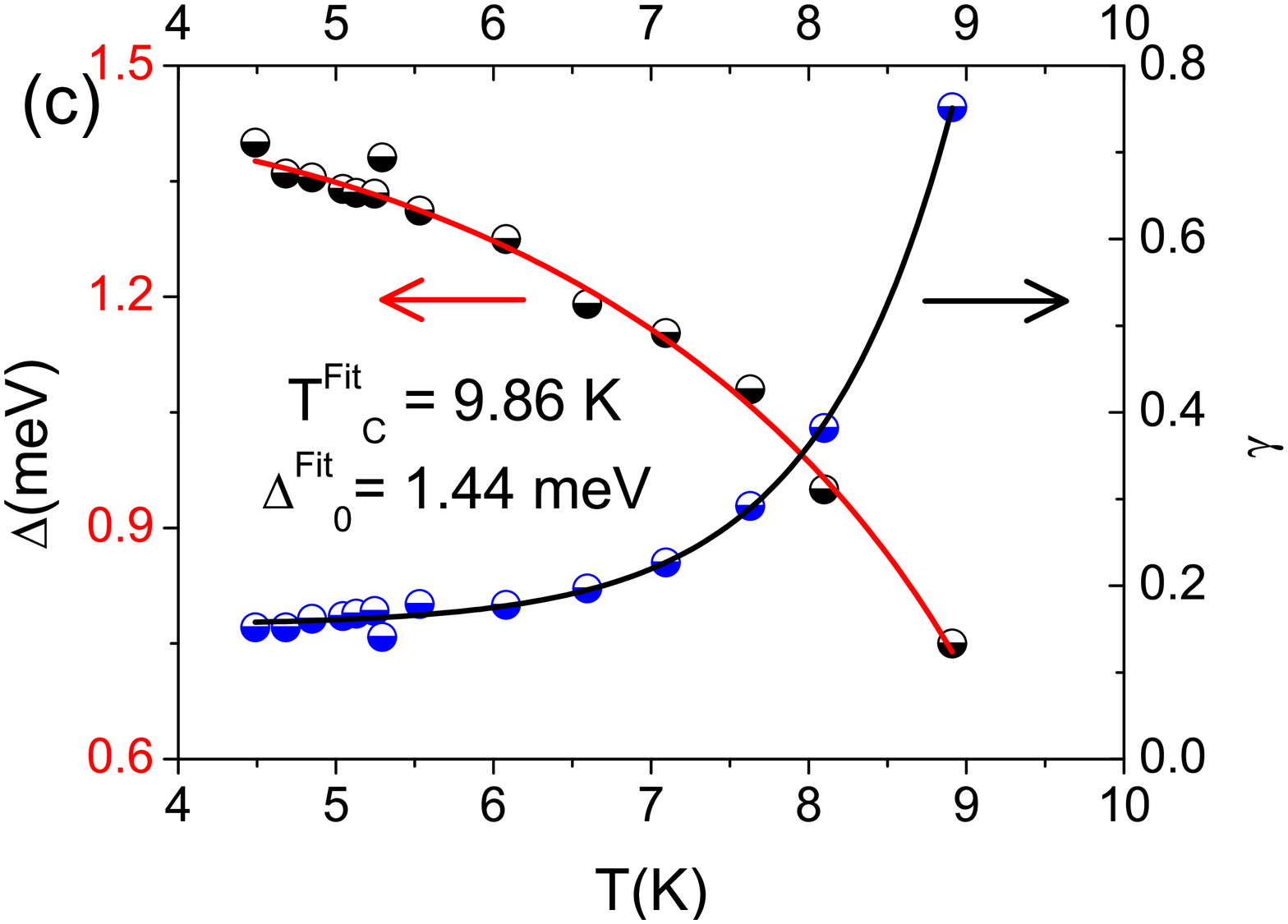}
 &

 \includegraphics[width= 6 cm]{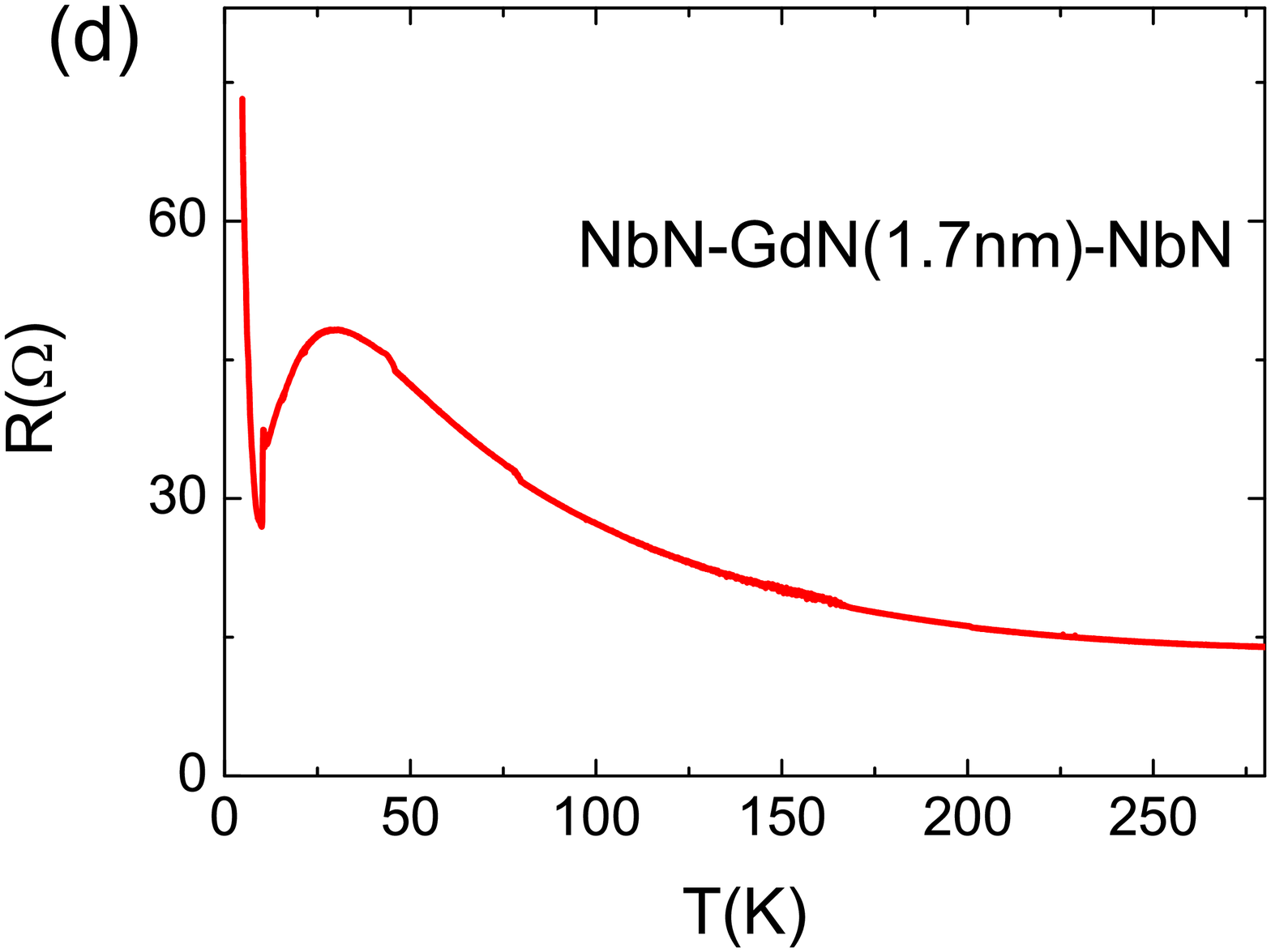}\\

\end{tabular}
\caption{(a) Temperature evolution of normalized conductance
spectra $G(V)/G_N$ of NbN-GdN(1.7 nm)-NbN tunnel junction. The
conductance spectra were measured with a standard lock-in
technique.(b)I-V and normalized conductance spectra $G(V)/G_N$ of
the same junction at 4.2 K. Red solid line is the fit to Eq. (4)
with fitting parameter shown. Temperature dependence of the
fitting parameter $\Delta$ and $\gamma$. A BCS model; $ \Delta (T)
= \Delta (0)\tanh (1.74\sqrt {(T_C  - T)/T} ) $ gave $\Delta (0)$
= 1.44 meV and $T_C$ = 9.86 K. (d) Temperature dependence of
resistance $R(T)$ of the same junction.}
\end{figure}

\begin{figure}[!h]
\begin{tabular}{ll}

  \centering
  \includegraphics[width= 6 cm]{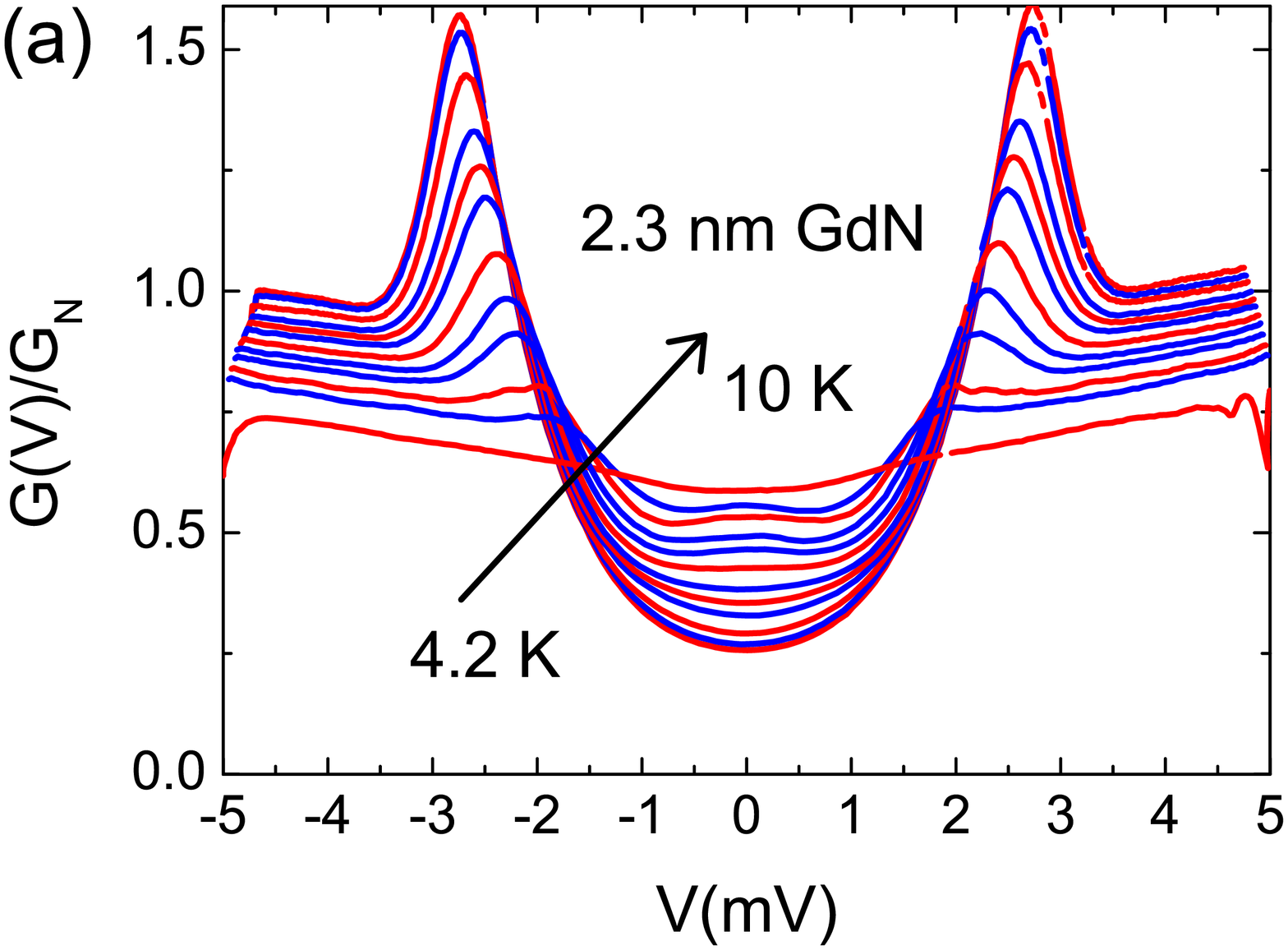}
&

  \includegraphics[width= 6 cm]{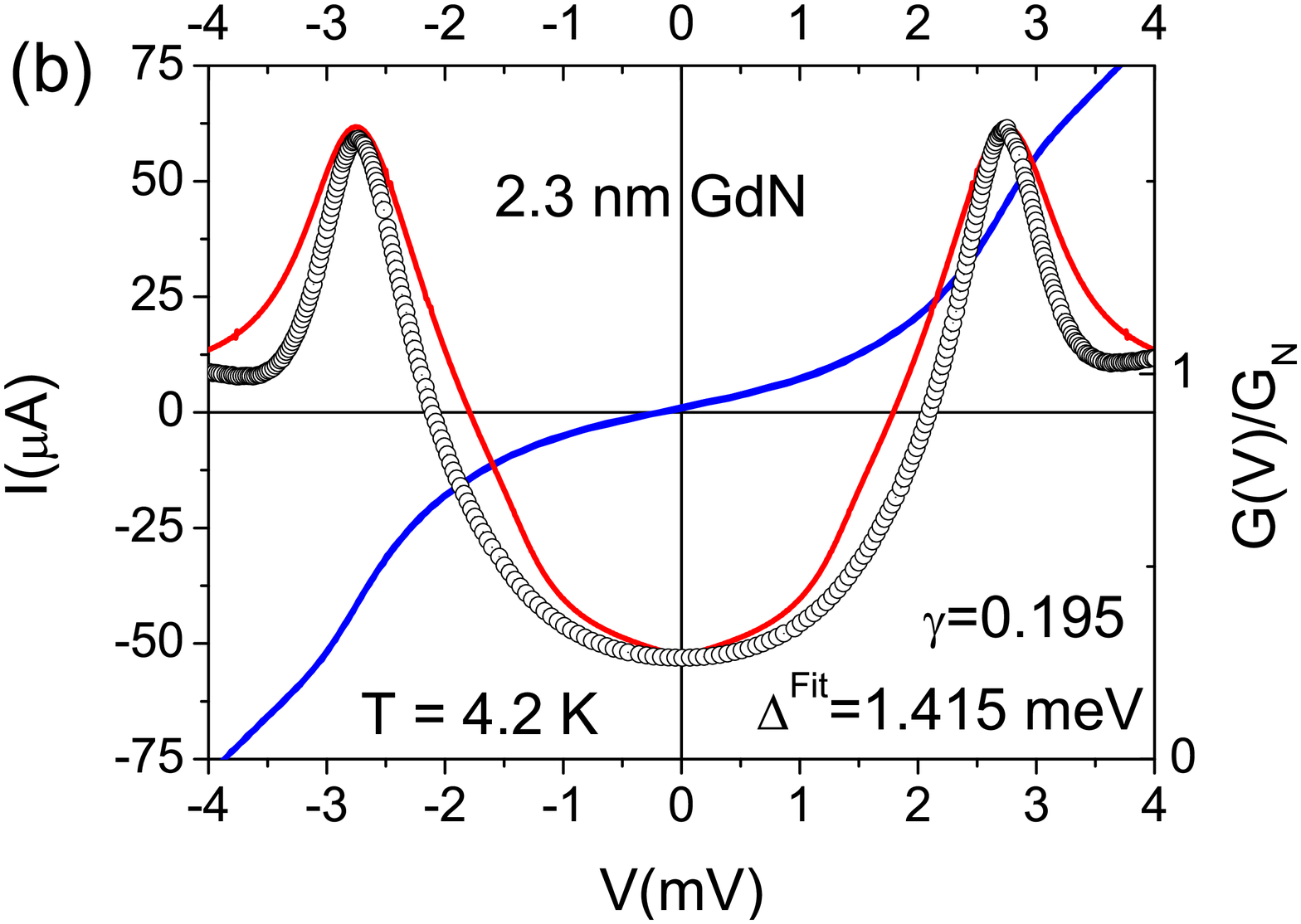}\\

 \centering
 \includegraphics[width= 6 cm]{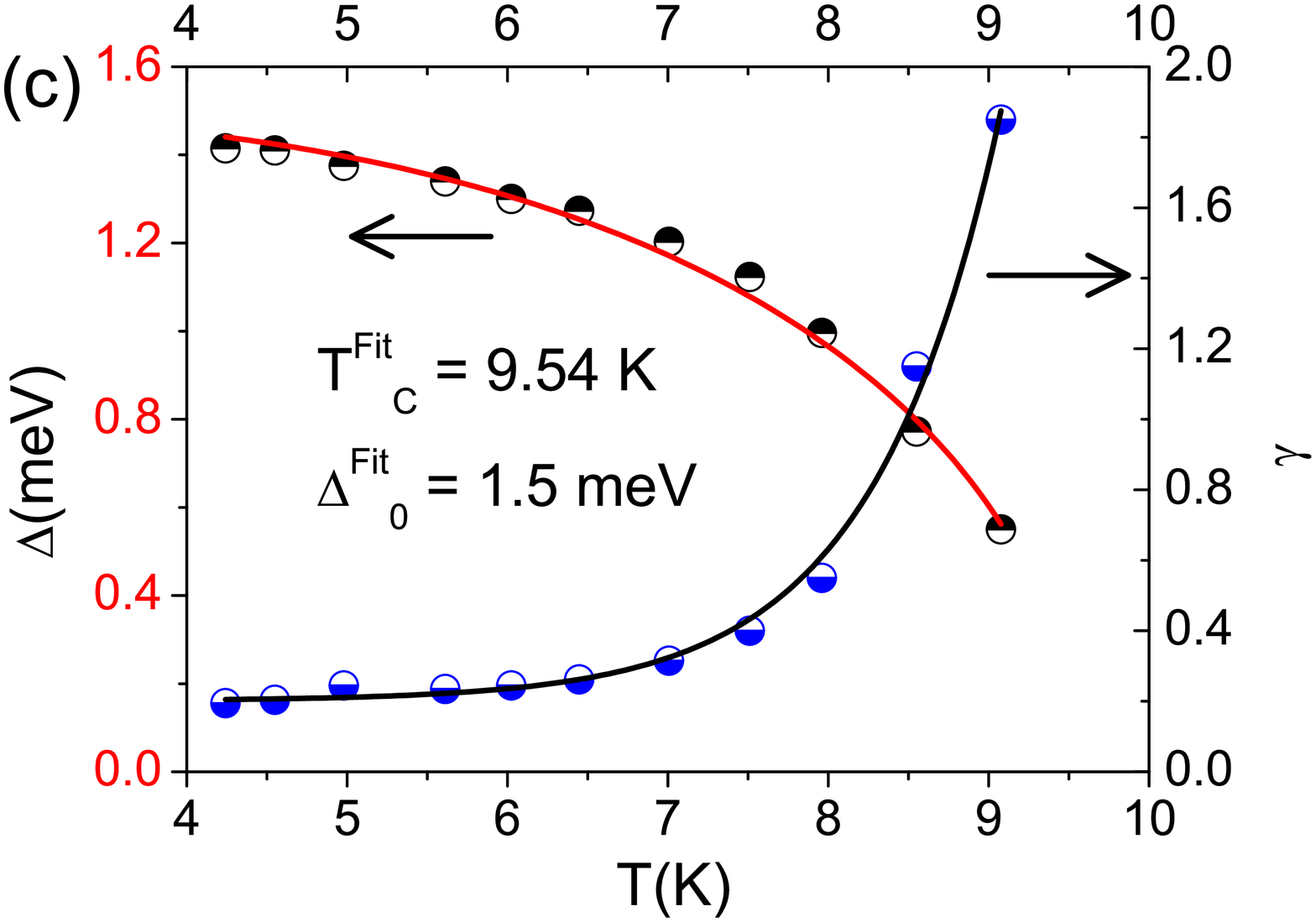}
 &

 \includegraphics[width= 6 cm]{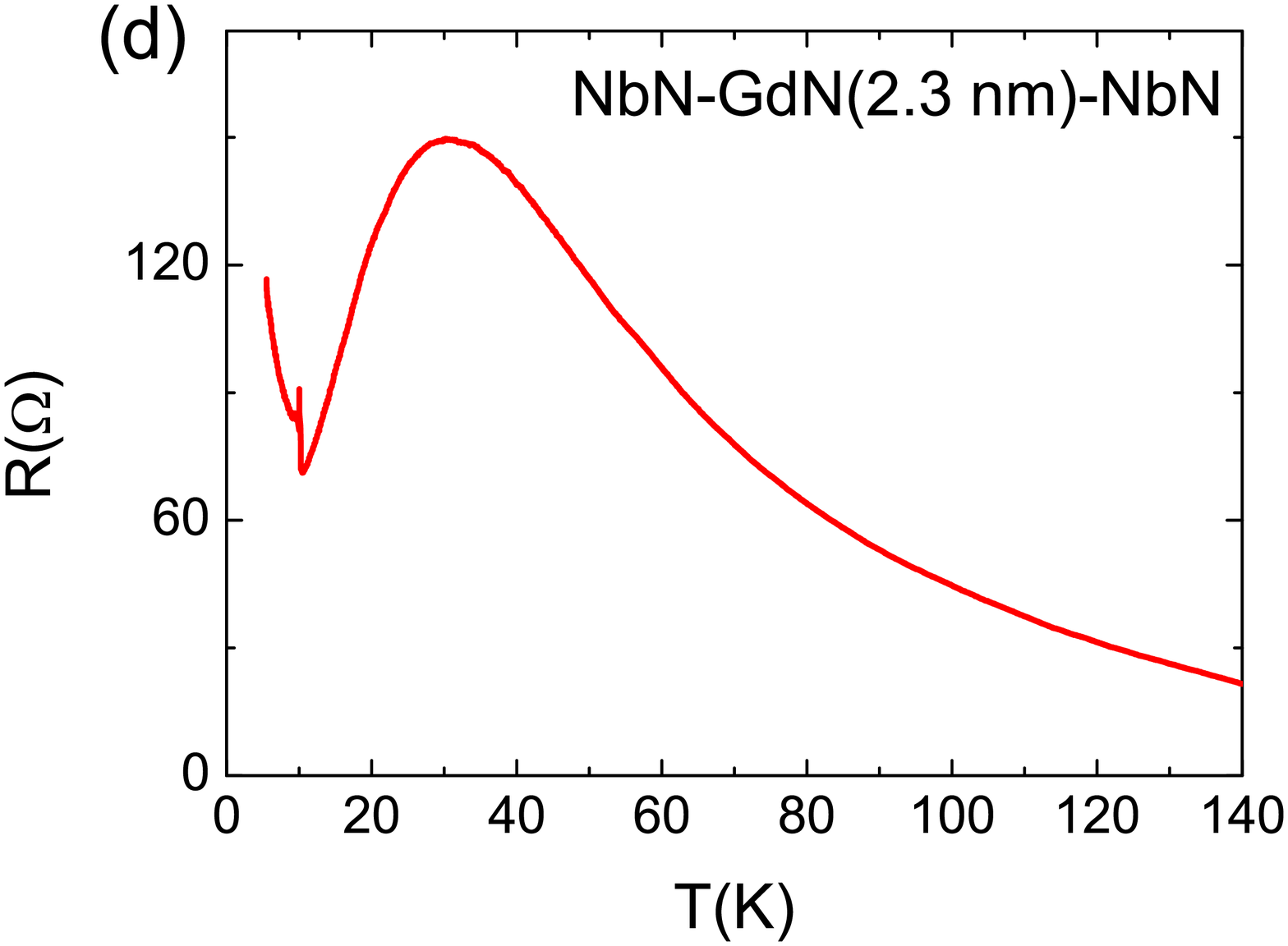}\\

\end{tabular}
\caption{(a) Temperature evolution of normalized conductance
spectra $G(V)/G_N$ of NbN-GdN(2.3 nm)-NbN tunnel junction. The
conductance spectra were measured with a standard lock-in
technique.(b)I-V and normalized conductance spectra $G(V)/G_N$ of
the same junction at 4.2 K. Red solid line is the fit to Eq. (4)
with fitting parameter shown. Temperature dependence of the
fitting parameter $\Delta$ and $\gamma$. A BCS model; $ \Delta (T)
= \Delta (0)\tanh (1.74\sqrt {(T_C  - T)/T} ) $ gave $\Delta (0)$
= 1.5 meV and $T_C$ = 9.54 K. (d) Temperature dependence of
resistance $R(T)$ of the same junction.}
\end{figure}

\begin{figure}[!h]
\begin{center}
\abovecaptionskip -10cm
\includegraphics [width=8 cm]{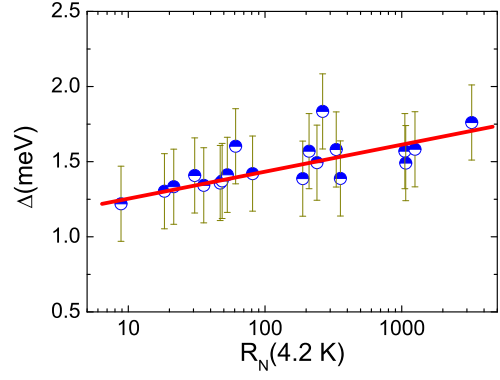}
\end{center}
\caption{\label{fig1} (Color online) Superonducting gap $\Delta$
of 20 NbN(50 nm)-GdN($d$)-NbN(50 nm) tunnel junctions plotted
against resistance of the junction measured at 4.2 K ($R_N(4 mV)$)
with bias voltage 4 mV ($>$ 2$\Delta$). Maximum error of 0.25 mV
was assumed due to smeared gap edges. Conductance spectra of each
junction is shown in figures SFig. 14-16 . All the 20 NbN-GdN-NbN
tunnel junctions were not prepared in the same deposition.
Therefore, in this graph $R_N(4 mV)$ is plotted against $\Delta$
instead of thickness of GdN layer $d$ vs $\Delta$. Note that in a
tunnel junction $R \propto e^{ - \kappa d}$ with $\kappa  =  -
\frac{2}{\hbar }\sqrt {2m\Phi }$; where $\Phi$ is the barrier
height and $m$ is electron mass.
  }
\end{figure}

\newpage
\begin{widetext}

\begin{figure}[!h]
\begin{tabular}{ll}
  \centering
  \includegraphics[width= 6 cm]{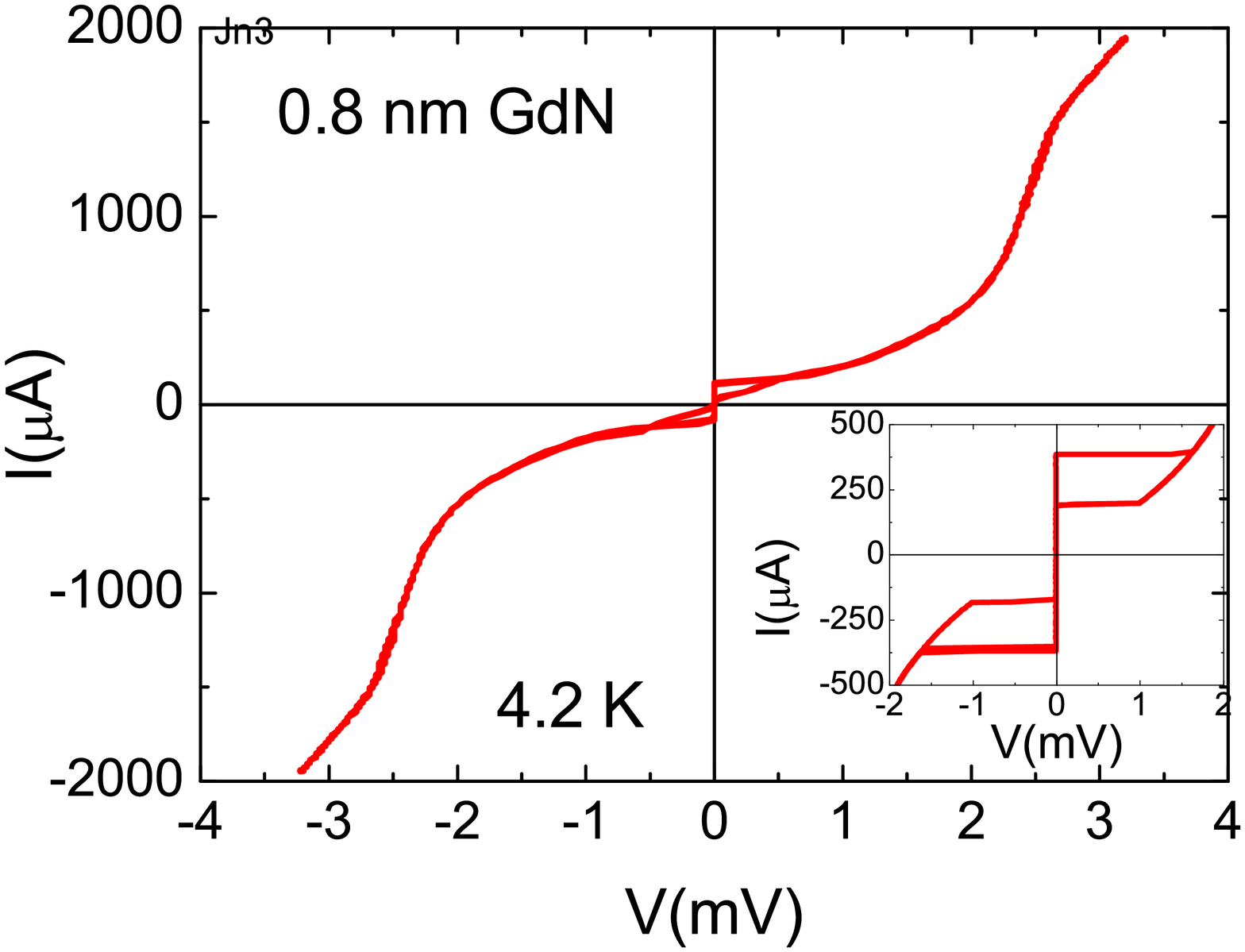}
&

  \includegraphics[width= 6 cm]{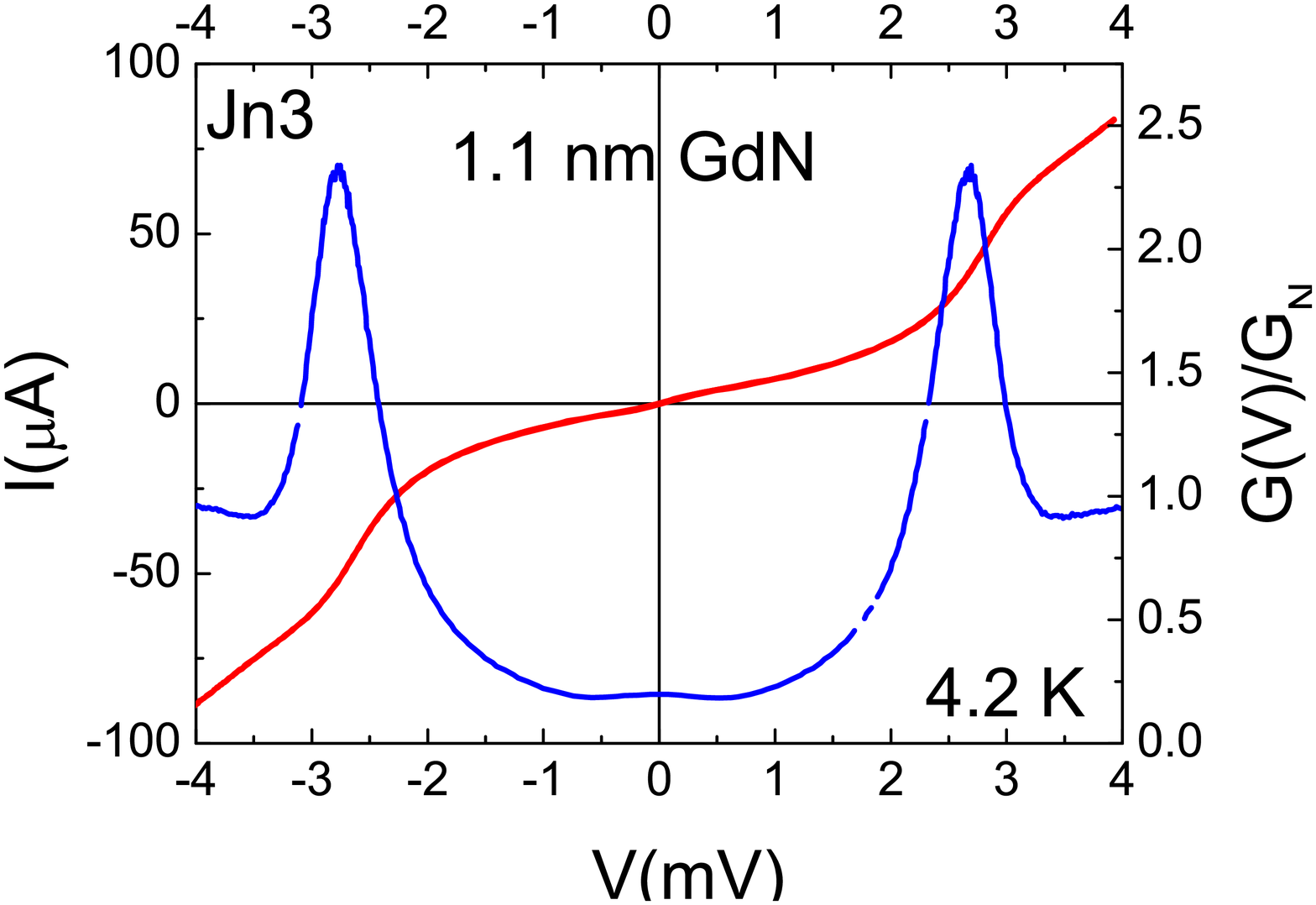}\\

 \centering
 \includegraphics[width= 6 cm]{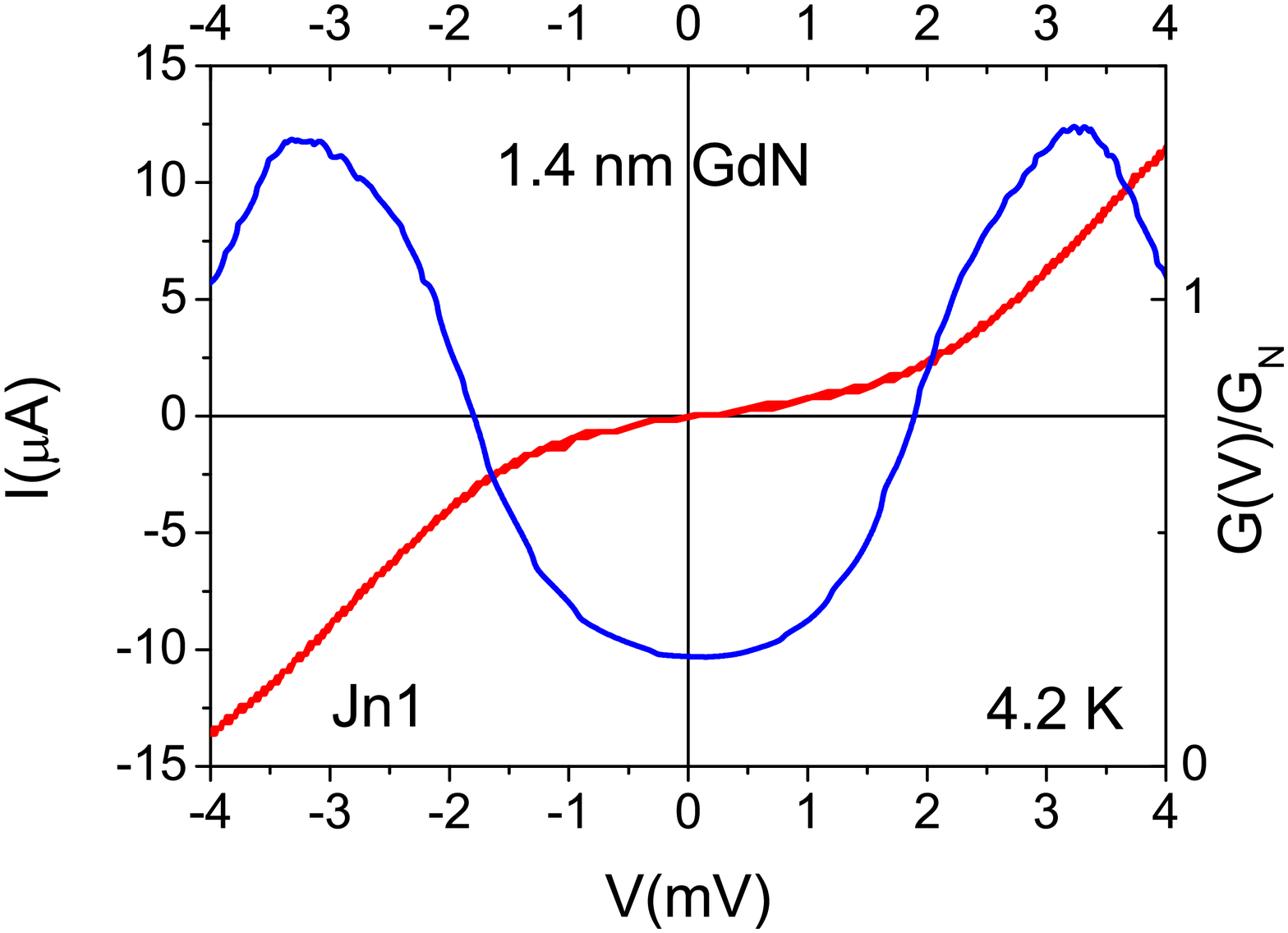}
&

 \includegraphics[width= 6 cm]{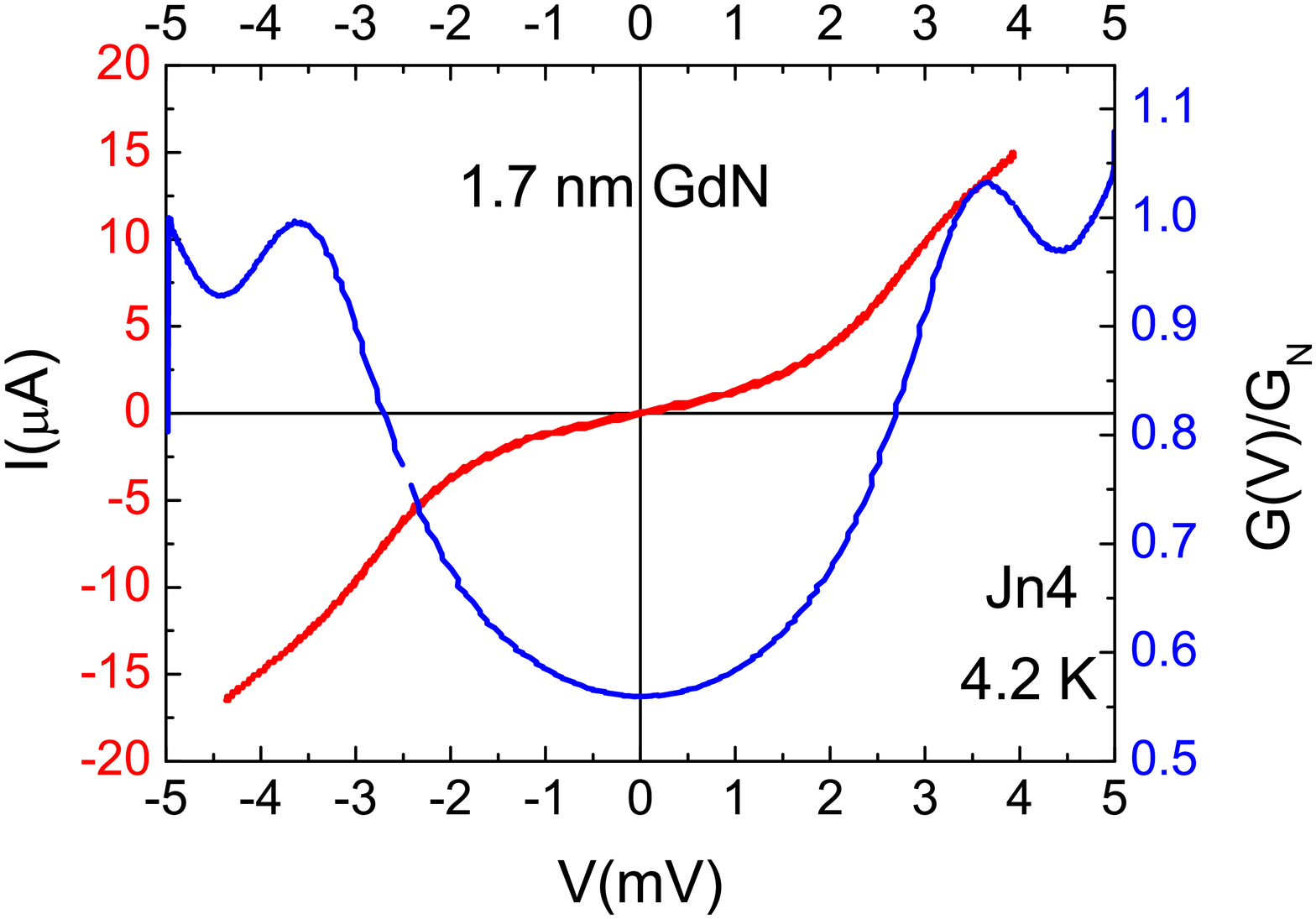}\\

 \includegraphics[width= 6 cm]{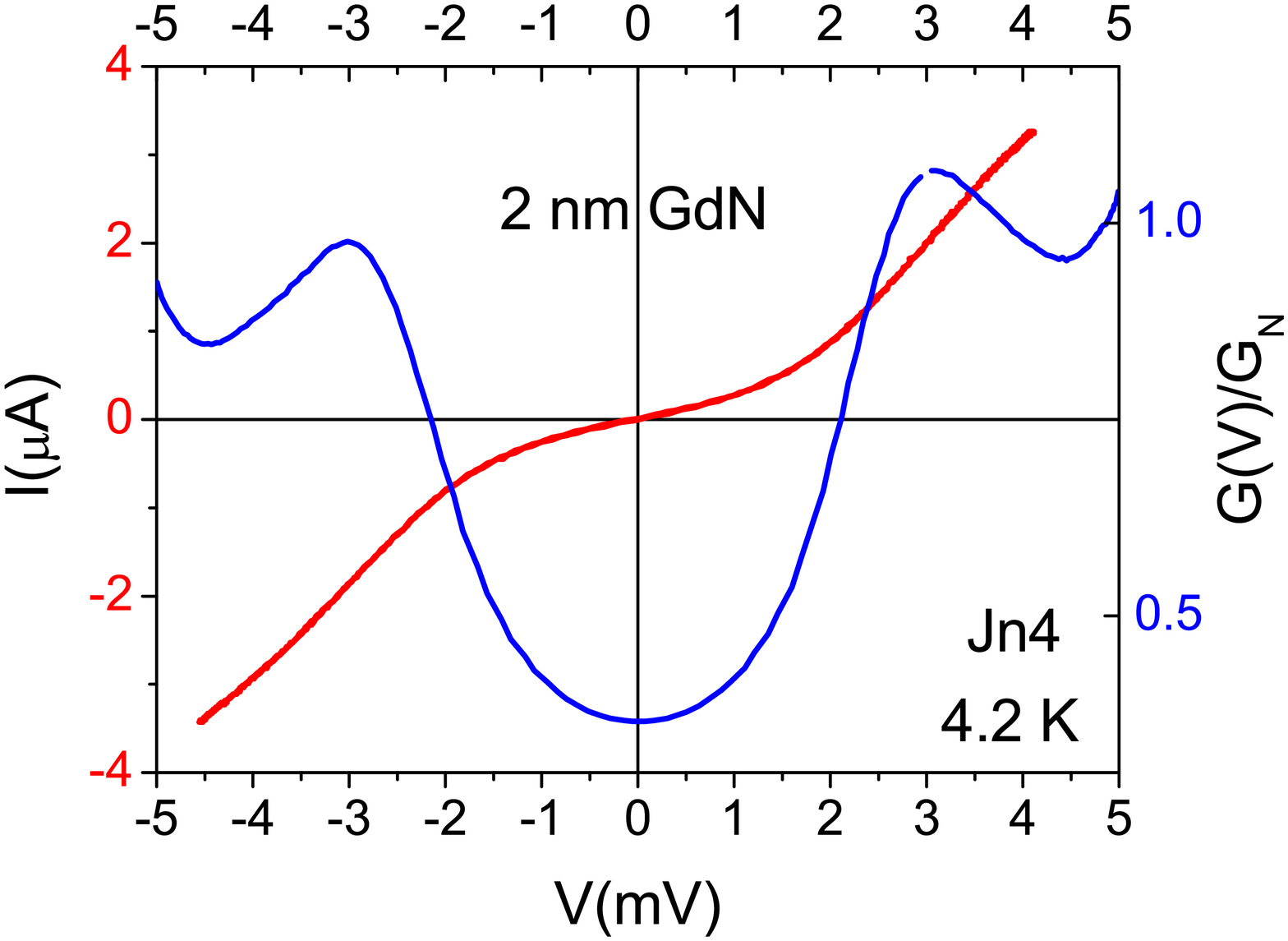}
&
 \includegraphics[width= 6 cm]{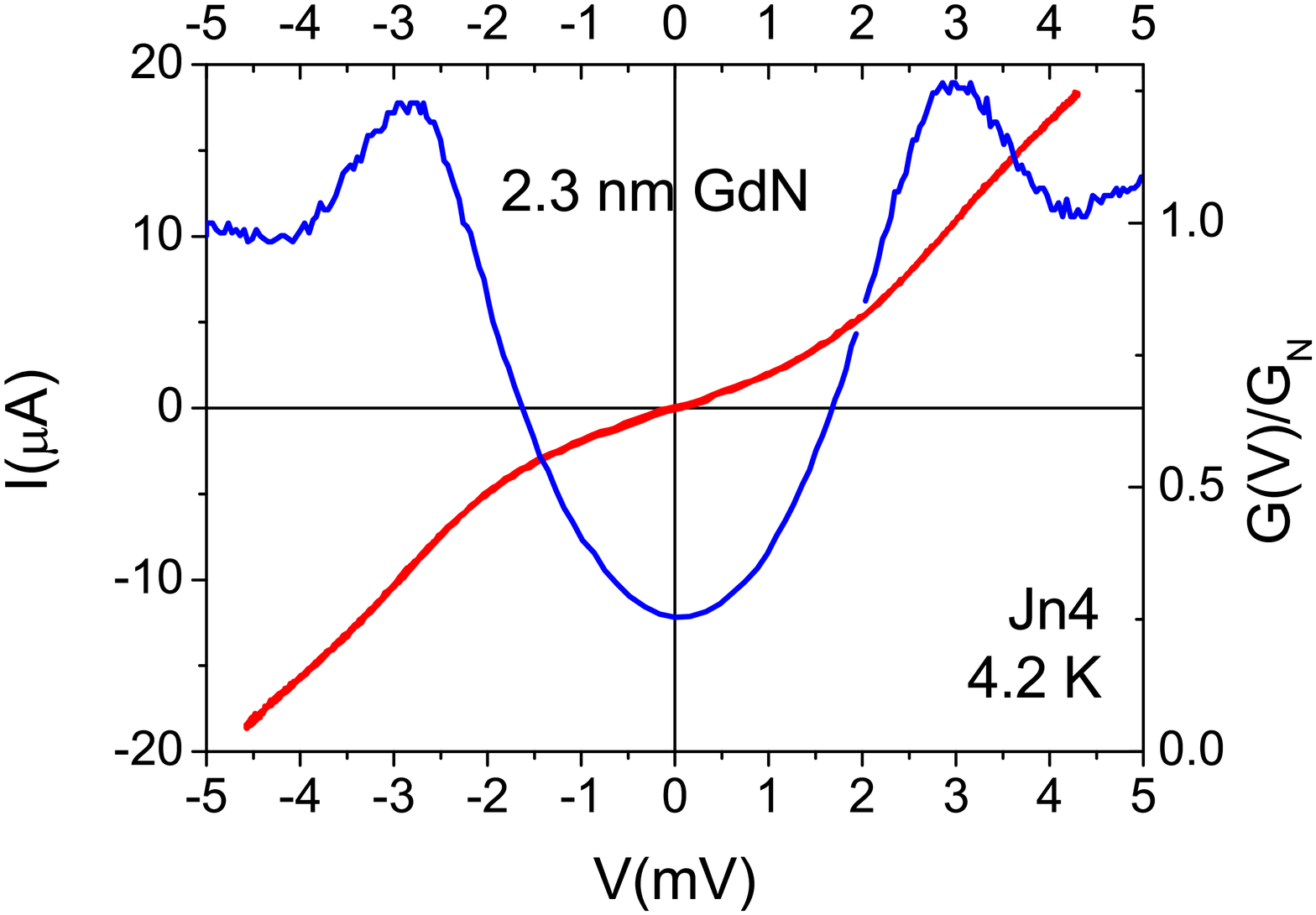}
\end{tabular}
\caption{SERIES-1 (23431):I-V and normalized conductance spectra
$G(V)/G_N$ of the NbN(50 nm)-GdN($t$)-NbN(50 nm) spin-filter
device with GdN thickness in the range 0.8-2.3 nm. Conductance
$G(V)$  measurement was not done  in junctions with critical
current due to divergence at the origin. All the tunnel junctions
were prepared from the trilayer stack deposited at the same time.
}
\end{figure}
\end{widetext}

\newpage

\begin{widetext}

\begin{figure}[!h]
\begin{tabular}{ll}
  \centering
  \includegraphics[width= 6 cm]{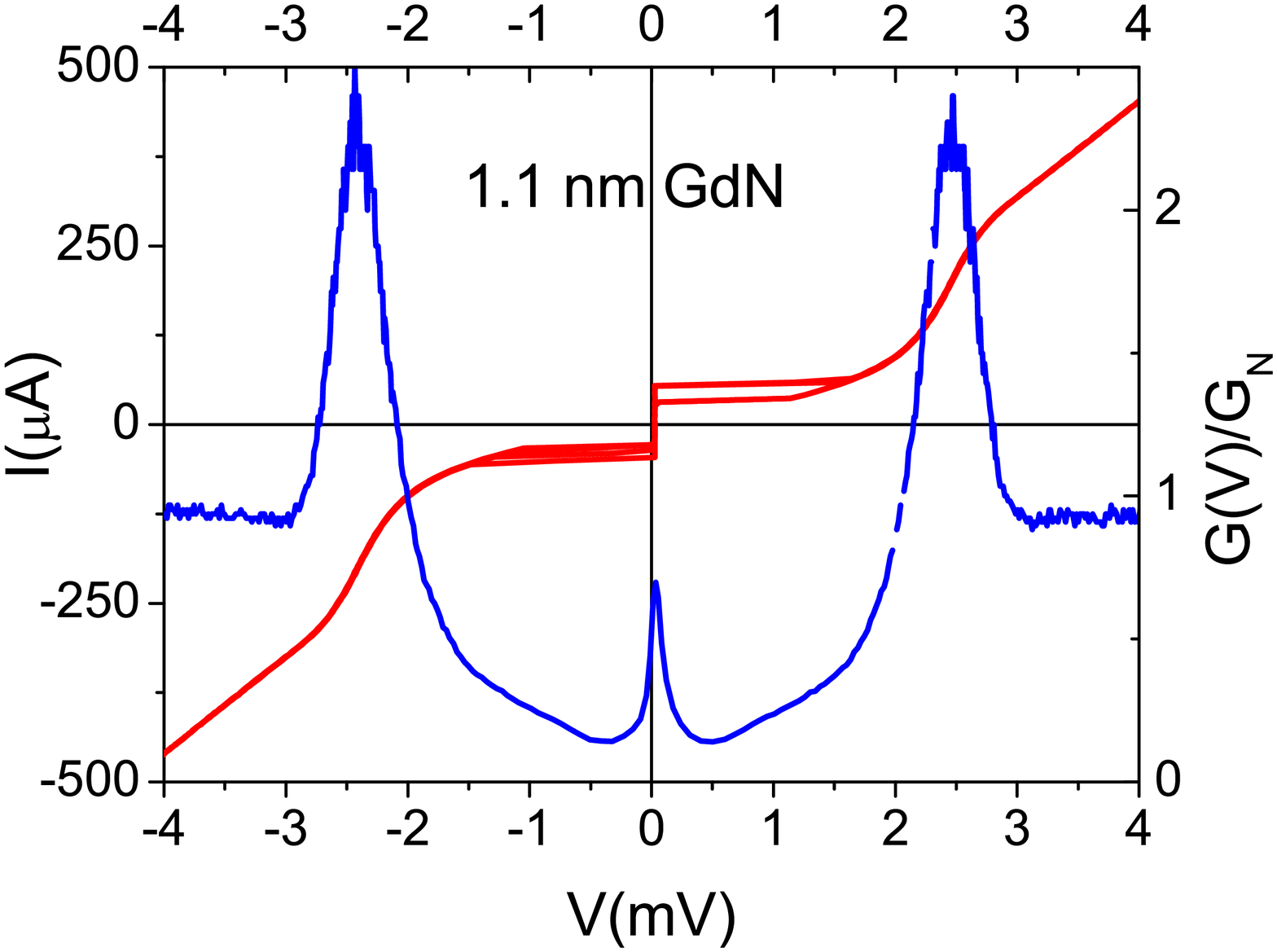}
&

  \includegraphics[width= 6 cm]{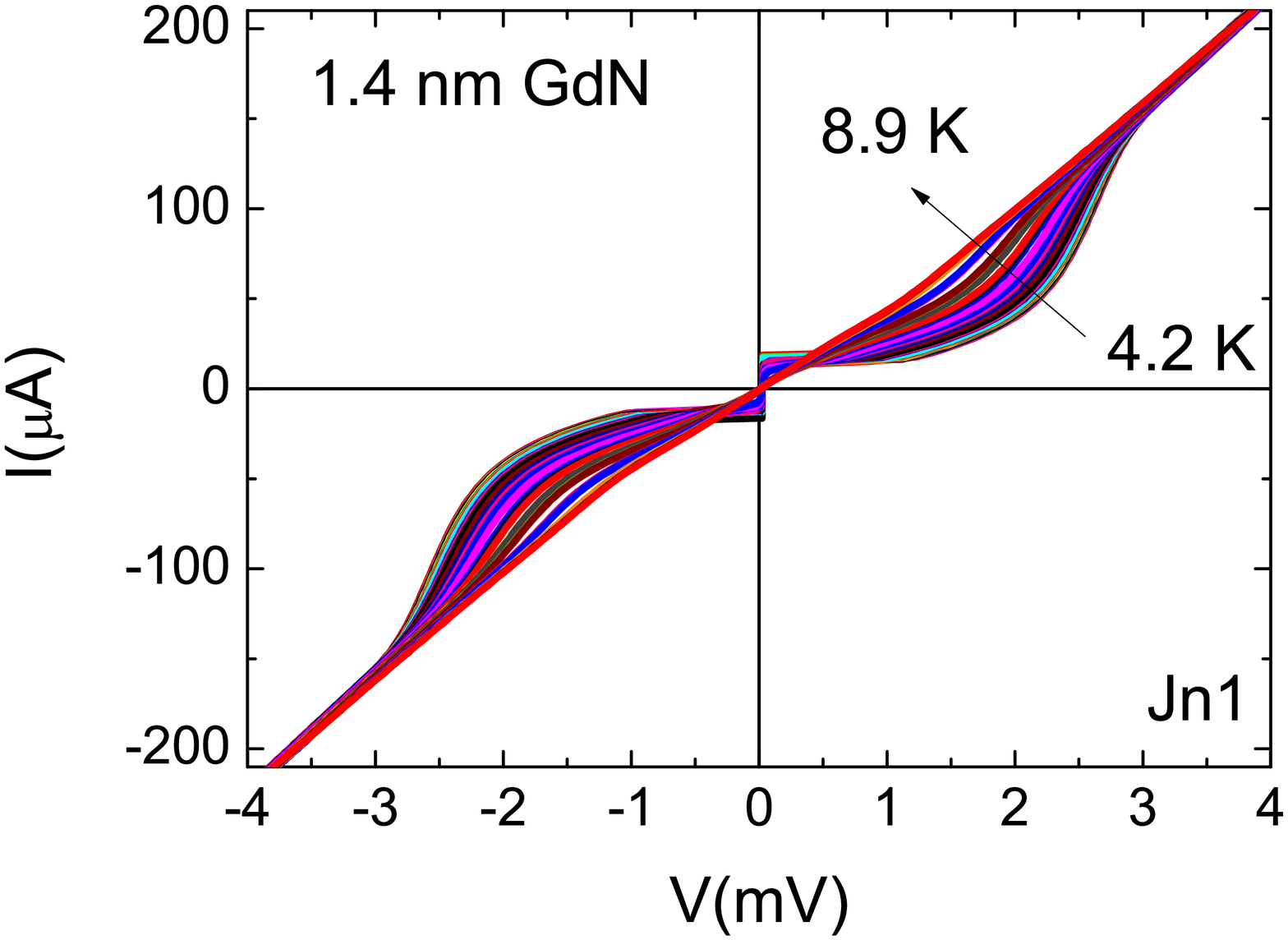}\\

 \centering
 \includegraphics[width= 6 cm]{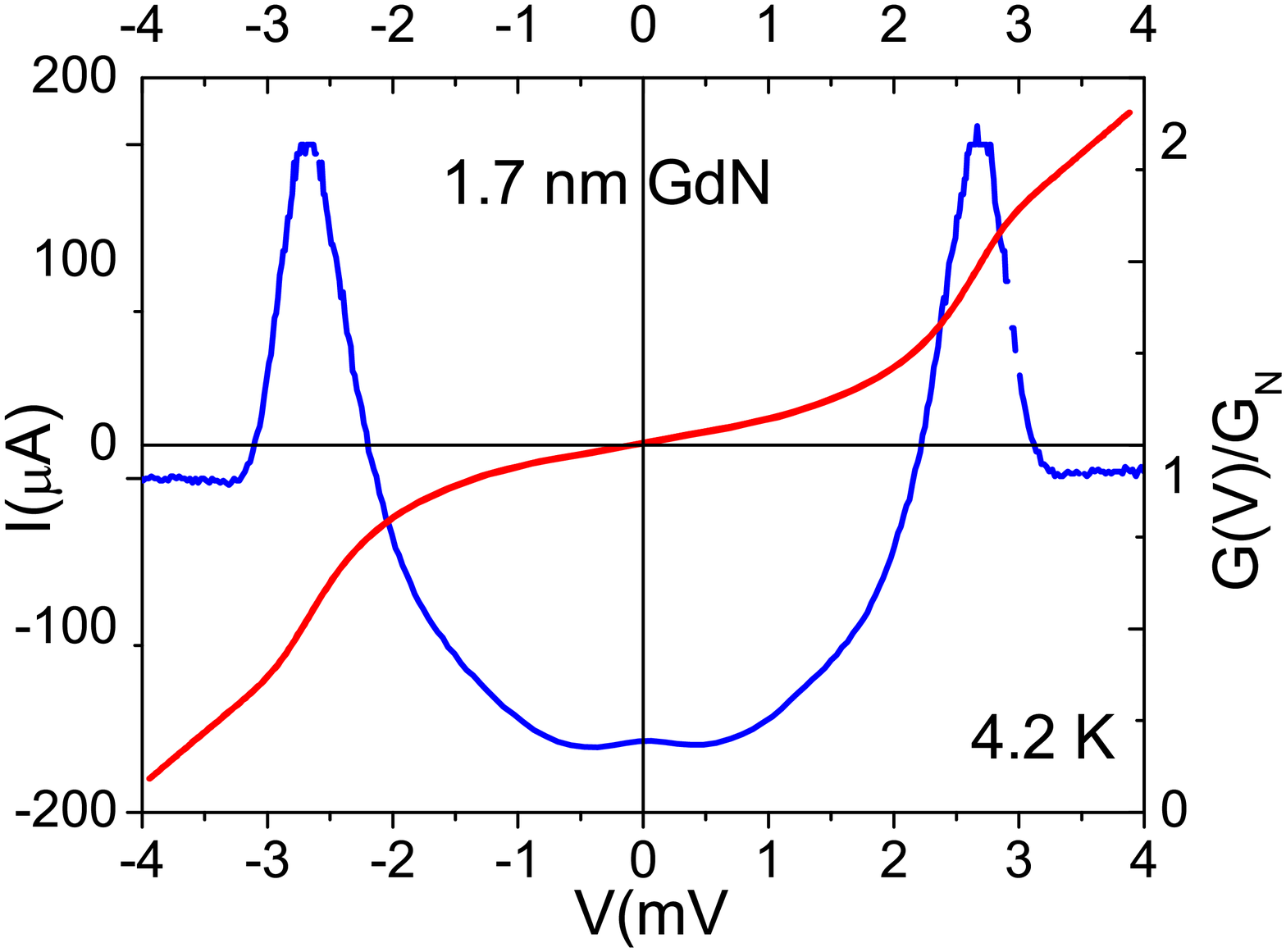}
&

 \includegraphics[width= 6 cm]{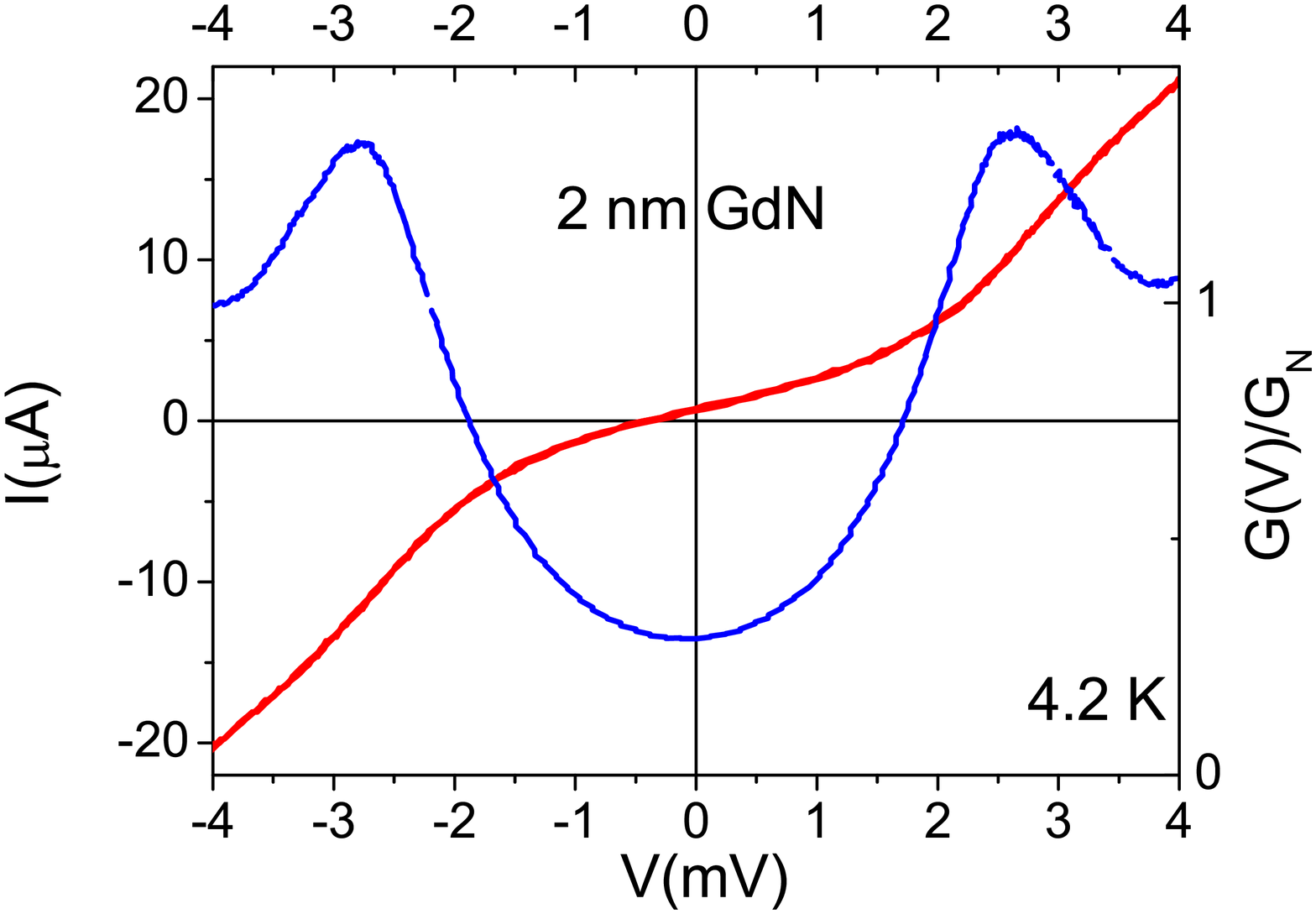}\\

 \includegraphics[width= 6 cm]{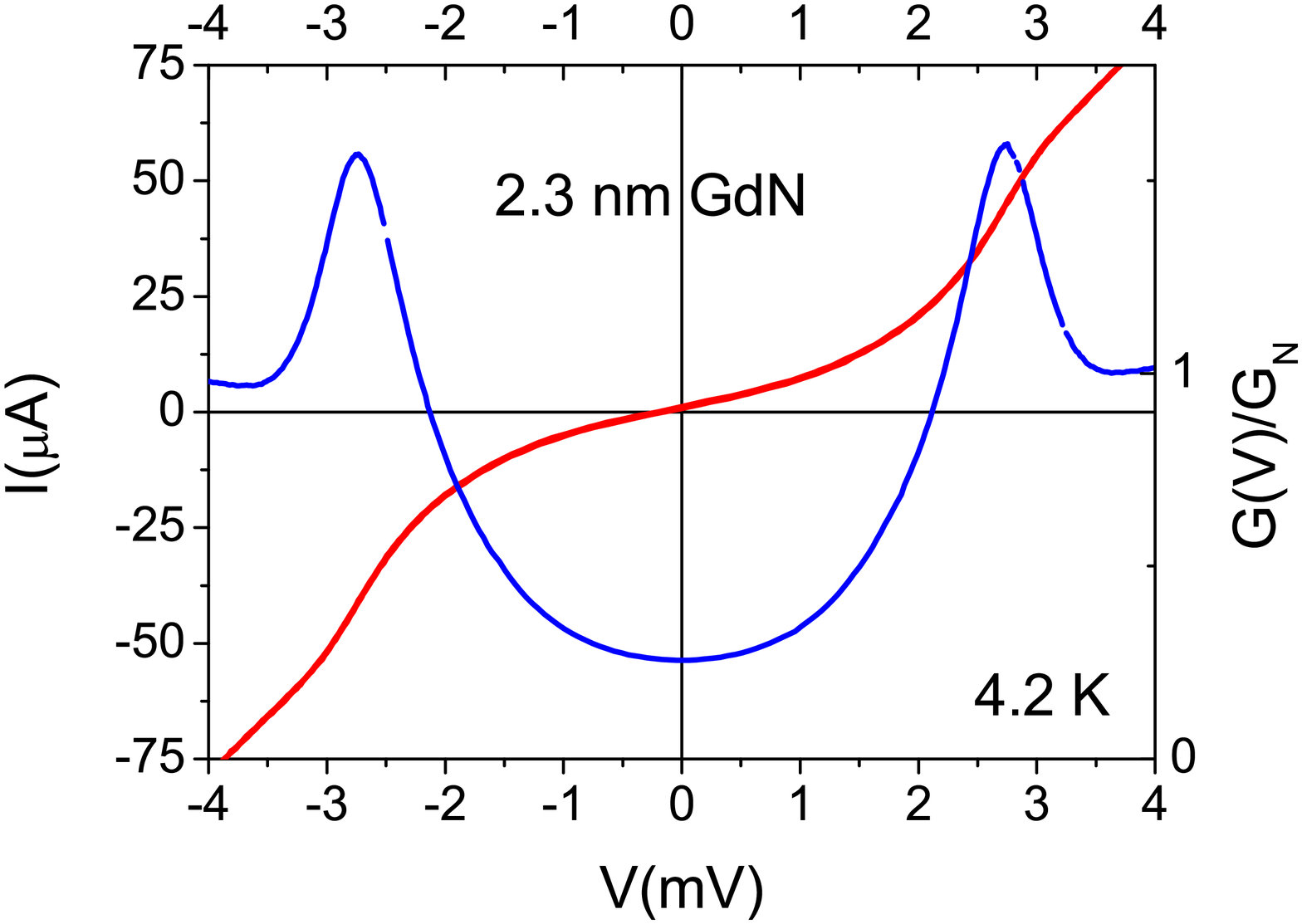}
&
 \includegraphics[width= 6 cm]{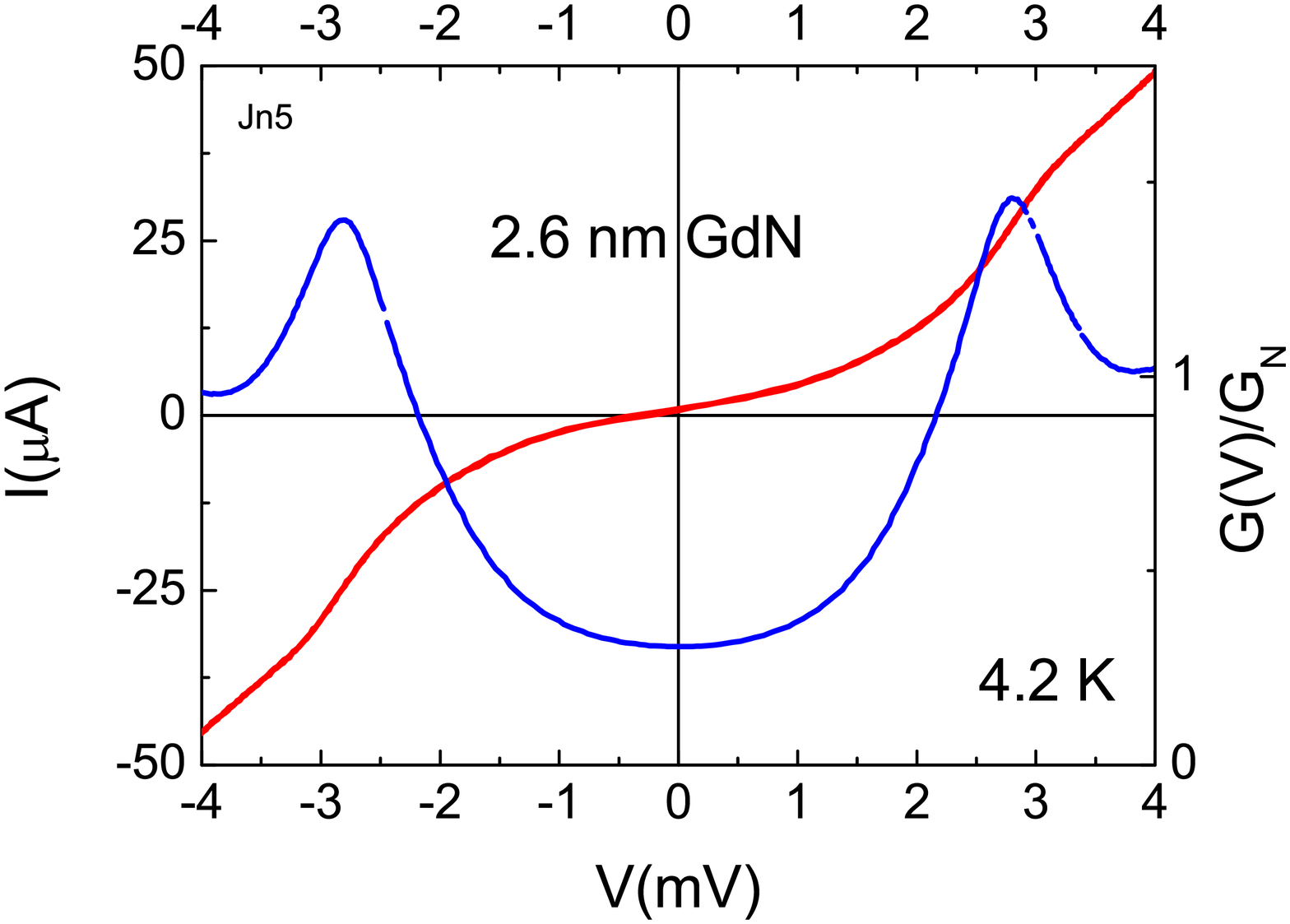}\\
 \centering
 \includegraphics[width= 6 cm]{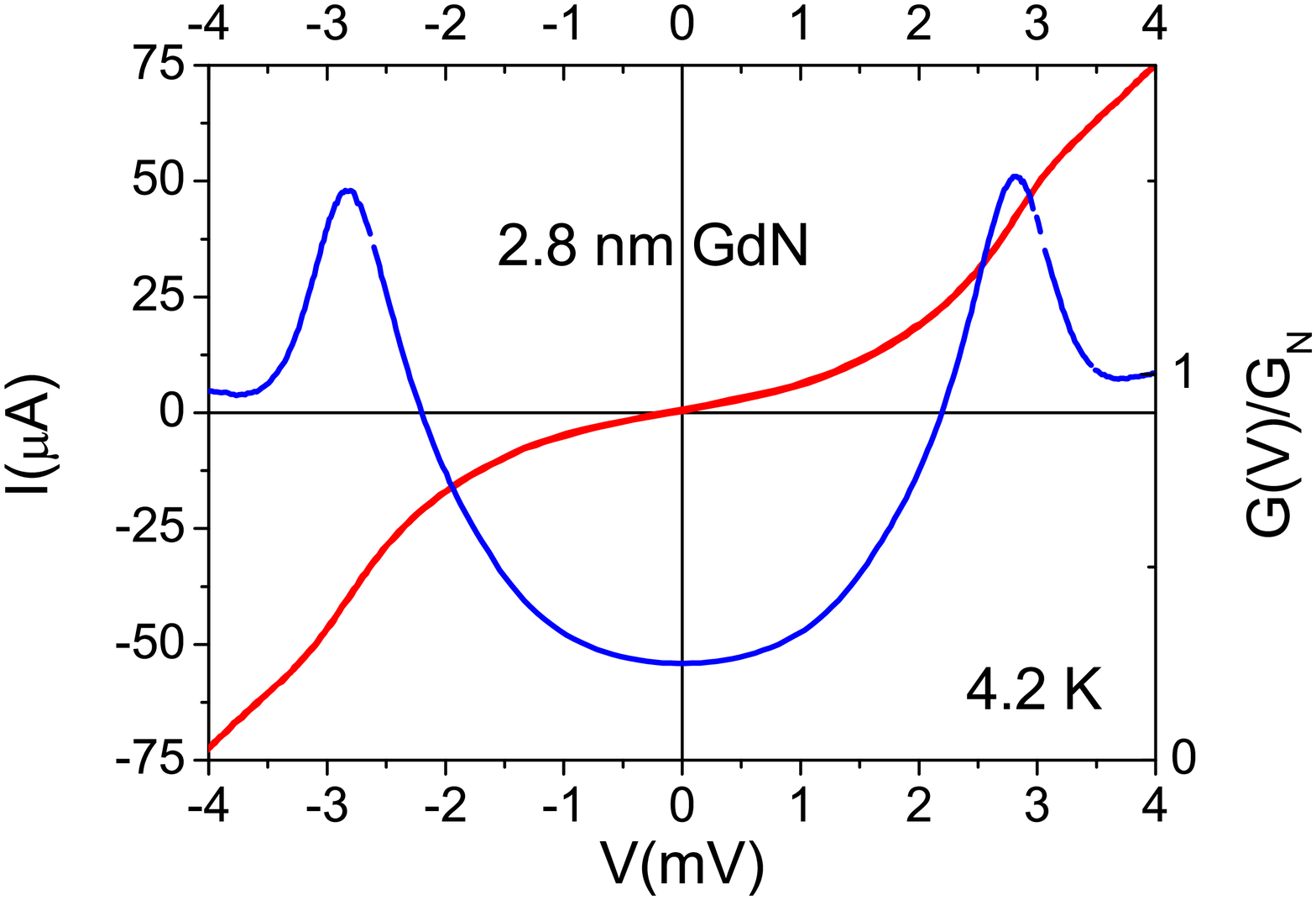}
&
 \includegraphics[width= 6 cm]{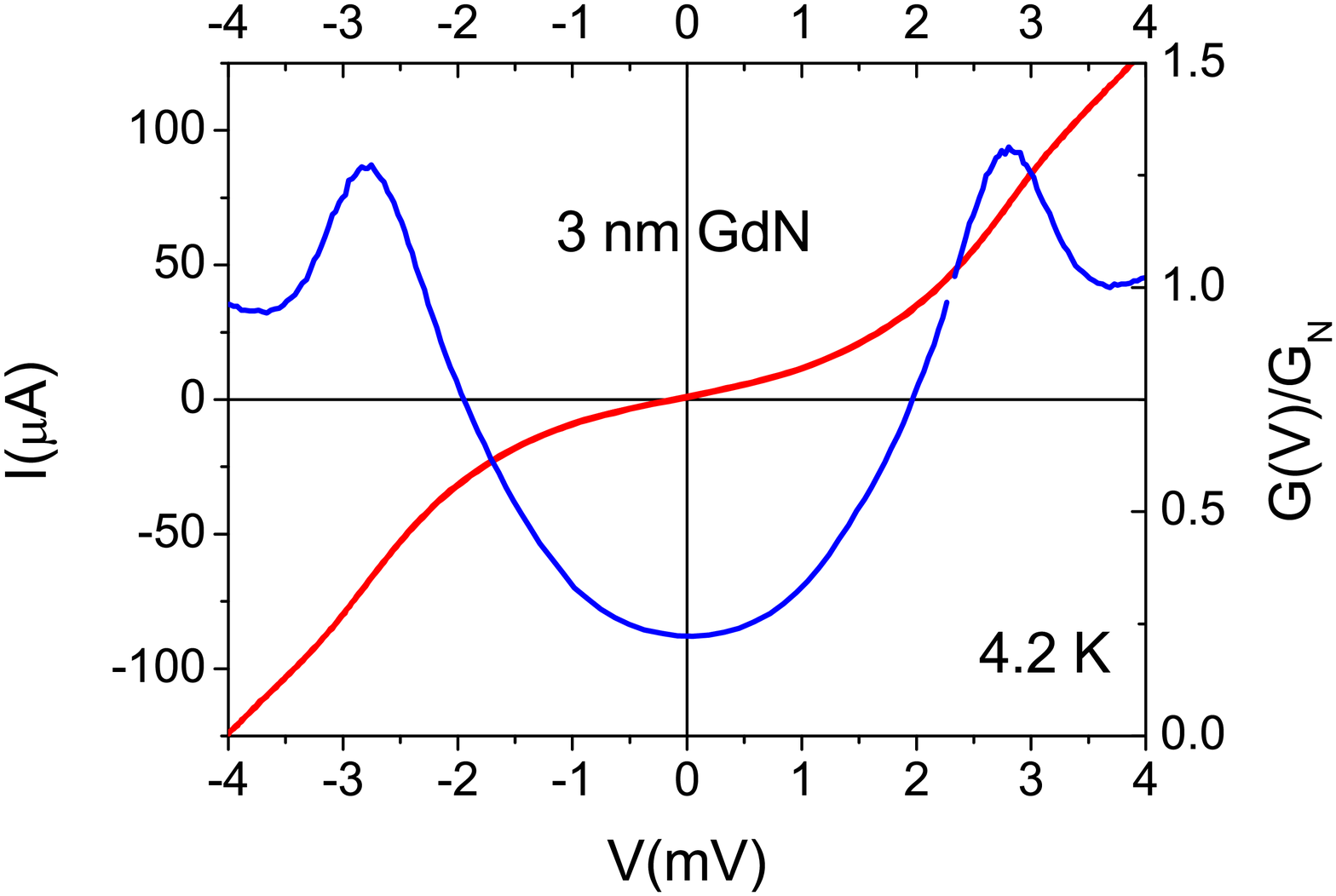}

\end{tabular}
\caption{SERIES-2 (23432):I-V and normalized conductance spectra
$G(V)/G_N$ of the NbN(50 nm)-GdN($t$)-NbN(50 nm) spin-filter
device with GdN thickness in the range 1.1-3.3 nm. Conductance
$G(V)$  measurement was not done  in junctions with critical
current due to divergence at the origin. All the tunnel junctions
were prepared from the trilayer stack deposited at the same time.}
\end{figure}
\end{widetext}

\newpage
\begin{widetext}

\begin{figure}[!h]
\begin{tabular}{ll}
  \centering
  \includegraphics[width= 6 cm]{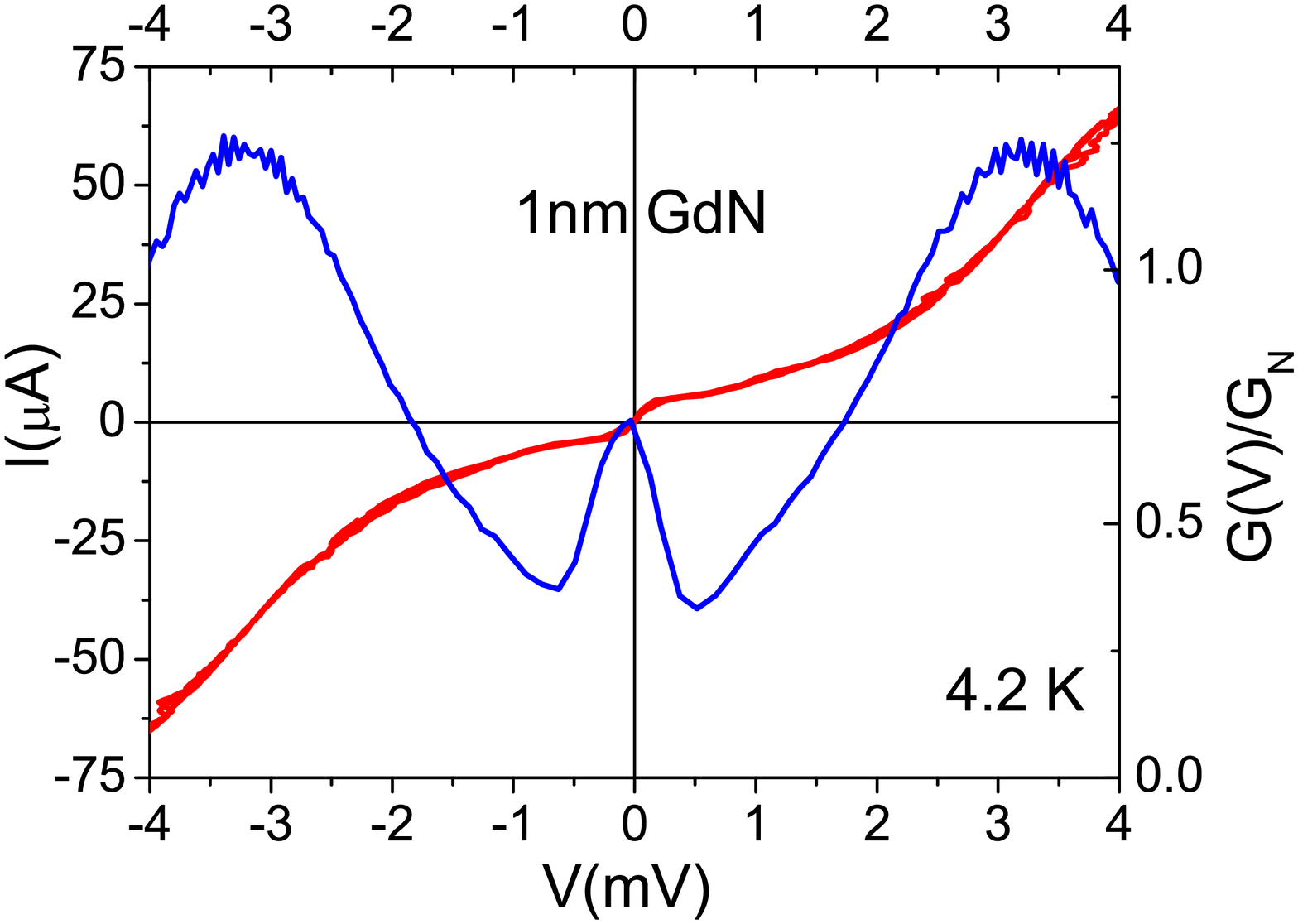}
&

  \includegraphics[width= 6 cm]{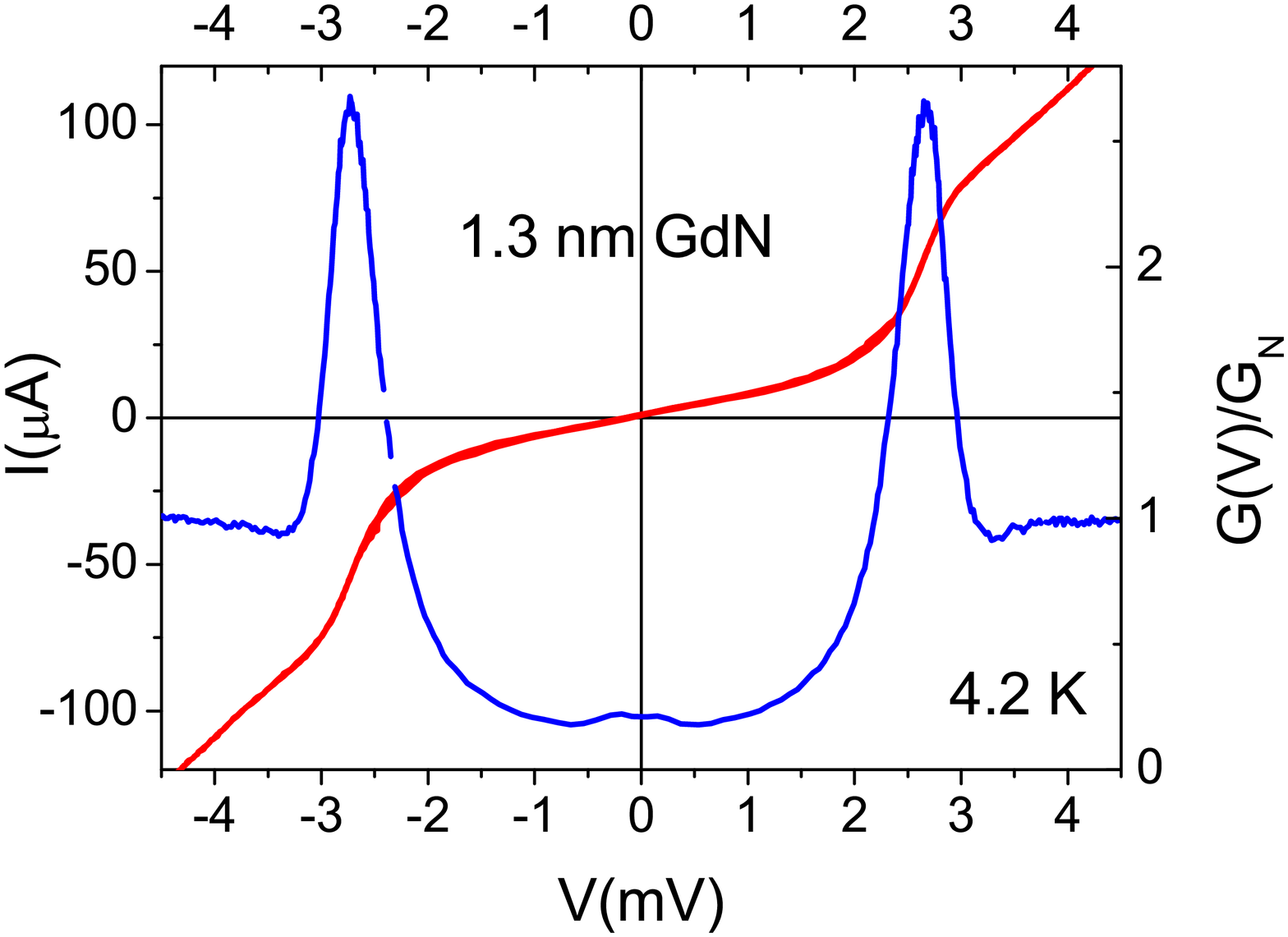}\\

 \centering
 \includegraphics[width= 6 cm]{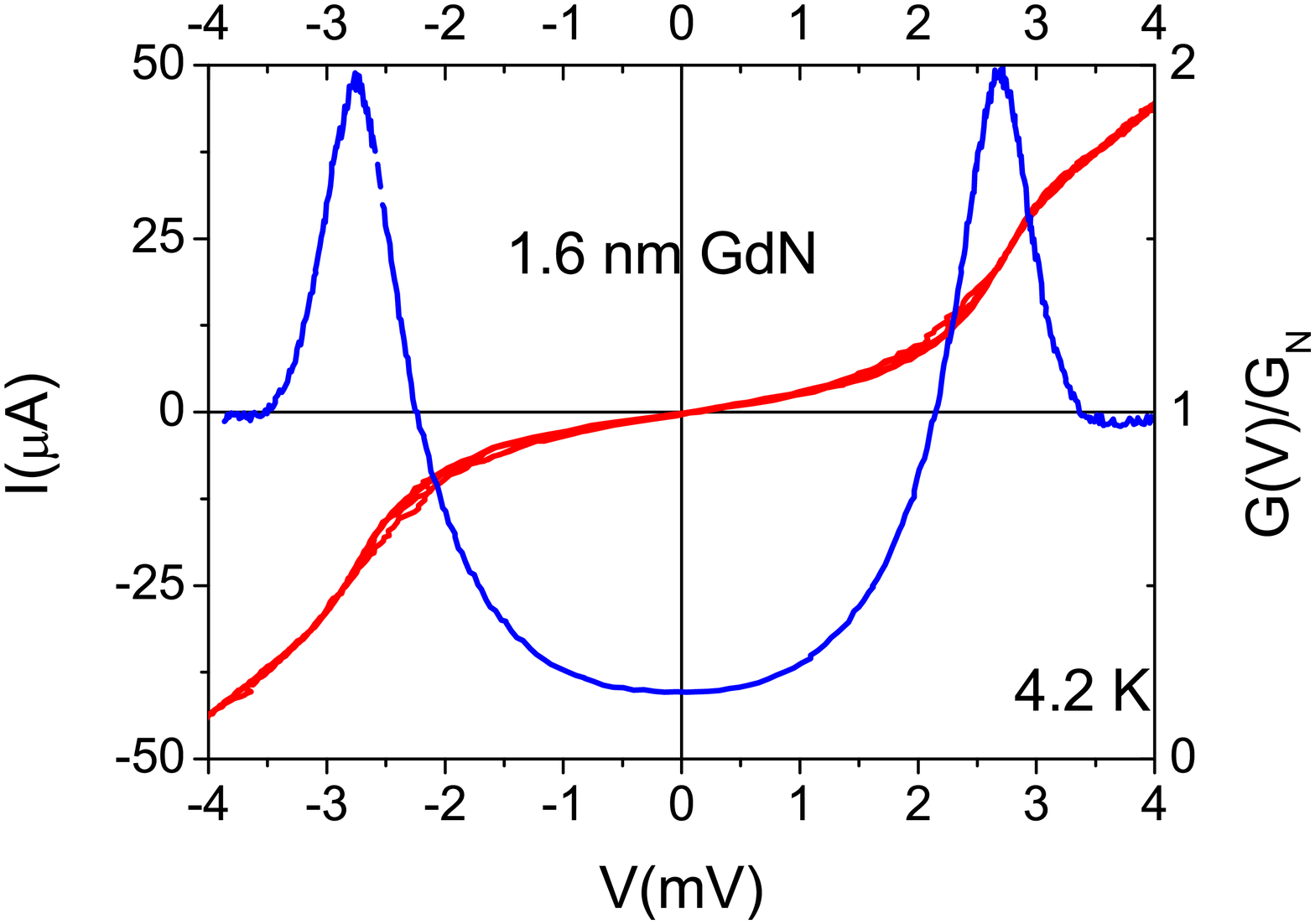}
&

 \includegraphics[width= 6 cm]{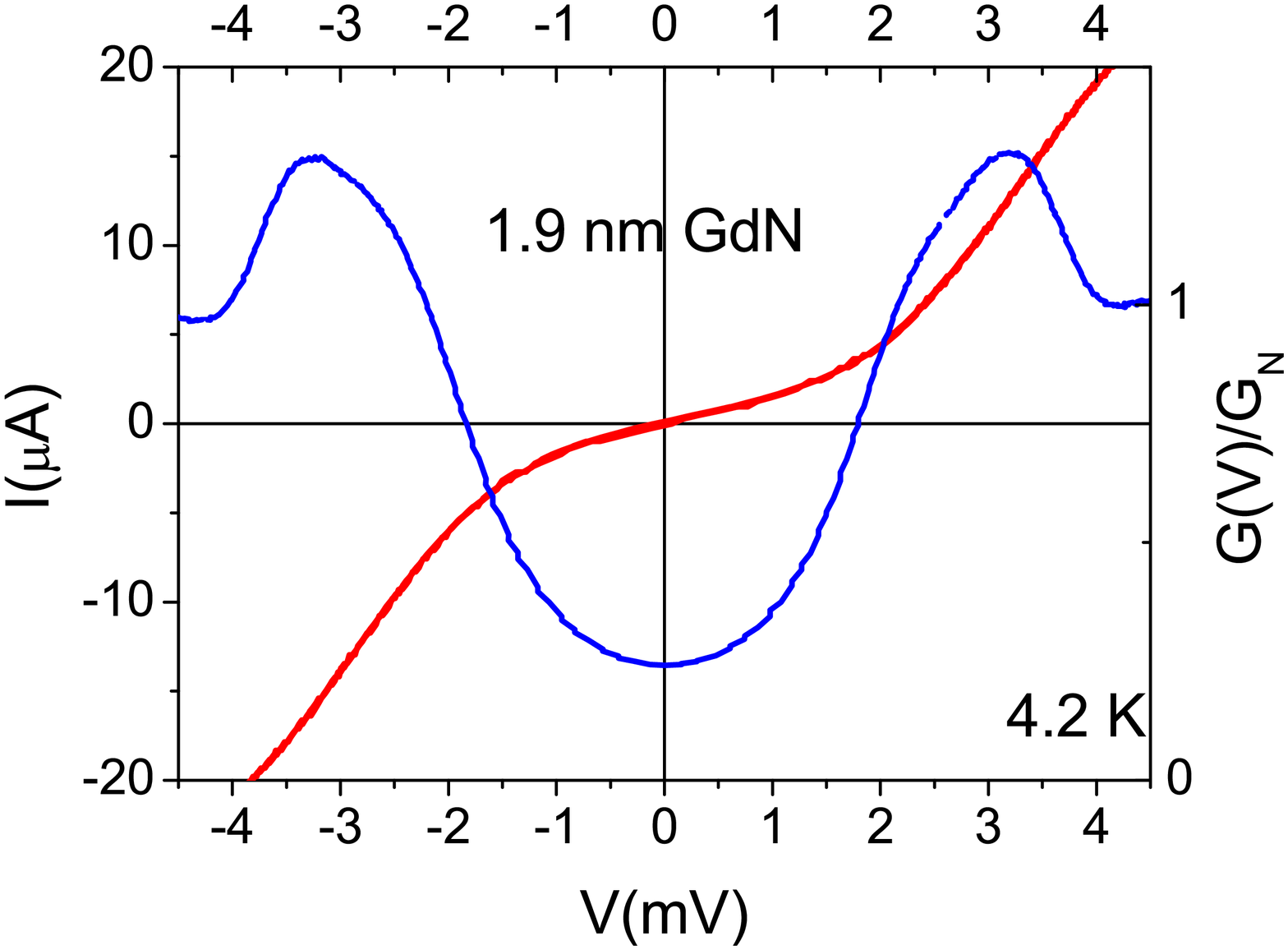}\\

 \centering
 \includegraphics[width= 6 cm]{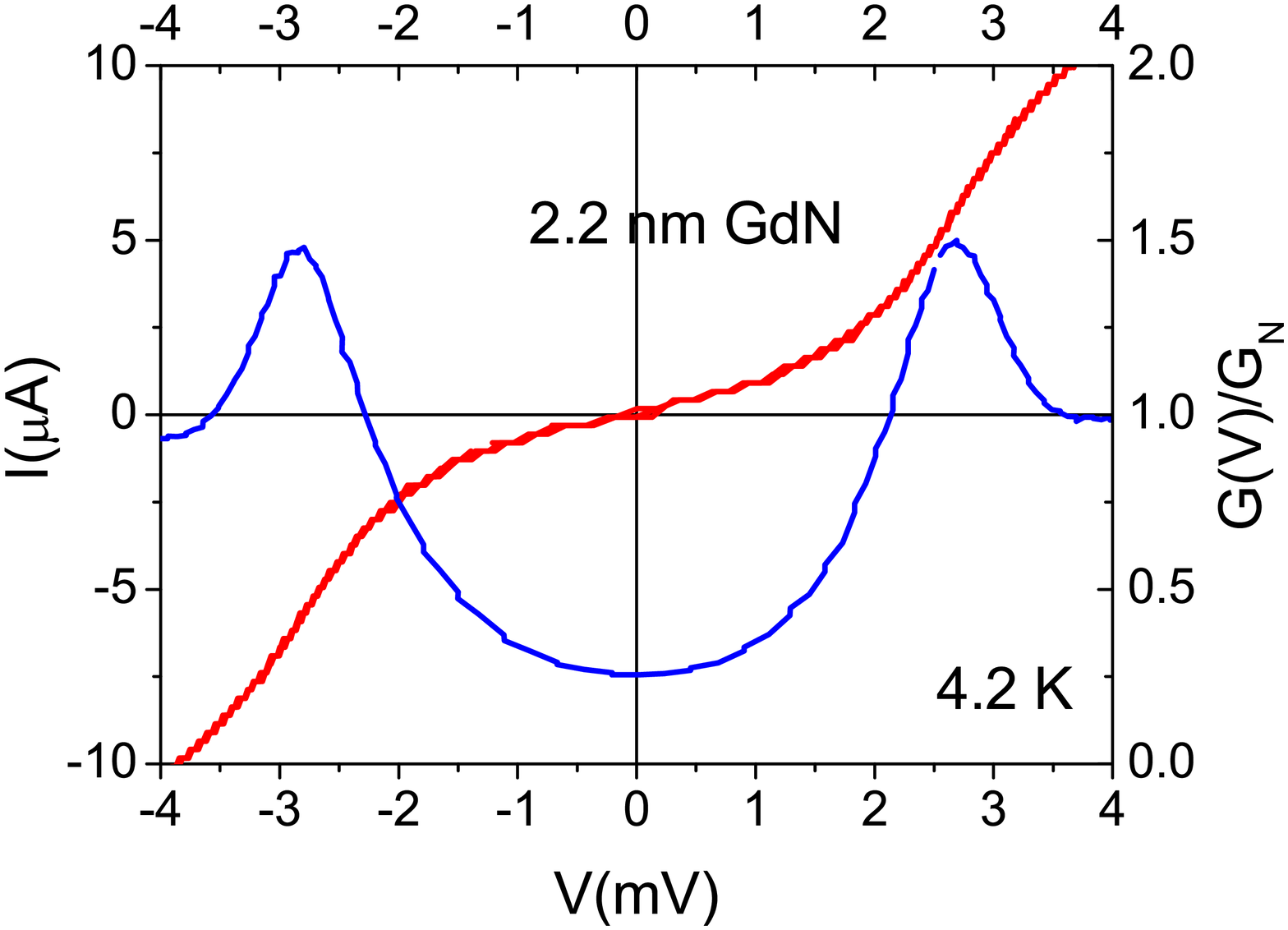}
&
 \includegraphics[width= 6 cm]{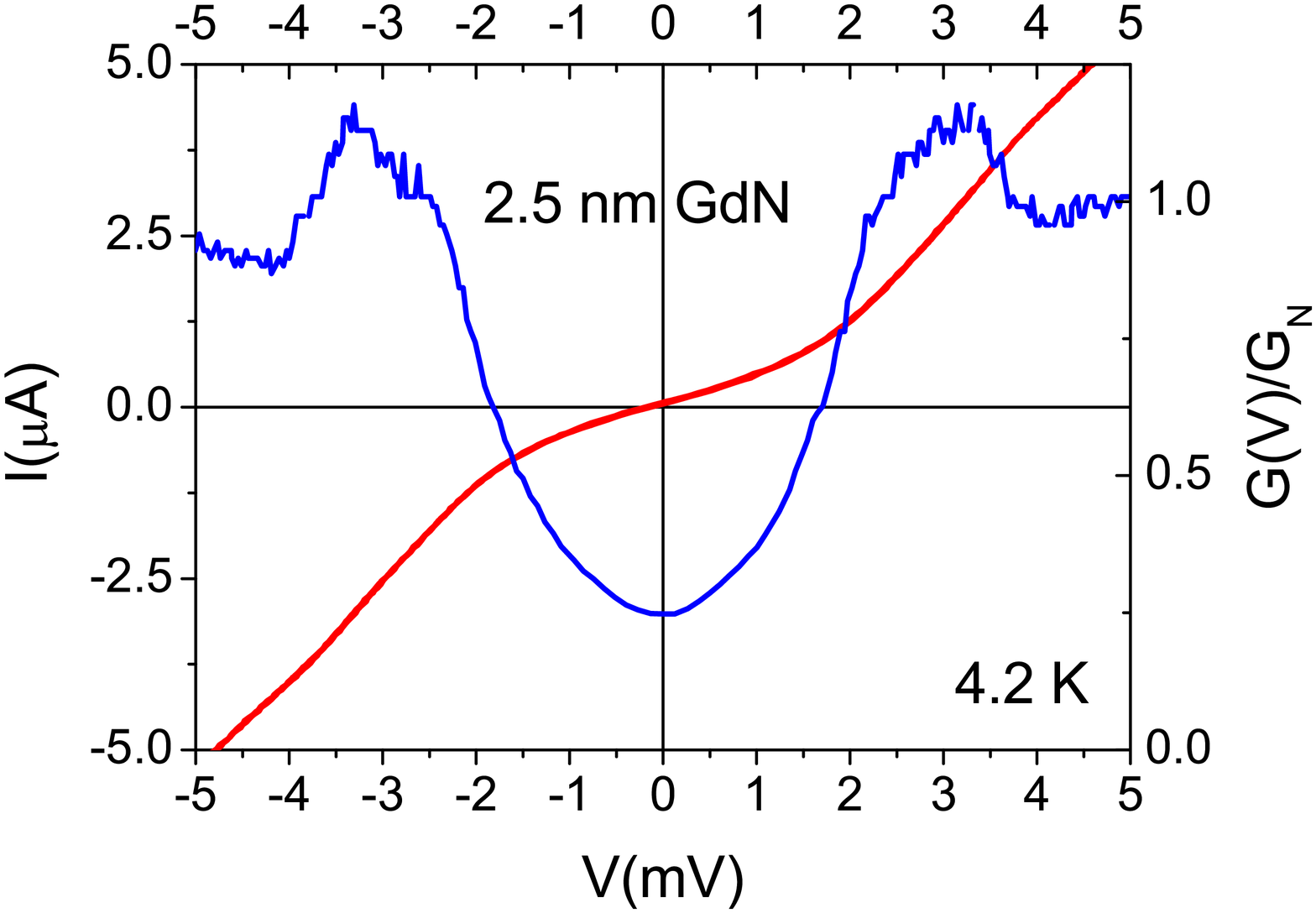}\\
 \centering
 \includegraphics[width= 6 cm]{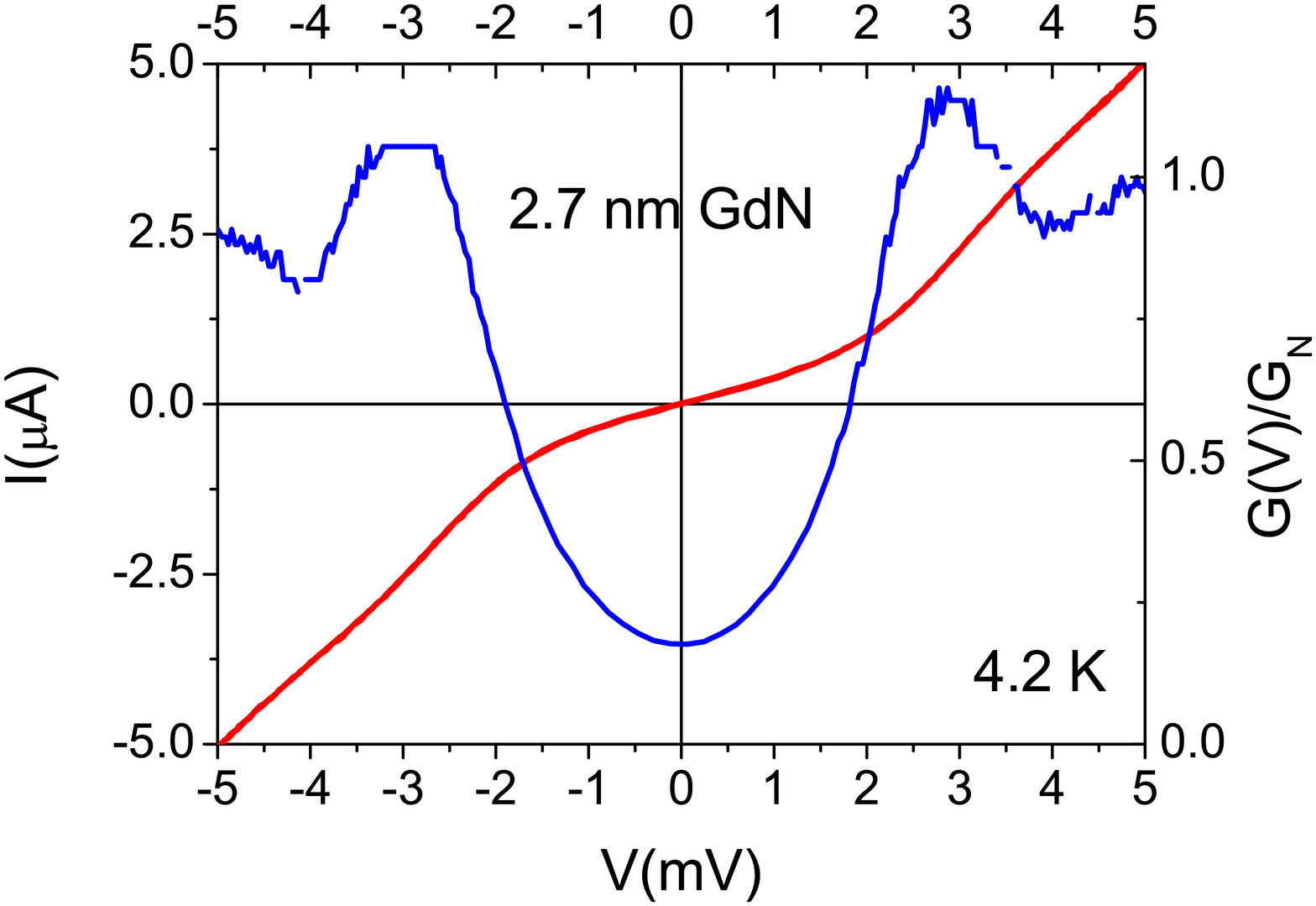}
&

 \includegraphics[width= 6 cm]{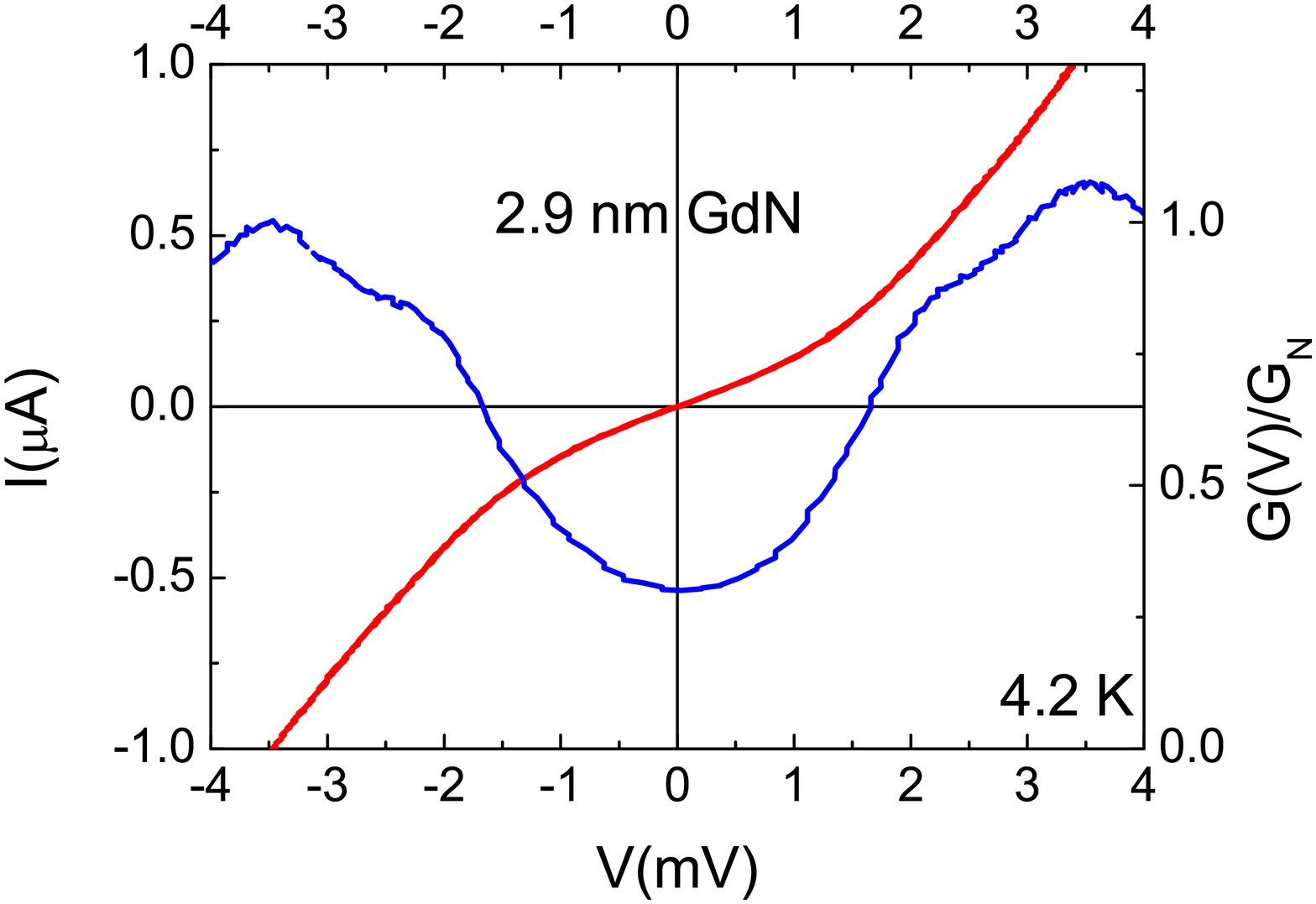}\\

\end{tabular}
\caption{SERIES-3 (23434):I-V and normalized conductance spectra
$G(V)/G_N$ of the NbN(50 nm)-GdN($t$)-NbN(50 nm) spin-filter
device with GdN thickness in the range 1.0-2.9 nm. All the tunnel
junctions were prepared from the trilayer stack deposited at the
same time.}
\end{figure}

\end{widetext}
\end{widetext}
\end{widetext}


\begin{thebibliography}{100}
\bibitem{note}
All the conductance spectra shown in Fig. 2 are obtained by
numerically differentiating measured I-V curves in a closed-cycle
refrigirator. The conductance spectra shown in Fig. 3(a,b,c) are
obtained directly by lock-in technique in a custom made dip-stick.
We have found that the conductance spectra measured in the
closed-cycle refrigirator showed slightly redued gap compared to
the measurements done in the dip-stick as the junction gets
coupled to different electromagnetic enviroment.
\end{thebibliography}
\end{document}